\documentclass[%
 reprint,
superscriptaddress,
 amsmath,amssymb,
 aps,
]{revtex4-2}

\usepackage{graphicx}
\usepackage{dcolumn}
\usepackage{bm}

\usepackage{physics}
\usepackage{bm}
\usepackage{mathtools}
\usepackage{physics}
\usepackage[linesnumbered,ruled,lined]{algorithm2e}
\usepackage{algpseudocode}
\SetKwInput{KwInput}{Input}
\SetKwInput{KwOutput}{Output}
\usepackage[export]{adjustbox}
\usepackage{makecell}
\usepackage{multirow}

\makeatletter
\newcommand{\thealgorithm}{\arabic\algocf@float}
\newcommand{\AlgoCaptionFormat}{}
\newcommand{\SetAlgoCaptionFormat}[1]{\def\AlgoCaptionFormat{#1}}
\renewcommand{\algocf@makecaption@ruled}[2]{%
  \global\sbox\algocf@capbox{\hskip\AlCapHSkip%
    \setlength{\hsize}{\columnwidth}
    \addtolength{\hsize}{-2\AlCapHSkip}
    \vtop{\AlgoCaptionFormat\algocf@captiontext{#1}{#2}}}
}%
\makeatother
\SetAlgoCaptionFormat{\raggedright}

\begin{document}


\title{Algorithm-Oriented Qubit Mapping for Variational Quantum Algorithms}

\author{Yanjun Ji}
\email{quantumji@hotmail.com}
\affiliation{Institute of Computer Architecture and Computer Engineering, University of Stuttgart, Pfaffenwaldring 47, 70569 Stuttgart, Germany}

\author{Xi Chen}
\email{xi.chen@csic.es}
\affiliation{Instituto de Ciencia de Materiales de Madrid (CSIC), Cantoblanco, E-28049 Madrid, Spain}

\author{Ilia Polian}
\email{ilia.polian@informatik.uni-stuttgart.de}
\affiliation{Institute of Computer Architecture and Computer Engineering, University of Stuttgart, Pfaffenwaldring 47, 70569 Stuttgart, Germany}

\author{Yue Ban}
\email{yue.ban@csic.es}
\affiliation{Instituto de Ciencia de Materiales de Madrid (CSIC), Cantoblanco, E-28049 Madrid, Spain}

\date{\today}

\begin{abstract}

Quantum algorithms implemented on near-term devices require qubit mapping due to noise and limited qubit connectivity. In this paper we propose a strategy called algorithm-oriented qubit mapping (AOQMAP) that aims to bridge the gap between exact and scalable mapping methods by utilizing the inherent structure of algorithms. While exact methods provide optimal solutions, they become intractable for large circuits. Scalable methods, like SWAP networks, offer fast solutions but lack optimality. AOQMAP bridges this gap by leveraging algorithmic features and their association with specific device substructures to achieve depth-optimal and scalable solutions. The proposed strategy follows a two stage approach. First, it maps circuits to subtopologies to meet connectivity constraints. Second, it identifies the optimal qubits for execution using a cost function and performs postselection among execution results across subtopologies. Notably, AOQMAP provides both scalable and optimal solutions for variational quantum algorithms with fully connected two qubit interactions on common subtopologies including linear, T-, and H-shaped, minimizing circuit depth. Benchmarking experiments conducted on IBM quantum devices demonstrate significant reductions in gate count and circuit depth compared to Qiskit, Tket, and SWAP network. Specifically, AOQMAP achieves up to an 82\% reduction in circuit depth and an average 138\% increase in success probability. This scalable and algorithm-specific approach holds the potential to optimize a wider range of quantum algorithms.

\end{abstract}

\maketitle


\section{Introduction}

Recent strides in variational quantum algorithms (VQAs)~\cite{cerezo2021variational, bharti2022noisy} have demonstrated considerable potential in solving complex problems, such as combinatorial optimization~\cite{farhi2014quantum} and quantum simulation of materials~\cite{ma2020quantum}, surpassing the efficiency of classical algorithms. However, existing constraints of quantum processing units (QPUs) pose a significant obstacle to realizing the full capabilities of VQAs. A major challenge is the presence of noise, limited number of qubits, and restricted connectivity between qubits. These constraints impede the scalability and applicability of VQAs in tackling larger and more intricate problems. Efforts to overcome these challenges are critical for unlocking the complete potential of VQAs in practical applications.
Specifically, the execution of quantum algorithms requires compilation before being deployed on a quantum device, which involves addressing connectivity constraints by introducing $\mathrm{SWAP}$ gates and decomposing algorithms into native hardware basis gates. Moreover, optimizing circuits to minimize the influence of noise is crucial to ensure accurate performance of algorithms~\cite{hu2022greedy}.

A crucial step in compilation is mapping logical qubits to physical qubits available on quantum devices, a complex problem commonly referred to as qubit mapping~\cite{li2019tackling, cowtan2019qubit, zhu2022an, niu2020hardware, zhang2021time}. The main objective of qubit mapping is to minimize the number of inserted $\mathrm{SWAP}$ gates or circuit depth and to maximize circuit fidelity. This problem is identified as NP-hard~\cite{siraichi2018qubit}, underscoring the necessity for efficient and effective methods to tackle it.
Qubit mapping can be expressed as a mathematical optimization problem and solved using constraint satisfaction techniques. These approaches are called exact methods and have been investigated in various studies~\cite{bhattacharjee2017depth, ShafaeiSP14, murali2019noise, siraichi2019qubit, tan2020optimal}. While exact methods provide high quality and stable solutions, their compilation time increases exponentially with problem size. In contrast, heuristic approaches~\cite{siraichi2019qubit, zulehner2018efficient, childs19, li2020qubits, niu2020hardware, murali2019noise} prioritize efficiency by providing fast solutions without guaranteeing optimality. Another approach involves constructing swap layers~\cite{harrigan2021quantum, kivlichan2018quantum, o2019generalized, weidenfeller2022scaling, hashim22optimized} aimed at providing scalable solutions. However, similar to heuristic methods, optimality is not guaranteed.

This paper proposes an efficient, two-stage approach to qubit mapping, prioritizing both optimality and scalability. In the first stage, quantum algorithm is mapped to the target QPU's subtopologies to address connectivity constraints. Here, we provide optimal and scalable solutions with minimal circuit depth for VQAs with all-to-all connected two qubit interactions on common subtopologies including linear, T-, and H-shaped. For VQAs with partially connected two qubit interactions, solutions can be obtained by optimizing the initial qubit order to minimize CX gate count. These solutions are applicable to various noisy intermediate scale quantum (NISQ) devices, such as Google's Sycamore, IBM's QPUs, and Rigetti's processors. After addressing connectivity constraints, the second stage focuses on determining optimal qubits for execution and postselecting the execution outcomes associated with different subtopologies. Each subtopology-adapted circuit is mapped onto the QPU taking into account real-time noise characteristics. After executing all subtopology-adapted circuits on the quantum device, the results corresponding to different subtopology types are postselected to ensure high performance of algorithms.

Unlike conventional methods that usually map predefined circuits, our approach begins directly from the algorithm's Hamiltonian, allowing for optimization during the conversion process from Hamiltonian to circuit representation. This codesign strategy generates mappings tailored to the algorithm's structure and hardware constraints. This methodology promises not only optimality but also enhanced adaptability and scalability in tackling the qubit mapping challenge. Benchmarks on six IBM QPUs with 7, 27, and 127 qubits demonstrate significant performance gains. In particular, we achieve an average reduction of 31\% (up to 82\%) in circuit depth and an average increase of 138\% in success probability compared to Qiskit~\cite{qiskit}, Tket~\cite{sivarajah2020tket}, and SWAP network~\cite{harrigan2021quantum}.

The paper is structured as follows. Section~\ref{sec:backg} provides background on NISQ devices and the qubit mapping problem. Section~\ref{sec:method} details our proposed approach, including the identification of subtopologies, analysis of routing solutions, introduction of mapping strategies, and discussion on optimality and scalability. Section~\ref{sec:applic} demonstrates practical applications of our method to VQAs, in particular the quantum approximate optimization algorithm (QAOA) and variational quantum eigensolver (VQE), as well as the broader implications for executing other algorithms on various quantum devices. Section~\ref{sec:bench} presents benchmarking results comparing our technique to existing methods using both simulators and real quantum hardware. Finally, Sec.~\ref{sec:concl} concludes.

\section{Background}
\label{sec:backg}

\subsection{Near-term quantum devices}

\begin{figure*}[tb]
\centering
\includegraphics[width=0.25\linewidth]{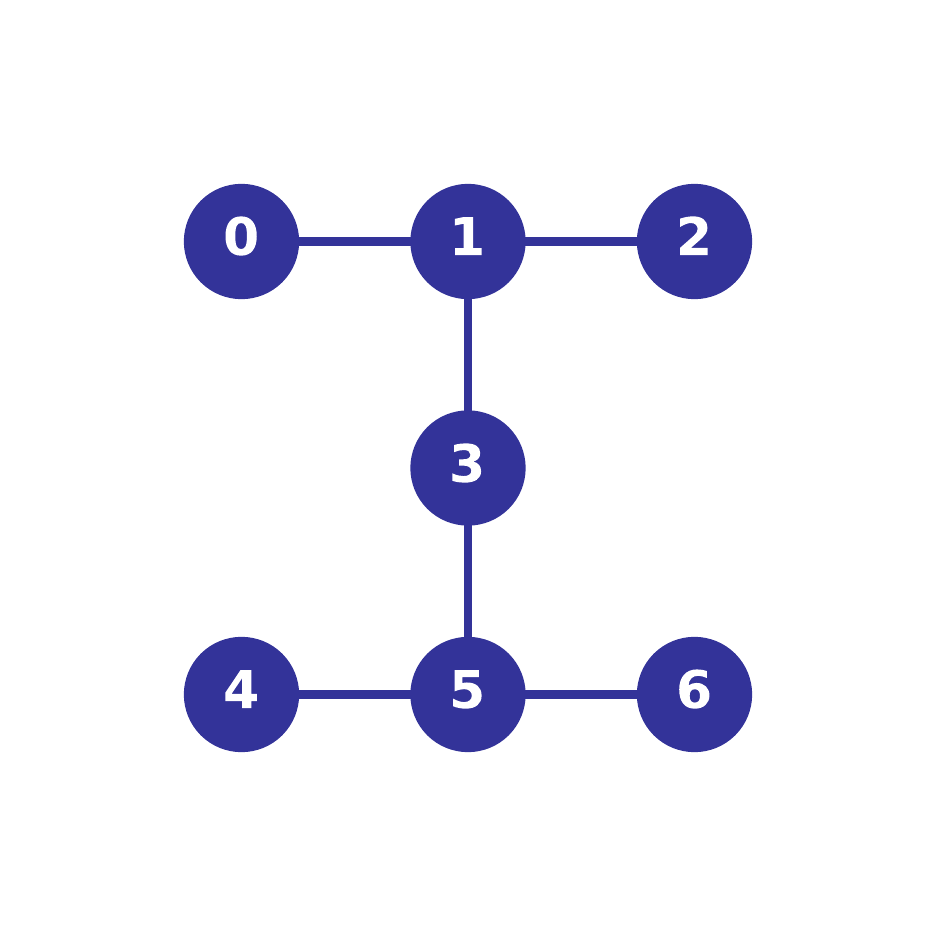}
  \includegraphics[width=0.43\linewidth]{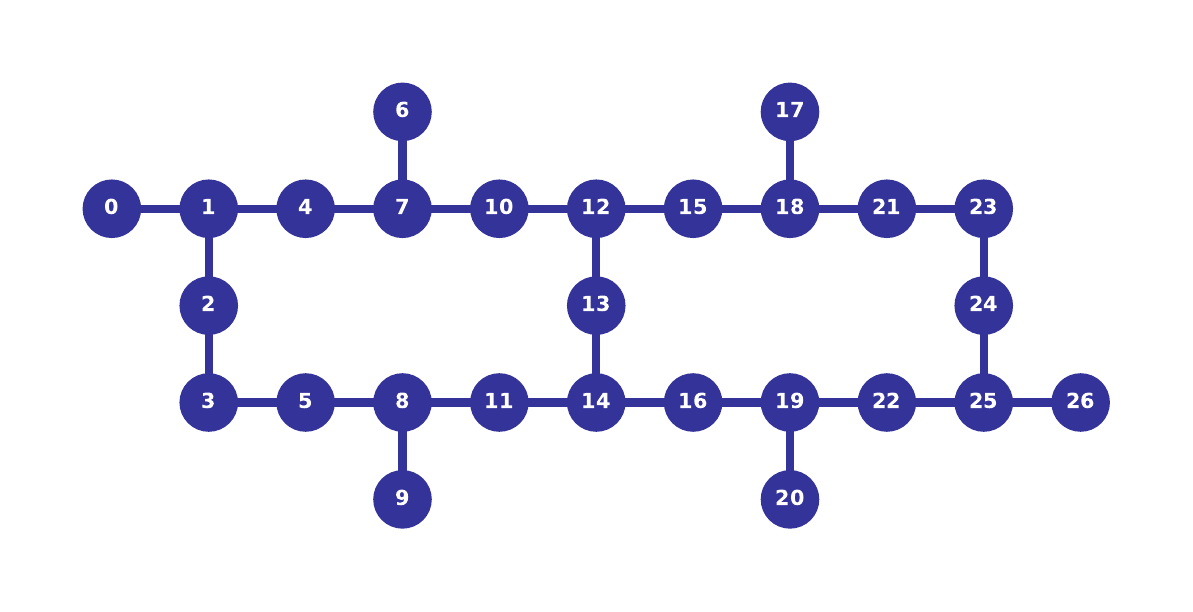}
  \includegraphics[width=0.3\linewidth]{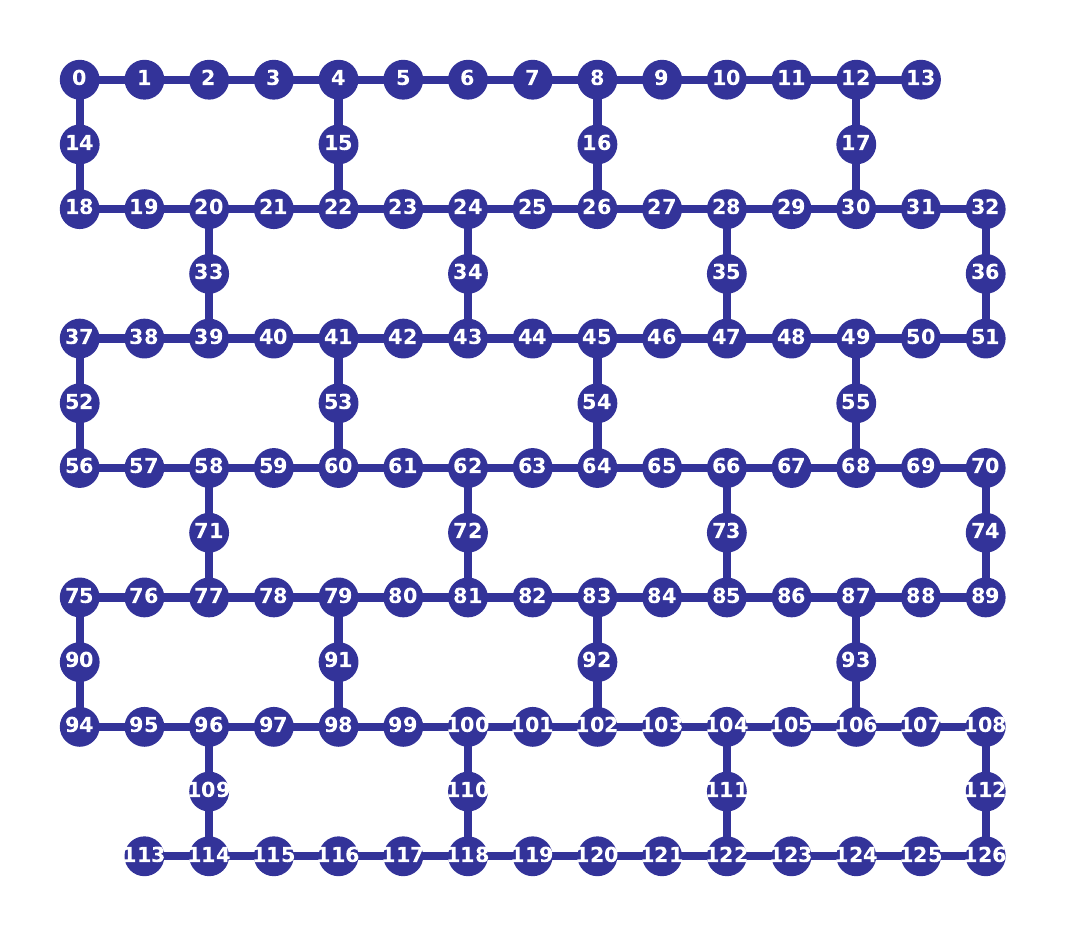}
    \put(-495,105){{\textbf{(a)}}}
    \put(-372,105){{\textbf{(b)}}}
    \put(-160,120){{\textbf{(c)}}}
    \vspace{-6pt}
  \caption{Topologies of IBM QPUs with (a) 7, (b) 27, and (c) 127 qubits. Circles denote individual qubits. Lines connecting the circles represent available two-qubit gates between corresponding qubit pairs.}
  \label{fig:topology_qpu}
\end{figure*}

Prominent near-term quantum platforms exhibit distinctive characteristics. Superconducting qubits demonstrate increased connectivity that extends beyond nearest neighboring qubits~\cite{rosenberg2020solid, niedzielski2019silicon, rahamim2017double}, while trapped ions showcase relatively long coherence times~\cite{wang2021single}, and photons exhibit low noise~\cite{brida2012an}. In the era of NISQ computing~\cite{preskill2018quantum}, the development of practical algorithms encounters substantial constraints due to the inherent limitations of quantum hardware.
For instance, the 7-, 27-, and 127-qubit IBM QPUs illustrated in {Fig.~\ref{fig:topology_qpu}}, exhibit restricted connectivity. Specifically, each qubit can only directly interact with up to three neighboring qubits. Additionally, the noise in quantum systems varies over time, leading to temporal fluctuations in errors. These hardware limitations pose challenges for implementing quantum algorithms that rely on long-range qubit interactions or high precision.
Current endeavors concentrate on enhancing the control of qubits by improving coherence times, elevating gate fidelities, and expanding qubit connectivity. Moreover, compiling quantum algorithms to reduce circuit depth and mitigate errors is essential to fully realizing the potential of these devices.

To address the limitations of NISQ devices, researchers have developed VQAs~\cite{cerezo2021variational}, which are hybrid algorithms combining classical optimization techniques with quantum resources. However, implementing VQAs on NISQ devices presents challenges. One significant hurdle is selecting an appropriate classical optimization algorithm, which can profoundly impact the success of VQAs. Moreover, mapping algorithms onto physical qubits of quantum devices, especially for larger circuits, is difficult. Strategies to tackle these challenges have been developed, such as employing machine learning techniques to optimize parameters~\cite{chandarana2023}, developing new ansatz to capitalize on qubit connectivity~\cite{kandala2017hardware}, and implementing efficient compilation processes~\cite{ji2022calibration, gokhale2020optimized, smith2022summary, gokhale2019partial, shi2019optimized, gokhale2021faster}.
In this study, our focus is on efficient qubit mapping, which is a crucial step in implementing VQAs. We examine the challenges and opportunities associated with qubit mapping and propose an approach to improve scalability and optimality. We assess the effectiveness of our approach and demonstrate its application on real quantum hardware.

\subsection{Qubit mapping problem}

The primary objective of qubit mapping is to minimize errors inherent in the implementation of algorithms on QPUs. This task is critical in mitigating errors arising from the noisy nature of qubits, particularly in the context of high error rates associated with two-qubit gates such as $\mathrm{CNOT}$ or $\mathrm{CX}$ that can significantly impact the overall performance of algorithms.
Optimizing the qubit mapping process requires a careful balance between two crucial factors. First, the insertion of $\mathrm{SWAP}$ gates introduces errors when connecting two qubits that are not directly linked, as the implementation of a SWAP gate necessitates three $\mathrm{CX}$ gates. Second, the quality of qubits and qubit pairs can vary, necessitating the identification of the most suitable qubits for circuit execution. While searching for solutions of qubit mapping on a specific subtopology can markedly reduce computational complexity compared to exploring solutions across the entire topology, particularly for larger topologies with hundreds of qubits, it's crucial to strike the right balance between the number of $\mathrm{SWAP}$ gates inserted on specific topologies and the quality of qubits on those topologies.

\section{Methodology and preliminaries}
\label{sec:method}

\begin{figure*}[tb]
\centering
  \includegraphics[width=0.9\linewidth]{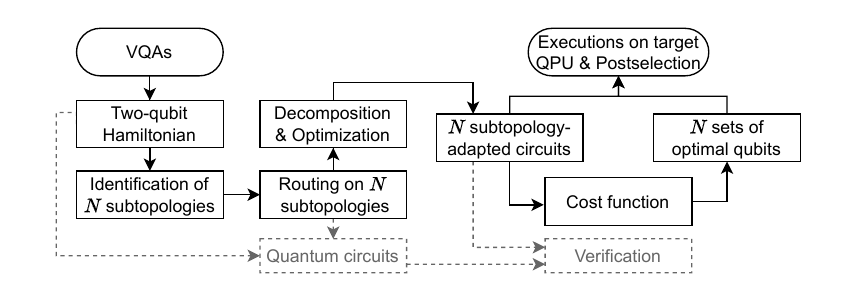}
  \caption{Algorithm-oriented qubit mapping (AOQMAP) flow is outlined for mapping VQAs onto a target QPU. The AOQMAP approach differs from traditional qubit mapping methods in that it starts from a two-qubit Hamiltonian, rather than a predefined circuit. The process of AOQMAP involves two main steps: (1) adapting the VQA to $N$ subtopologies of target QPU; and (2) selecting $N$ optimal mapping schemes to implement such subtopology-adapted circuits, followed by executions on the QPU and postselection. The adaptation process ensures algorithm excitability, while the mapping process guarantees the use of high-quality qubits for execution. By first adapting circuits and then choosing an optimal mapping scheme, AOQMAP minimizes errors and optimizes algorithm performance. Finally, the $N$ results produced by the QPU, each associated with a different subtopology, are postselected by minimizing the expectation value of the problem Hamiltonian. Dashed lines represent the circuit verification process, which is optional but recommended for small circuits. A more detailed discussion of circuit verification can be found in Appendix~\ref{appen:verif}.
  }
  \label{fig:aoqmap_flow}
\end{figure*}

This section details our methodology. Figure~\ref{fig:aoqmap_flow} illustrates the algorithm oriented qubit mapping (AOQMAP) flow. The process begins with decomposing the input VQA into a two-qubit Hamiltonian. After identifying $N$ subtopologies within the target QPU, the Hamiltonian is routed onto subtopologies by introducing $\mathrm{SWAP}$ gates, ensuring compliance with hardware connectivity constraints. Each of these routed circuits is then decomposed into native basis gates of the target QPU, followed by an optimization step to reduce redundant gates in circuit and improve fidelity. In this work, the decomposition and optimization process is performed using Qiskit transpiler~\cite{qiskit} with default settings (optimization level 1). Finally, a cost function is utilized to select the optimal set of qubits for execution.
After executing $N$ circuits on the target QPU, a postselection process can be applied to determine results associated with different subtopologies by minimizing the expectation value of the problem Hamiltonian, as will be detailed later.
Verification is a technique used to protect against implementation errors that can occur during processes such as routing, decomposition, and optimization. While optional, the verification process, represented by dashed lines, is particularly recommended for small circuits.
For verification, a reference circuit is constructed directly from the two-qubit Hamiltonian, matching parameters and gate sequences of the routed circuit. The Hellinger distance between output distributions of the reference and routed circuits can be employed to assess correctness. Additional details on the verification process are provided in Appendix~\ref{appen:verif}.
This comprehensive flow guarantees effective adaptation and optimization of quantum circuits within constraints imposed by target hardware topology.

Below, we detail each step. Beginning with the identification of three common subtopologies, we present optimal and scalable solutions for routing VQAs on these topologies, focusing on fully connected two-qubit interactions. Additionally, we employ a qubit selection strategy to map subtopology-adapted circuits onto high-quality qubits of the QPU and introduce the postselection process to determine the results obtained from different types of subtopologies. Finally, we conclude by analyzing the optimality and scalability of our proposed methods.

\subsection{Identification of subtopologies\label{subsec:ident_of_subtop}}

\begin{figure}[tb]
\centering
  \includegraphics[width=0.9\linewidth]{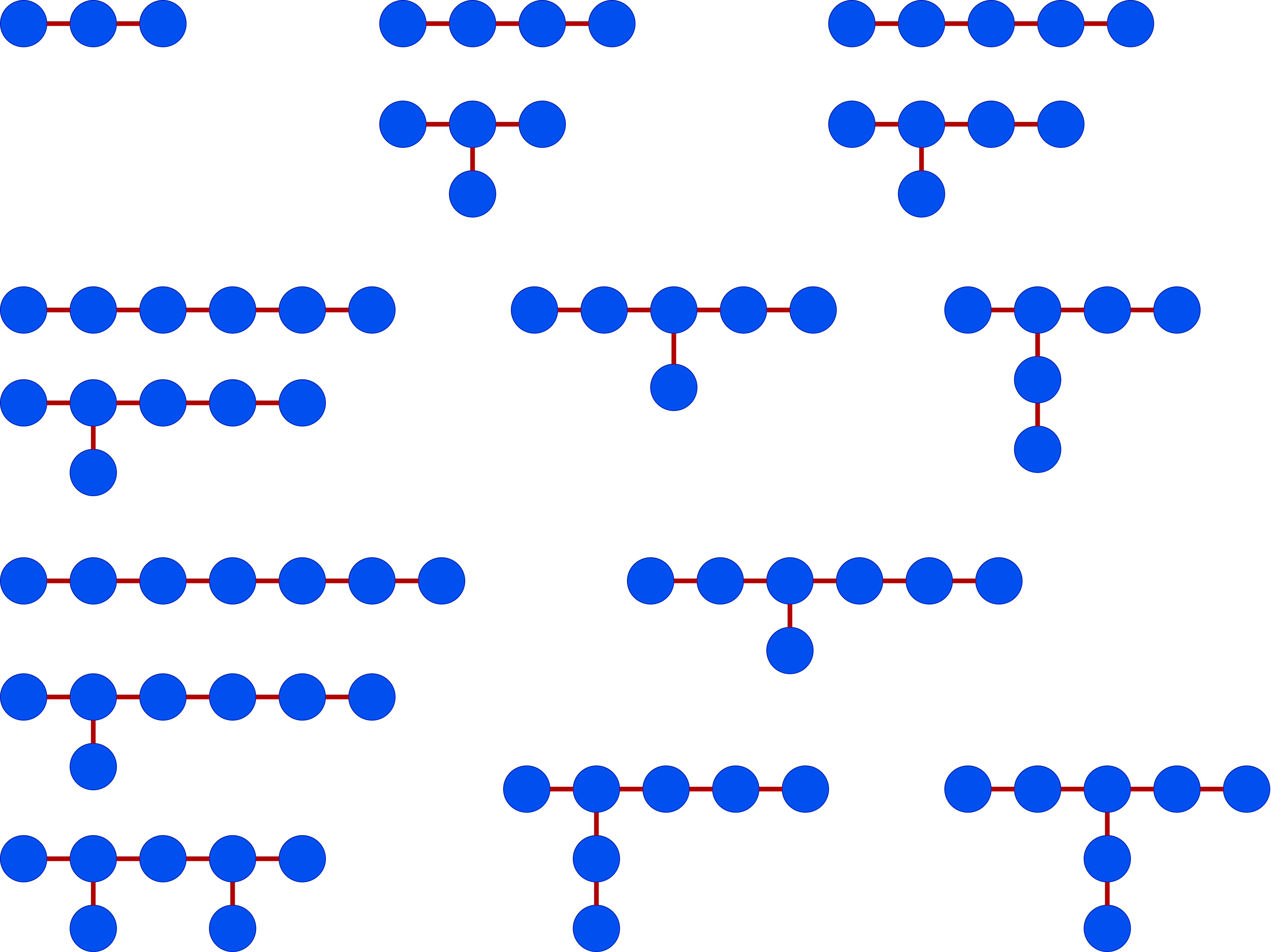}
  \put(-237,160){\textbf{(a)}}
  \put(-170,160){\textbf{(b)}}
  \put(-170,142){\textbf{(c)}}
  \put(-93,160){\textbf{(d)}}
  \put(-93,142){\textbf{(e)}}
  \put(-237,110){\textbf{(f)}}
  \put(-237,95){\textbf{(g)}}
  \put(-148,110){\textbf{(h)}}
  \put(-71,110){\textbf{(i)}}
  \put(-235,62){\textbf{(j)}}
  \put(-237,42){\textbf{(k)}}
  \put(-235,15){\textbf{(l)}}
  \put(-132,62){\textbf{(m)}}
  \put(-149,27){\textbf{(n)}}
  \put(-72,27){\textbf{(o)}}
  \vspace{6pt}
  \caption{All possible subtopologies within the heavy-hex topology for varying numbers of qubits, ranging from 3 to 7. The most common subtopology on a 27-qubit topology (Fig.~\ref{fig:topology_qpu}(b)) is linear, followed by T-shaped and then H-shaped.}
  \label{fig:layout}
\end{figure}
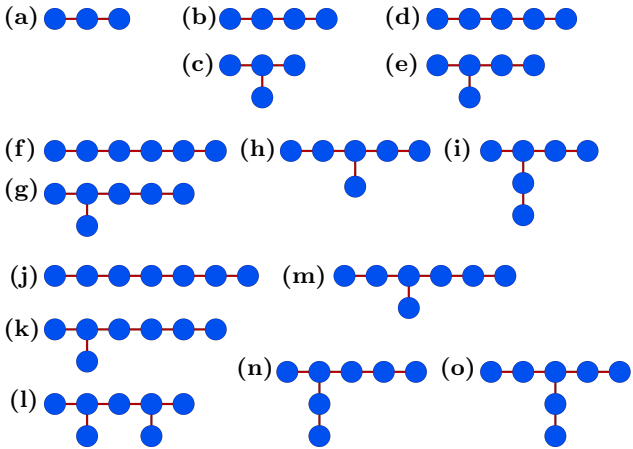

We identify $N=3$ prevalent subtopologies within IBM QPUs: linear, T-, and H-shaped configurations. Given the NP-hard nature of qubit mapping, we focus on symmetric subtopologies to facilitate the development of optimal and scalable solutions. Linear configurations serve as a fundamental topology, while T- and H-shaped topologies (formally defined in Sec.~\ref{subsec:subtop_circ_adapt}) represent the simplest symmetric extensions.
Moreover, to achieve optimality and scalability, we will employ exact methods for small-scale problems and then develop scalable approaches (see Sec.~\ref{subsec:subtop_circ_adapt}). The symmetric T- and H-shaped subtopologies enable efficient analysis of solutions and offer potential for extending these solutions to other topologies. Future research will explore these extensions.
Additionally, our analysis of 27-qubit IBM QPUs reveals that these subtopologies are dominant for the problem sizes considered. Specifically, we exhaustively identify all possible subtopologies of up to seven qubits and calculate their corresponding layouts within the QPU using mapomatic~\cite{nation2023suppressing}. Figure~\ref{fig:layout} illustrates these subtopologies, and Table~\ref{tab:num_subtopologies} summarizes the numbers of their corresponding layouts. As shown in Table~\ref{tab:num_subtopologies}, linear topologies exhibit the highest number of layouts, followed by T- and H-shaped configurations.

\begin{table}[tb]
 \caption{Number of layouts within a 27-qubit QPU that enable the execution of circuits satisfying the connectivity constraints of subtopologies shown in Figs.~\ref{fig:layout}(a-o).}
 \centering
 \label{tab:num_subtopologies}
 \begin{ruledtabular}
  \begin{tabular}{@{}llllll@{}}
    Topology & 3 & 4 & 5 & 6 & 7 \\
    \hline
    Linear  & 74 (a) & 80 (b) & 100 (d) & 104 (f) & 132 (j) \\
    T-shaped  &  & 48 (c) & 36 (e) & 64 (g) & 48 (k) \\
    H-shaped  &  &  &  &  & 56 (l) \\
    T-variant  &  &  &  & 24 (h-i) & 44 (m-n), 12 (o) \\
  \end{tabular}
\end{ruledtabular}
\end{table}

To identify the optimal subtopology among the three candidates, a postselection strategy is employed. This approach is essential when the optimal subtopology is not immediately evident. The cost function, which will be detailed in Sec.~\ref{subsec:mapping_sub_circ}, incorporates gate fidelity to select the optimal set of qubits for executing subtopology-adapted circuits. In addition to enabling the selection of qubits for a given topology, this cost function allows for the distinction between various types of subtopologies. However, each subtopology exhibits unique crosstalk effects that are not accounted for in the calibration data of IBM QPUs.
Furthermore, as will be demonstrated subsequently, increased qubit connectivity reduces the number of two-qubit gates required but increases circuit depth, creating a trade-off between circuit fidelity and execution time. The cost function based solely on gate fidelity is insufficient for accurately assessing the quality of a circuit.
To overcome these limitations, a postselection procedure is performed after executing circuits corresponding to each subtopology on QPUs. The criterion for postselection is based on minimizing the expectation value of the problem Hamiltonian, which allows us to identify the most effective subtopology. This will be investigated in Sec.~\ref{subsec:demons_on_ibm_quant}. The method is also demonstrated by applying AOQMAP to a digitized counterdiabatic quantum optimization algorithm in Ref.~\cite{ji2023improving}.

\subsection{Subtopology-aware circuit adaptation}
\label{subsec:subtop_circ_adapt}

This section introduces the methodology for subtopology aware circuit adaptation on NISQ devices. We focus on linear, T-, and H-shaped configurations, and develop strategies for adapting VQAs to each subtopology.

In this study, we concentrate on the QAOA applied to dense portfolio optimization problems, where each qubit necessitates interaction with all other qubits. As detailed in Ref.~\cite{brandhofer2023benchmarking}, the problem Hamiltonian for $n$ asset portfolio optimization is expressed as
\begin{equation}
 {H_c} = \sum_{i=1}^{n-1}\sum_{j=i+1}^{n} c_{ij} {Z_iZ_j} + \sum_{i=1}^{n} c_{i} {Z_i} + c_0.
\label{eq:qaoa_prob_hamil}
\end{equation}
Here, ${Z_i Z_j}$ represents ${ZZ}$ interaction between qubits $i$ and $j$, defined by $ {ZZ}(\theta) = e^{-i \frac{\theta}{2} {Z}\otimes {Z}}$ with rotation angle $\theta$. ${Z_i}$ denotes Pauli ${Z}$ operator acting on qubit $i$. The coefficients $c_{ij}$, $c_i$, and $c_0$ are real numbers determined by parameters in the portfolio optimization problem, where $c_0$ is a constant term. The coefficients $c_{ij}$ and $c_i$ are defined as follows
\begin{align}
    &c_{i,j}=\frac{\lambda}{2}(q\sigma_{ij}+A),\\
    &c_i=\frac{\lambda}{2} \left[A (2B-n)+(1-q)\mu_i-q \sum_{j=1}^n \sigma_{ij}\right],
\end{align}
where $\lambda$ is the global scaling factor, $q$ represents risk preference, $\sigma_{ij}$ is the covariance between assets $i$ and $j$, $A$ is the penalty factor, $B$ is the number of assets to be selected, and $\mu_i$ is the expected return of asset $i$.

We consider the mixing operator in QAOA with the form
\begin{equation}
{H_m} = \sum_{i=1}^{n} {X_i},
\label{eq:qaoa_mixer}
\end{equation}
where ${X_i}$ is Pauli ${X}$ operator acting on qubit $i$. The QAOA commences by selecting an initial state, denoted by $\ket{\bm{\psi_0}}$, which is the eigenstate of mixer Hamiltonian ${H_m}$. Subsequently, problem and mixer Hamiltonians are applied alternately to $\ket{\bm{\psi_0}}$ in a parameterized quantum circuit of depth $p$. This yields a quantum state
\begin{equation}
\ket{\bm{\psi_{\gamma, \beta}}} = \prod_{k=1}^{p} e^{-i{\beta}_k {H_m}} e^{-i{\gamma}_k {H_c}} \ket{\bm{\psi_0}},
\label{eq:qaoa_psi_p}
\end{equation}
where $\bm{\beta}$ and $\bm{\gamma}$ are $p$-dimensional vectors of rotation angles that control the evaluation of $ {H_c}$ and $ {H_m}$ in each layer $k$. The $2p$ parameters $(\bm{\gamma}, \bm{\beta})$ can be optimized using a classical solver to minimize the expectation value $\bra{\bm{\psi_{\gamma, \beta}}} {H_c}\ket{\bm{\psi_{\gamma, \beta}}}$. This tuning process adjusts $\ket{\bm{\psi_{\gamma, \beta}}}$ to approximate the ground state of $ {H_c}$, offering a well-approximated solution to the optimization problem encoded in $ {H_c}$.

To achieve optimal qubit mapping solutions, a comprehensive analysis of problem and mixer Hamiltonians described in Eqs.~\eqref{eq:qaoa_prob_hamil} and \eqref{eq:qaoa_mixer} is essential. Notable observations guide our approach:
(i) The two-qubit gates in Eq.~\eqref{eq:qaoa_prob_hamil} exclusively involve $ {ZZ}$ interactions. Their summation form implies that the order of application does not affect the outcome, allowing flexibility in mapping $ {H_c}$; 
(ii) The single rotation gates $\mathrm{R_Z}$ and $\mathrm{R_X}$ in $ {H_c}$ and $ {H_m}$ can be independently applied to each qubit without affecting others. Hence, qubit mapping is unnecessary for these gates; 
(iii) For high depths, as depicted in Eq.~\eqref{eq:qaoa_psi_p}, a fixed gate arrangement for two-qubit gates is required at each depth to maintain an equivalent $ {H_c}$. However, since all $\mathrm{ZZ}$ gates commute with each other, they can be assigned in each depth with an arbitrary gate arrangement. This implies that the $\mathrm{ZZ}$ gate arrangement needed for implementing QAOA at each depth can be arbitrary, providing additional flexibility in qubit mapping. By leveraging these observations, we can formulate a tailored mapping method for quantum algorithms to optimize their performance, minimizing the number of $\mathrm{CX}$ gates, reducing circuit depth, and thereby maximizing the algorithm's overall performance.

In the upcoming sections, we explore qubit mapping solutions on three subtopologies, emphasizing QAOA with all-to-all connected two-qubit gate interactions. Although our analysis focuses on this specific instance of VQAs, the implications extend to algorithms with analogous features. Specifically, we underscore the relevance of our findings to Hamiltonian with partially connected interactions, VQE, and other NISQ devices in Sec.~\ref{sec:applic}.

\subsubsection{Linear subtopology}

In quantum computing, linear subtopology arranges qubits sequentially in a one-dimensional configuration, forming a linear chain or arrangement. To determine optimal mapping on this subtopology, we initially employ an exact method aimed at minimizing the circuit depth~\cite{tan2020optimal}. However, instead of mapping the entire algorithm, we focus solely on mapping the $\mathrm{ZZ}$ gates in Eq.~\eqref{eq:qaoa_prob_hamil}, treating them as single entities and avoiding the decomposition into two $\mathrm{CX}$ gates to streamline the mapping process. Additionally, we narrow our attention to the first QAOA depth, significantly reducing the effort required to identify a solution.

\begin{figure}[tb]
    \centering
    \includegraphics[width=\linewidth]{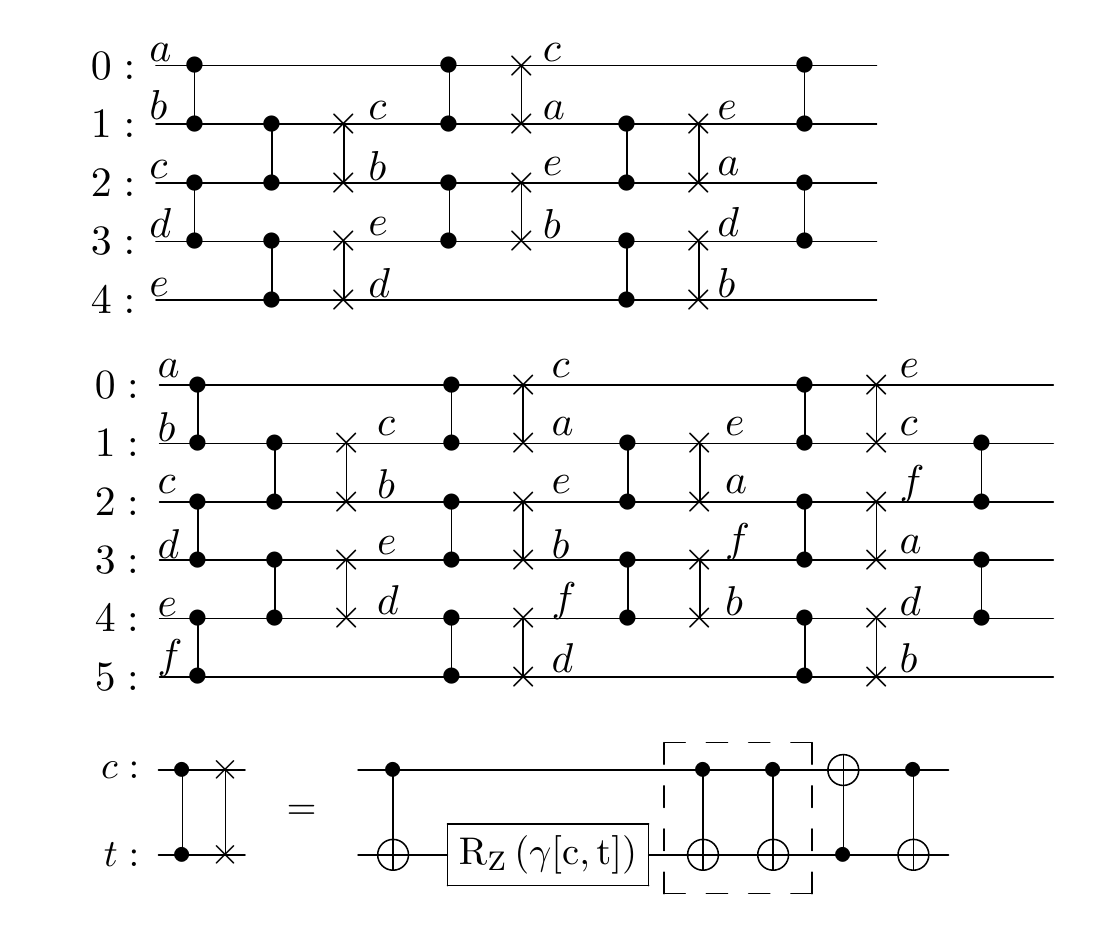}
    \put(-240,200){{\textbf{(a)}}}
     \put(-240,130){{\textbf{(b)}}}
     \put(-238,42){{\textbf{(c)}}}
     \vspace{-6pt}
    \caption{Routing solutions of $\mathrm{ZZ}$ gates in QAOA at depth $p=1$ with (a) five and (b) six qubits. (c) $\mathrm{ZZ}$ gate on qubit pair $(c,t)$ with rotation angle $\gamma[c,t]$, followed by a $\mathrm{SWAP}$ gate, with the box showing $\mathrm{CX}$ gate cancellation between $\mathrm{ZZ}$ and $\mathrm{SWAP}$ gates.}\label{fig:qubit_mapping_linear}
\end{figure}

\begin{algorithm}[htb]
	\caption{AOQMAP for QAOA on linear subtopology} 
	\label{alg:aoqmap_qaoa_linear}
	\DontPrintSemicolon
	\KwInput{Number of qubits $n$, QAOA depth $p$, Parameters $\bm{\gamma}[c,t]$, $\bm{\alpha}[i]$, and $\bm{\beta}[j]$ of gates $\rm{ZZ}$ on qubit pair $(c,t)$, $\rm{R_{Z}}$ on qubit $i$, and $\rm{R_{X}}$ on qubit $j$, respectively, where $c,t,i,j \in \{0, ..., n-1\}$ and $c<t$}
	\KwOutput{Circuit satisfying connectivity constraints}
        \SetKwFunction{ApplyZZGate}{ApplyZZGate}
	\SetKwFunction{ApplyZZSWAPGate}{ApplyZZSWAPGate}
        \SetKwProg{Fn}{Function}{}{end}
        \Fn{\ApplyZZGate{$O$, $i$, $j$}}{
          Apply $\rm{ZZ}$$(\bm{\gamma}$$[O[i], O[j]])$ on ($i$, $j$)
        }
        \SetKwFunction{ApplyZZSWAPGate}{ApplyZZSWAPGate}
        \Fn{\ApplyZZSWAPGate{$O$, $i$, $j$}}{
        Apply $\rm{ZZ}$$(\bm{\gamma}$$[O[i], O[j]])$ on ($i$, $j$)\;
            Apply $\rm{SWAP}$ on ($i$, $j$)\;
            $O[i] \leftrightarrow O[j]$ \tcp{Exchange qubit order}
        }
        $O \gets \{0, 1, ..., n-1\}$\tcp{Initialize the qubit order}
	Prepare the initial state $| 0 \rangle^{\otimes n}$\;
        Apply $\rm{H}^{\otimes n}$\;
        \While{$p>0$}{
		$s \gets 0$\;
            \While{$s < n$}{
            \For{$q \coloneqq 0$ to $n-1$ step 2}{
                \eIf{$s==0$ or $s==n-1$}{
                    \ApplyZZGate{$O$, $q$, $q+1$}
                }{
                    \ApplyZZSWAPGate{$O$, $q$, $q+1$}
                }
            }
            $s \gets s+1$\;
            \If{$s < n$}{
                \For{$q\coloneqq 1$ to $n-1$ step 2}{
                    \eIf{$s==0$ or $s==n-1$}{
                          \ApplyZZGate{$O$, $q$, $q+1$}
                    }{
                          \ApplyZZSWAPGate{$O$, $q$, $q+1$}
                    }
                }
            }
            $s \gets s+1$\;
            }
            \For{$k\coloneqq 0$ \KwTo $n$}{
                Apply $\rm{R_Z}$$(\bm{\alpha}[O[k]])$ on $k$\;
		      Apply $\rm{R_X}$$(\bm{\beta}[O[k]])$ on $k$\;
            }
            $p \gets p-1$\;
        }
     Measure the qubits ($[0, ..., n-1]$ $\rightarrow$ $[O[0], ..., O[n-1]]$)\;
\end{algorithm}

\begin{figure*}[tb]
  \includegraphics[width=\linewidth]{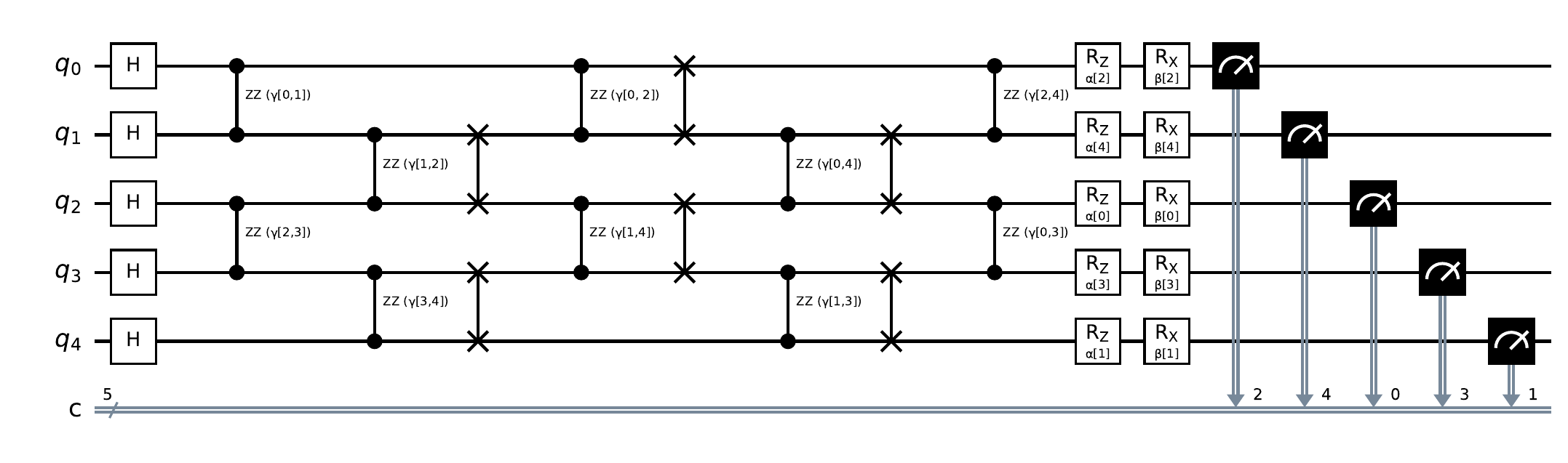}
  \vspace{-9pt}
  \caption{Resulting circuit of five-qubit QAOA for portfolio optimization on a linear subtopology using AOQMAP approach. This circuit is further decomposed and optimized for the target QPU. The optimal mapping scheme is then selected for execution based on the QPU's noise information.}
  \label{fig:5q_qaoa_solution}
\end{figure*}

The routing solution for $\mathrm{ZZ}$ gates in a five-qubit QAOA on linear topology is illustrated in {Fig.~\ref{fig:qubit_mapping_linear}(a)}. Assuming an arbitrary initial qubit order $[a, b, c, d, e]$, after each swap layer, qubit orders evolve as follows: $[a, c, b, e, d]$, $[c, a, e, b, d]$, and $[c, e, a, d, b]$. Since each qubit needs to interact with all other qubits, $\mathrm{ZZ}$ gate acts on following qubit pairs: for qubit $a$, $(a, b)$, $(a, c)$, $(a, d)$, and $(a, e)$; for qubit $b$, $(b, c)$, $(b, d)$, and $(b, e)$; for qubit $c$, $(c, d)$, and $(c, e)$; and for qubit $d$, $(d, e)$. Our analysis reveals that all ten $\mathrm{ZZ}$ gates required in a five-qubit QAOA can be executed on a linear subtopology, regardless of the initial qubit order. This observation is highly beneficial for achieving scalability, as it facilitates straightforward extension to high QAOA depths by repeating these swap layers. Similarly, as depicted in Fig.~\ref{fig:qubit_mapping_linear}(b), the fifteen $\mathrm{ZZ}$ gates required for six-qubit QAOA can be implemented using four swap layers.
We observe that in an $n$-qubit QAOA, $\mathrm{ZZ}$ gates are organized into $n$ layers to minimize circuit depth. The swap layers, excluding the first and last layers, are positioned after each ZZ layer. As shown in Fig.~\ref{fig:qubit_mapping_linear}(c), due to the cancellation of $\mathrm{CX}$ gates between $\mathrm{ZZ}$ and $\mathrm{SWAP}$ gates, each $\mathrm{SWAP}$ gate following a $\mathrm{ZZ}$ gate introduces only one additional $\mathrm{CX}$ gate. The optimal routing solutions of $\mathrm{ZZ}$ gates in QAOA on linear subtopology exhibit a structure similar to the SWAP network presented in Ref.~\cite{harrigan2021quantum}, with the absence of the first and last swap layers. This similarity enables the extension of solutions to an arbitrary number of qubits. The dispensability of the first and last swap layers stems from the flexibility to adjust initial qubit order and measurement order to eliminate $\mathrm{SWAP}$ gates in the first and last ZZ layers, leveraging commutativity of $\mathrm{ZZ}$ and $\mathrm{SWAP}$ gates. Once two-qubit gates are mapped, we can construct the entire QAOA circuit.

Algorithm~\ref{alg:aoqmap_qaoa_linear} presents pseudocode for mapping QAOA on linear subtopology. The reconstruction process starts with an initialized qubit order $O = \{0, 1, ...i, ..., j , n-1\}$ for $n$ qubits. We first prepare an initial state $\ket{\bm{\psi_{0}}}$ by applying a Hadamard gate to each qubit. Then, we apply $\mathrm{ZZ}$ or $\mathrm{ZZ}$-$\mathrm{SWAP}$ gate layer continuously according to the solutions presented in Fig.~\ref{fig:qubit_mapping_linear}. The order of qubits $i$ and $j$ is exchanged only when a $\mathrm{SWAP}$ gate is applied to qubit pair $(i,j)$. To improve algorithm efficiency, it is crucial to optimize $\mathrm{ZZ}$ and $\mathrm{ZZ}$-$\mathrm{SWAP}$ gates using strategies such as gate cancellation or pulse level optimization techniques~\cite{earnest2021pulse,ji2023optimizing}. Finally, $\mathrm{R_Z}$ gates in problem Hamiltonian and  $\mathrm{R_X}$ gates in mixer Hamiltonian are implemented on qubits with appropriate parameters. Figure~\ref{fig:5q_qaoa_solution} shows the resulting circuit of a five-qubit QAOA at depth $p=1$. All $\mathrm{ZZ}$, $\mathrm{R_Z}$, and $\mathrm{R_X}$ gates are executed according to the current qubit order, followed by measurement of qubits.

As demonstrated, an arbitrary qubit order can implement all required $\mathrm{ZZ}$ gates in QAOA with $n$ qubits using $n-2$ swap layers. A solution of depth $p$ can be obtained by repeating $p$ times swap layers in depth $p=1$ circuit.
Inserting these swap layers consecutively, the initial order returns after $2n$ swap layers. Since each QAOA depth introduces $n-2$ swap layers, circuit repeats every $\textrm{lcm}(2n, n-2)/(n-2)$ depths, where $\textrm{lcm}(2n, n-2)$ is the least common multiple of $2n$ and $n-2$.
For odd numbers of qubits, one repetition of swap layers in depth $p=1$ circuit demonstrates symmetry, and at depth two, the final qubit order returns directly to the initial qubit order. For even numbers of qubits, one repetition of such swap layers demonstrates alternating odd and even-numbered swap layers, resulting in circuits that repeat a subcircuit with the same gate arrangement every $\textrm{lcm}(2n, n-2)/(n-2)$ layers. Since all $\mathrm{ZZ}$ gates commute with each other, these two constructions of high depth solutions are equivalent. However, we can utilize mirror symmetry of swap layers to construct high depth circuits for even numbers of qubits and obtain circuits repeating every two depths. Additionally, mirror symmetry can also be utilized to construct algorithms with partially connected two qubit gates, which we discuss in Sec.~\ref{sec:applic}.

\subsubsection{T-shaped subtopology}

\begin{figure*}[tb]
    \centering
    \includegraphics[width=0.7\linewidth]{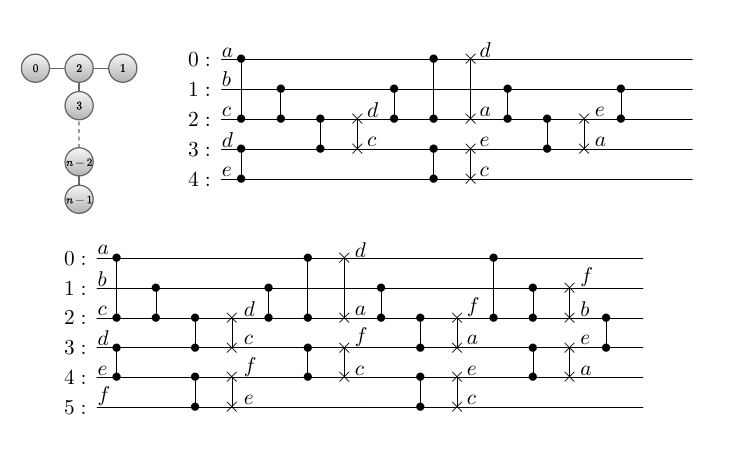}
    \put(-0.712\linewidth,0.384\linewidth){{\textbf{(a)}}}
    \put(-0.552\linewidth,0.384\linewidth){{\textbf{(b)}}}
    \put(-0.67\linewidth,0.2\linewidth){{\textbf{(c)}}}
    \vspace{-15pt}
    \caption{Definition of T-shaped topology and corresponding routing solutions of $\mathrm{ZZ}$ gates in QAOA at depth $p=1$. (a) T-shaped topology defined for $n$ qubits with qubit 2 as the center qubit. The connectivity of this subtopology for $n$ qubits is given by $\{ (0,2), (1,2), (2,3), (3,4), ..., (n-2, n-1) \}$. (b) Five-qubit routing solution. (c) Six-qubit routing solution.}\label{fig:qubit_mapping_tshaped}
\end{figure*}

We now analyze solutions obtained by exact method~\cite{tan2020optimal} for two-qubit gates in QAOA with five and six qubits on T-shaped subtopology. A T-shaped subtopology features a central qubit at the apex of ``T" shape, serving as the primary qubit. The arms of T-shaped topology consist of one or more qubits that are linearly connected to the central qubit. The qubits in arms typically do not have direct interactions with each other. We note that the minimum number of qubits on a T-shaped subtopology is four.
Figure~\ref{fig:qubit_mapping_tshaped}(a) shows the definition of T-shaped topology for $n$ qubits with qubit 2 as the center qubit. Consider an arbitrary initial qubit order $[a, b, c, d, e]$ for the solution of five-qubit, as shown in {Fig.~\ref{fig:qubit_mapping_tshaped}(b)}. Following each swap layer, the qubit order evolves sequentially: $[a, b, d, c, e]$, $[d, b, a, e, c]$, and $[d, b, e, a, c]$. This sequence allows us to execute all required ten $\mathrm{ZZ}$ gates for five-qubit QAOA. Similarly, the four swap layers in six-qubit QAOA are capable of performing the necessary fifteen $\mathrm{ZZ}$ gates, as shown in {Fig.~\ref{fig:qubit_mapping_tshaped}(c)}.

\begin{algorithm}[tb]
\caption{Swap layers on T-shaped subtopology}
\label{alg:swap_layers_tshaped}
\KwInput{Number of qubits $n$, Connectivity of T-shaped topology defined in Fig.~\ref{fig:qubit_mapping_tshaped}(a)}
\KwOutput{List of swap layers $S$}
\Begin{
$n_{\text{odd}} \gets (n-1)-1+(n-1)\bmod 2$\;
$n_{\text{even}} \gets (n-1)-(n-1)\bmod 2$\;
$S \gets$ empty List\;
$j \gets 0$\;
\While{$j < n$}{
$S[j] \gets S[j] \cup [(2,3), (4,5), ..., (n_{\text{odd}}-1, n_{\text{odd}})]$\;
\lIf{$\texttt{++}j \geq n$}{{break}}
$S[j] \gets S[j] \cup [(0,2), (3,4), ..., (n_{\text{even}}-1, n_{\text{even}})]$\;
\lIf{$\texttt{++}j \geq n$}{{break}}
$S[j] \gets S[j] \cup [(2,3), (4,5), ..., (n_{\text{odd}}-1, n_{\text{odd}})]$\;
\lIf{$\texttt{++}j \geq n$}{{break}}
$S[j] \gets S[j] \cup [(1,2), (3,4), ..., (n_{\text{even}}-1, n_{\text{even}})]$\;
\lIf{$\texttt{++}j \geq n$}{{break}}
}
}
\end{algorithm}

Algorithm~\ref{alg:swap_layers_tshaped} outlines procedure for generating $n$ swap layers for $n$-qubit QAOA on T-shaped subtopology. As depicted in Fig.~\ref{fig:qubit_mapping_tshaped}(c), the initial swap layer starts at center qubit 2 and consists of $\mathrm{SWAP}$ gates on qubit pairs $\{(2, 3), (4, 5), ..., (n_{\text{odd}}-1, n_{\text{odd}})\}$ with $n_{\text{odd}} = (n-1)-1+(n-1)\bmod 2$. This layer alternates with another layer that starts from two different qubits in short arms of a T-shaped structure, forming two distinct layers $\{(0, 2), (3, 4), ..., (n_{\text{even}}-1, n_{\text{even}})\}$ and $\{(1, 2), (3, 4), ..., (n_{\text{even}}-1, n_{\text{even}})\}$ with $n_{\text{even}} = (n-1)-(n-1)\bmod 2$.
After constructing $n$ swap layers, all required $\mathrm{ZZ}$ gates can be implemented on connected qubit pairs. Algorithm~\ref{alg:aoqmap_qaoa_tshaped} presents pseudocode for AOQMAP applied in $n$-qubit QAOA at depth $p=1$ on T-shaped subtopology. We begin the process by initializing a qubit order $O = \{0, 1, ..., i, ..., j, n-1\}$ for $n$ qubits, with qubit 2 being center qubit. This initial qubit order enables executing one layer of $\mathrm{ZZ}$ gates. One possible option is gates on qubit pairs of the center qubit and two qubits in short arm, and gates on qubit pairs starting with the first qubit in long arm. For instance, the first ZZ layer in six-qubit QAOA consists of $\{(0,2), (1,2), (3,4)\}$. The second layer starts with center qubits and proceeds to qubits on long arm to form $\mathrm{ZZ}$ gates on qubit pairs $\{(2,3), (4,5)\}$, followed by the first swap layer that introduces new qubit order and provides opportunity for additional gate implementation. We first execute remaining $\mathrm{ZZ}$ gates that are not on qubit pairs of the second swap layer. Then, we implement remaining $\mathrm{ZZ}$ gates on qubit pairs of the second swap layer, followed by $\mathrm{SWAP}$ gates. This process ensures that $\mathrm{ZZ}$ gate is positioned immediately before $\mathrm{SWAP}$ gate on the corresponding qubit pair, thereby enabling CX gate cancellation. We maintain this iterative process until all $\mathrm{ZZ}$ gates are implemented.

\begin{algorithm}[tb]
	\caption{AOQMAP for QAOA at depth $p=1$ on T-shaped subtopology} \label{alg:aoqmap_qaoa_tshaped}
		\DontPrintSemicolon
		\KwIn{Number of qubits $n$, Parameters $\bm{\gamma}[c,t]$, $\bm{\alpha}[i]$, and $\bm{\beta}[j]$ of gates ZZ on qubit pair $(c,t)$, $\rm{R_{Z}}$ on qubit $i$, and $\rm{R_{X}}$ on qubit $j$, respectively, where $c,t,i,j \in \{0, ..., n-1\}$ and $c<t$}
		\KwOut{Circuit satisfying connectivity constraints}
        $O \gets \{0, 1, ..., n-1\}$\tcp{Initialize the qubit order}
        $L \gets$ List of all $\rm{ZZ}$ gates\;
        $E \gets$ List of connected edges on T-shaped subtopology\;
        $E_{O} \gets [(O[r], O[s])~\text{for}~(r, s) \in E]$\;
        $S \gets$ List of $n$ swap layers \tcp{Algorithm~\ref{alg:swap_layers_tshaped}}
	$S_{O} \gets [(O[r], O[s])~\text{for}~(r, s) \in S[k]]$\;
        \SetKwFunction{ApplyZZGate}{ApplyZZGate}
        \SetKwProg{Fn}{Function}{}{end}
        \Fn{\ApplyZZGate{$O$, $c$, $t$}}{
          Apply ZZ($\bm{\gamma}$$[c,t]$) on $(O.\FuncSty{index}(c),O.\FuncSty{index}(t))$\;
        }
        \SetKwFunction{ApplySWAPGate}{ApplySWAPGate}
        \Fn{\ApplySWAPGate{$O$, $i$, $j$}}{
        Apply $\rm{SWAP}$ on ($i,j$)\;
        $O[i] \leftrightarrow O[j]$ \tcp{Exchange qubit order}
        }
        Prepare the initial state $| 0 \rangle^{\otimes n}$\;
        Apply $\rm{H}^{\otimes n}$\;
        \For{$k\coloneqq 0$ to $n$}{
                \lIf{$L$ \rm{is} empty}{continue}
                \ForEach{\rm{ZZ}($\bm{\gamma}$$[c,t]$) $\in L$}{
                    \If{$(c, t) \in E_{O}$ \rm{and} $(c, t) \notin S_{O}$}{
                            \ApplyZZGate{$O$, $c$, $t$}\;
                    $L$.\FuncSty{remove}(\rm{ZZ}($\bm{\gamma}$$[c,t]$))\;
                    }
                }
                    \lIf{$L$ \rm{is} empty}{continue}
                \ForEach{$(i, j) \in S[k]$}{
                        $c, t = O[i], O[j]$\;
                        \If{$c>t$}{$c \leftrightarrow t$ \tcp{Exchange}}
                    \If{\rm{ZZ}($\bm{\gamma}$$[c,t]$) $\in L$}{
                            \ApplyZZGate{$O$, $c$, $t$}\;
                        $L$.\FuncSty{remove}(\rm{ZZ}($\bm{\gamma}$$[c,t]$))\;
                        \lIf{$L$ \rm{is} empty} {continue}
                            \ApplySWAPGate{$O$, $i$, $j$}
                    }
                }
        }
           \ForEach{\rm{SWAP on $(i,j)$ located at the circuit end}}{
              Remove SWAP\;
              $O[i] \leftrightarrow O[j]$\;
           }
           \For{$k \gets 0$ \KwTo $n$}{
              Apply $\rm{R_Z}(\bm{\alpha}[O[k]])$ on $k$\;
              Apply $\rm{R_X}(\bm{\beta}[O[k]])$ on $k$\;
           }
   Measure the qubits ($[0, ..., n-1]$ $\rightarrow$ $[O[0], ..., O[n-1]]$)\;
\end{algorithm}

Similar to linear subtopology, there are two solutions for higher depths: repeating swap layers in depth $p=1$ circuit and leveraging mirror symmetry of swap layers. Alternating swap layers and their mirrors causes gates in QAOA at depth $p$ to invert gates at the previous depth. For commuting gates such as the $\mathrm{ZZ}$ gates used in QAOA, this reversal has no impact on algorithm performance. However, for noncommuting gates, this inversion may suppress Trotter errors~\cite{childs2019faster,faehrmann2022randomizing}, making AOQMAP promising for optimizing other algorithms.
As with the linear topology, T-shaped subtopology also requires $n-2$ swap layers for $n$-qubit QAOA, which we will discuss in more detail in Sec.~\ref{subsec:optim_scala}. It is worth noting that SWAP gates at the end of a circuit can be eliminated, as no remaining two-qubit gates require the introduction of new qubit orders.

The T-shaped subtopology affords enhanced qubit connectivity for the central qubit, reducing required $\mathrm{SWAP}$ gates but increasing circuit depth. This implies that T-shaped subtopology is advantageous when the algorithm's fidelity is a primary factor affecting performance. Conversely, linear subtopology is more effective for larger circuit sizes where schedule duration is paramount.

\subsubsection{H-shaped subtopology}

The H-shaped subtopology shares similarities with T-shaped, but has two central qubits at each end instead of one. The horizontal segments of ``H" serve as a bridge connecting central qubits. To implement an H-shaped subtopology, a minimum of six qubits is required. We define the connectivity of an H-shaped subtopology with center qubits 2 and $n-3$ as $\{(0,2), (1,2), (2,3), ..., (n-4, n-3), (n-3, n-2), (n-3,n-1)\}$ for $n$ qubits.

Algorithm~\ref{alg:swap_layers_hshaped} presents pseudocode for generating $n$ swap layers that enable the implementation of Hamiltonian with fully connected two-qubit gates on H-shaped topology. For odd numbers of qubits, the first and third swap layers differ in initial connections of $(0,2)$ and $(1,2)$, which alternate. Similarly, the second and fourth layers differ in final connections of $(n_{\text{even}}-2, n_{\text{even}}-1)$ and $(n_{\text{even}}-2, n_{\text{even}})$. For even numbers of qubits, the first and third layers are identical, which include connections between two center qubits. In contrast, the second and fourth layers differ in the first and last connections, where (1,2) and $(n_{\text{odd}}-2, n_{\text{odd}}-1)$ are for the second layer and (0,2) and $(n_{\text{odd}}-2, n_{\text{odd}})$ are for the fourth layer. This alternating pattern of connections between neighboring swap layers efficiently constructs the set of minimized $\mathrm{SWAP}$ gates for VQAs with arbitrary numbers of qubits mapped to H-shaped topology.

\begin{algorithm}[tb]
\caption{Swap layers on H-shaped subtopology}
\label{alg:swap_layers_hshaped}
\KwInput{Number of qubits $n$}
\KwOutput{List of swap layers $S$}
\Begin{
$n_{\text{odd}} \gets (n-1)-1+(n-1)\bmod 2$\;
$n_{\text{even}} \gets (n-1)-(n-1)\bmod 2$\;
$S \gets$ empty List\;
$j \gets 0$\;
\eIf{$n\bmod 2 \neq$ 0}{
\While{$j < n$}{
$S[j] \gets S[j] \cup [(0,2), (3,4), ..., (n_{\text{odd}}-2, n_{\text{odd}}-1)]$\;
\lIf{$\texttt{++}j \geq n$}{{break}}
$S[j] \gets S[j] \cup [(2,3), (4,5), ..., (n_{\text{even}}-2, n_{\text{even}}-1)]$\;
\lIf{$\texttt{++}j \geq n$}{{break}}
$S[j] \gets S[j] \cup [(1,2), (3,4), ..., (n_{\text{odd}}-2, n_{\text{odd}}-1)]$\;
\lIf{$\texttt{++}j \geq n$}{{break}}
$S[j] \gets S[j] \cup [(2,3), ..., (n_{\text{even}}-4, n_{\text{even}}-3), (n_{\text{even}}-2, n_{\text{even}})]$\;
\lIf{$\texttt{++}j \geq n$}{{break}}
}
}{
\While{$j < n$}{
$S[j] \gets S[j] \cup [(2,3), (4,5), ..., (n_{\text{even}}-2, n_{\text{even}}-1)]$\;
\lIf{$\texttt{++}j \geq n$}{{break}}
$S[j] \gets S[j] \cup [(1,2), (3,4), ..., (n_{\text{odd}}-2, n_{\text{odd}}-1)]$\;
\lIf{$\texttt{++}j \geq n$}{{break}}
$S[j] \gets S[j] \cup [(2,3), (4,5), ..., (n_{\text{even}}-2, n_{\text{even}}-1)]$\;
\lIf{$\texttt{++}j \geq n$}{{break}}
$S[j] \gets S[j] \cup [(0,2), ..., (n_{\text{odd}}-4, n_{\text{odd}}-3), ..., (n_{\text{odd}}-2, n_{\text{odd}})]$\;
\lIf{$\texttt{++}j \geq n$}{{break}}
}}
}
\end{algorithm}

The procedure to construct a depth-one circuit with arbitrary qubit number $n$ on H-shaped subtopology is similar to Algorithm~\ref{alg:aoqmap_qaoa_tshaped} for T-shaped subtopology. The list of connected edges $E$ is updated to reflect the H-shaped connectivity, and the list of $n$ swap layers $S$ is generated according to Algorithm~\ref{alg:swap_layers_hshaped} for H-shaped subtopology. Similarly, repeating the same swap layers at depth $p=1$ circuit or leveraging mirror symmetry extends solutions to high depths.
Compared to linear and T-shaped topologies, H-shaped topology enables additional connections for two center qubits that reduce required $\mathrm{SWAP}$ gates but increase circuit depth. Additionally, the H-shaped subtopology requires $n-1$ swap layers to achieve full connectivity, compared to $n-2$ for linear and T-shaped subtopologies. Further details can be found in Sec.~\ref{subsec:optim_scala}.

\subsection{Mapping of subtopology-adapted circuits}
\label{subsec:mapping_sub_circ}

The next crucial step is to map these subtopology-aware circuits onto target quantum device by selecting an optimal qubit mapping scheme. To identify a high-quality qubit set, we utilize a cost function denoted as $C$, which takes into account the error rate of each gate and measurement in the circuit and is given by
\begin{equation}
C = 1 - \prod_{i=1}^{N_g} (1 - p_{g_i}) \prod_{j=1}^{N_m} (1 - p_{m_j}),
\label{eq:cost_func}
\end{equation}
where $N_g$ is number of gates in circuit, $p_{g_i}$ is error rate of the $i$th gate, $N_m$ is number of measurements, and $p_{m_j}$ is error rate of the $j$th measurement. A lower value of $C$ indicates a higher estimated fidelity. The error rates of gates and measurements can be obtained from device calibration data.
To select optimal qubits, we first utilize mapomatic~\cite{nation2023suppressing} to identify all layouts on QPU matching connectivity of adapted circuit. We evaluate circuit on each layout, choosing the one with the highest fidelity for execution.
Since IBM QPUs like Fig.~\ref{fig:topology_qpu} contain more linear and T-shaped subtopologies than H-shaped, we focus on mapping to linear and T-shaped in our demonstrations. Additionally, H-shaped subtopologies on IBM QPUs are limited to specific qubit numbers such as 7, 9, 11, 13, and 15, whereas linear and T-shaped provide more flexibility.

We adopt a postselection process to determine between linear and T-shaped mappings instead of relying solely on cost function evaluation. Specifically, we execute circuits on both subtopologies and use postselection to select the circuit corresponding to the minimum expectation value of the problem Hamiltonian.
While fidelity estimates from the cost function are valuable, factors such as gate scheduling can also impact algorithm performance~\cite{ji2023optimizing}. Moreover, different gate arrangements may be equivalent in the absence of noise but behave differently under actual hardware noise. Therefore, we further introduce an additional variant called AOQMAP-LS which performs AOQMAP on linear subtopology with mirror symmetric swap layers in depth $p=1$.

\subsection{Optimality and scalability\label{subsec:optim_scala}}

The optimal swap strategy on line topology with $n$ qubits has been proven to necessitate $n-2$ total swap layers to achieve fully connected two qubit gates~\cite{weidenfeller2022scaling}. This reduction of two swap layers has also been reported in previous work (e.g., Ref.~\cite{hagge2020optimal}). However, this paper reaches the same conclusion by analyzing solutions obtained from an exact solver for small-scale instances, providing independent verification and a potentially more generalizable perspective on minimizing swap layer overhead. These $n-2$ swap layers ensure scalability with respect to the number of qubits, and the approach can be extended to arbitrary depth $p$ by repeating the swap layers or alternating them with their symmetric counterparts.
For Hamiltonians with partially connected two-qubit interactions, scalability in terms of depth is preserved through mirror symmetry. However, with respect to the number of qubits, this study emphasizes the critical role of the initial qubit order in reducing the number of swap gates required. In particular, we can minimize the number of $\mathrm{CX}$ gates by optimizing the initial qubit mapping, which we will discuss in Sec.~\ref{subsec:qaoa_maxcut_nonc}, providing new insights into efficient routing strategies.

Similarly, we derive depth optimal and scalable solutions for VQAs with arbitrary depth $p$ and number of qubits on T- and H-shaped topologies. The depth optimality generated by AOQMAP-T and AOQMAP-H can be established by analyzing the interactions between their unique structures and shared linear chains. 
For the T-shaped subtopology, two overlapping linear chains are present:
\begin{equation}
    0 \leftrightarrow 2 \leftrightarrow \dots \leftrightarrow n-1, \quad \text{and} \quad 1 \leftrightarrow 2 \leftrightarrow \dots \leftrightarrow n-1, \nonumber
\end{equation}
which share the qubits $\{2, 3, \dots, n-1\}$, referred to as the shared chain. Each chain has $n-1$ qubits, requiring $n-3$ swap layers to achieve full connectivity. However, branching qubits 0 and 1, both connected to qubit 2, cannot simultaneously interact with the rest of the qubits. Sequential swaps involving $(0,2)$ and $(1,2)$ introduce an additional layer. After incorporating qubit 0 into the shared chain (which takes one timestep), the longest remaining distance is between logical qubits located on physical qubits 1 and $n-1$, corresponding to a chain of $n-1$ qubits, requiring $n-3$ additional swap layers. Thus, the T-shaped subtopology requires at least $n-2$ total swap layers to achieve depth optimality, validating the depth optimality of AOQMAP-T.

For the H-shaped subtopology, there are four overlapping linear chains that share the qubits \(\{2, 3, \dots, n-3\}\):
\begin{equation}
    \begin{aligned}
        0 &\leftrightarrow 2 \leftrightarrow \cdots \leftrightarrow (n-3) \leftrightarrow (n-2), \\\nonumber
        0 &\leftrightarrow 2 \leftrightarrow \cdots \leftrightarrow (n-3) \leftrightarrow (n-1), \\\nonumber
        1 &\leftrightarrow 2 \leftrightarrow \cdots \leftrightarrow (n-3) \leftrightarrow (n-2), \\\nonumber
        1 &\leftrightarrow 2 \leftrightarrow \cdots \leftrightarrow (n-3) \leftrightarrow (n-1).\nonumber
    \end{aligned}
\end{equation}
Each linear chain consists of $n-2$ qubits, which require $n-4$ swap layers to achieve full connectivity. However, the end-localized qubits $\{0, 1\}$, connected to qubit 2, and $\{n-2, n-1\}$, connected to qubit $n-3$, cannot interact with the rest of the network until swaps propagate along the shared chain $\{2, 3, \dots, n-3\}$.
As demonstrated in Algorithm~\ref{alg:swap_layers_hshaped}, three additional steps are required before the last end-localized qubit(s) are incorporated into the network for both even and odd numbers of qubits. These three steps account for the sequential propagation of swaps to fully integrate the branching qubits at both ends of the topology. Combining these three steps with the $n-4$ swap layers required for the $n-2$ qubit chain results in a total of $n-1$ swap layers, thereby establishing the depth optimality of the H-shaped subtopology. Furthermore, our investigation reveals that the final swap layer requires only a single SWAP gate to implement the last remaining ZZ gate.

The scalability of AOQMAP to large quantum systems, such as those comprising 100 qubits, is essential for its practical applicability. To compare scalability, we measure compilation time to obtain optimal mappings that minimize circuit depth using an exact approach proposed in Ref.~\cite{tan2020optimal}. The original work on the compiler in Ref.~\cite{tan2020optimal} focused on an entire topology rather than a specific substructure. Therefore, we map QAOA circuits onto the entire topology of a 27-qubit IBM QPU, as depicted in Fig.~\ref{fig:topology_qpu}.
For three-qubit QAOA circuits, the compilation time of an exact algorithm increases exponentially with higher depth $p$, expanding from seconds at depth 1 to over 5 days at depth 7. Similarly, four-qubit compilation requires 13 seconds for depth 1 but grows substantially to over 15 hours for depth 2 and more than a week for depth 3. Furthermore, increasing qubit number from 3 to 9 lengthens compilation from 3 seconds to over 41 hours. In contrast, our approach intrinsically generalizes solutions for arbitrary depth and number of qubits without requiring any computational effort.

AOQMAP achieves both optimality and scalability by analyzing the optimal solutions of small QAOA instances obtained using the exact method \cite{tan2020optimal} in conjunction with the scalable solutions provided by SWAPNK \cite{harrigan2021quantum}. Depth optimality is achieved by setting the objective to minimize circuit depth in the exact method \cite{tan2020optimal}, while scalability is ensured by preserving the structural properties of small-circuit solutions as the system size increases, leveraging symmetric subtopologies (linear, T-, and H-shaped). By integrating insights from both optimal and scalable approaches, AOQMAP provides depth-optimal and scalable solutions. This methodology not only bridges the gap between exact and scalable approaches but also offers significant insights for addressing other routing problems, advancing both theoretical understanding and practical applications.

While the mapping of Hamiltonians to circuits is inherently scalable, other stages, such as verification, qubit selection, and subtopology identification, require careful attention to ensure feasibility at larger scales.
As detailed in Appendix~\ref{appen:verif}, the verification process currently relies on classical simulation to validate the correctness of routing, decomposition, and optimization processes, which is effective for smaller circuits but is not scalable to larger systems due to the exponential growth in computational resources.
To overcome this limitation, the verification step can be validated on small circuits and subsequently omitted for larger ones. Alternatively, techniques such as ZX-calculus-based circuit verification~\cite{peham2023equivalence} provide a practical solution, having demonstrated the capability to handle systems with hundreds of qubits.
Qubit selection in AOQMAP is performed using the mapomatic framework, which has been demonstrated to scale to systems with hundreds of qubits ~\cite{nation2023suppressing}, ensuring its suitability for large problem instances. Postselection, the final step in the workflow, evaluates and selects the optimal outcome among the three identified subtopologies by minimizing the expectation value of the problem Hamiltonian. This process imposes minimal computational overhead, supporting the scalability of AOQMAP.

For subtopology selection, AOQMAP currently supports linear, T-, and H-shaped configurations. To meet the demands of larger or more complex systems, future work will extend these subtopologies to include variants with additional qubit connections, providing enhanced flexibility while maintaining computational efficiency. If no predefined subtopologies are sufficient, routing can still be performed on the available subtopologies using compilers such as Qiskit or Tket, which reduce computational complexity compared to routing on the entire topology while maintaining high quality results. The advantages of mapping onto subtopologies followed by postselection will be discussed in Sec.~\ref{subsec:advan_postsel_subtop}.
By integrating these strategies into the workflow, AOQMAP ensures scalability for large quantum systems. This efficient and adaptive process balances classical and quantum resources, making it well suited for real world problems on near-term quantum devices.

\section{Applications}
\label{sec:applic}

\subsection{QAOA for MaxCut on noncomplete graphs}
\label{subsec:qaoa_maxcut_nonc}

QAOA is a VQA designed specifically to address combinatorial optimization problems, such as the MaxCut problem for graphs~\cite{farhi2014quantum}. This problem involves partitioning nodes of a graph into two distinct groups to maximize the number of edges that connect these groups~\cite{mohar1990eigenvalues}. The problem Hamiltonian in QAOA for MaxCut problem is represented as
\begin{equation}
    {H}_p = \frac{1}{2} \sum_{i,j} (1- {Z}_i  {Z}_j),
\end{equation}
where $i$ and $j$ are two nodes of an edge. Compared to QAOA for portfolio optimization (Eq.~\eqref{eq:qaoa_prob_hamil}), QAOA for the MaxCut problem does not include Pauli $Z$ items, implying the absence of $\mathrm{R_Z}$ gates in quantum circuit.
QAOA for MaxCut starts with state preparation and then alternately applies the problem and mixer Hamiltonian to evolve the state toward an optimal solution. QAOA for MaxCut problem on complete graphs follows a similar approach to QAOA for portfolio optimization. However, in noncomplete graphs, the lack of connectivity necessitates the exclusion of corresponding $\mathrm{ZZ}$ gates, introducing additional challenges for qubit mapping.

For VQAs with partially connected two-qubit gates, swap layers obtained from fully connected interactions can still ensure that the circuit satisfies connectivity constraints. In such cases, $\mathrm{ZZ}$-$\mathrm{SWAP}$ gate corresponding to the missing edge is replaced by a $\mathrm{SWAP}$ gate, allowing for correct execution of quantum circuits on subtopologies. However, the presence of remaining $\mathrm{SWAP}$ gates due to the lack of two-qubit interactions significantly amplifies errors caused by noise. One solution is to optimize initial mapping or initial qubit order such that all $\mathrm{SWAP}$ gates are placed at the end of circuit. These end-located $\mathrm{SWAP}$ gates can then be removed by adjusting measurement order accordingly. Furthermore, if a $\mathrm{SWAP}$ gate is placed behind a $\mathrm{ZZ}$ gate in the first ZZ layer, it can be removed by adjusting the initial qubit order since $\mathrm{ZZ}$ and $\mathrm{SWAP}$ gates commute.

\begin{algorithm}[tb]
\caption{AOQMAP for QAOA at depth $p=1$ with partially connected $\rm{ZZ}$ interactions}
\label{alg:aoqmap_qaoa_maxcut_non}
\DontPrintSemicolon
\KwIn{Number of qubits $n$, Parameters $\bm{\gamma}[c,t]$ and $\bm{\beta}[j]$ of gates $\rm{ZZ}$ on qubit pair $(c,t)$ and $\rm{R_{X}}$ on qubit $j$, respectively, where $c,t,j \in \{0, ..., n-1\}$ and $c<t$}
\KwOut{Circuit satisfying connectivity constraints}
$O \gets \{0, 1, ..., n-1\}$\tcp{Initialize the qubit order}
$L \gets$ List of all existing $\rm{ZZ}$ gates\;
\SetKwFunction{MapTwoQubitGates}{MapTwoQubitGates}
\SetKwProg{Fn}{Function}{}{}
\Fn{\MapTwoQubitGates{$O$, $L$, $\bm{\gamma}$}}{
    $O_0 \gets O$ \tcp{Initial qubit order}
    $\rm{qc}_{\rm{zz}} \gets$ Quantum circuit initialized to $| 0 \rangle^{\otimes n}$
    \ForEach{$\rm{ZZ}$ in $L$}{Assign ZZ($\gamma[c,t]$) to $\rm{qc}_{\rm{zz}}$ according to $O_0$ and swap layers obtained for fully connected two qubit interactions and update current qubit order $O$ if SWAP is applied}
    $O_f \gets O$ \tcp{Final qubit order}
    \ForEach{\rm{SWAP on $(i,j)$ located at the end of $\rm{qc}_{\rm{zz}}$}}{
              Remove SWAP\;
              $O_f[i] \leftrightarrow O_f[j]$\;
           }
    \ForEach{\rm{SWAP on $(i,j)$ located at the end of the first ZZ layer of $\rm{qc}_{\rm{zz}}$}}{
              Remove SWAP\;
              $O_0[i] \leftrightarrow O_0[j]$\;
           }
    }
    \Return{$(\rm{qc}_{\rm{zz}}$$, O_f)$}
    
    ($\rm{qc}_{zz}^{\rm{opt}}$, $O_{f}$) = \MapTwoQubitGates($O$, $L$)\;
    $n_{\rm{cx}}^{\rm{opt}} \gets$ $\rm{CX}$ gate count of $\rm{qc}_{zz}^{\rm{opt}}$\;
    $\mathcal{O}_q \gets$ List of defined qubit orders\;
    \For{$O_i$ in $\mathcal{O}_q$}{
    ($\rm{qc}_{zz}$, $O_{f_i}$) = \MapTwoQubitGates($O_i$, $L$)\;
    $n_{\rm{cx}} \gets$ $\rm{CX}$ gate count of $\rm{qc}_{\rm{zz}}$\;
    \If{$n_{\rm{cx}} < n_{\rm{cx}}^{\rm{opt}}$}{
		$n_{\rm{cx}}^{\rm{opt}} = n_{\rm{cx}}$\;
		$\rm{qc}_{\rm{zz}}^{\rm{opt}} = \rm{qc}_{\rm{zz}}$\;
		$O_{f} = O_{f_i}$\;
		}
    }
    Prepare the initial state $| 0 \rangle^{\otimes n}$\;
    Apply $\rm{H}^{\otimes n}$\;
    Apply $\rm{qc}_{zz}^{\rm{opt}}$\;
    \For{$k \gets 0$ \KwTo $n$}{
              Apply $\rm{R_X}(\bm{\beta}[O_f[k]])$ on $k$}
   Measure the qubits ($[0, ..., n-1]$ $\rightarrow$ $[O_f[0], ..., O_f[n-1]]$)\;
\end{algorithm}

Algorithm~\ref{alg:aoqmap_qaoa_maxcut_non} presents pseudocode for mapping QAOA for MaxCut on noncomplete graphs by optimizing initial qubit order. For $n$ qubits, there are $n!/2$ distinguished permutations due to symmetry. Different initial qubit orders introduce different gate arrangements, resulting in different numbers of $\mathrm{CX}$ gates. By minimizing additional $\mathrm{CX}$ gates, we can obtain an optimized qubit mapping. Practically, we can calculate the optimal or minimum number of $\mathrm{CX}$ gates in the resulting circuit. This optimal solution arranges all existing $\mathrm{ZZ}$ gates in consecutive layers until all gates are implemented. Then, we remove every $\mathrm{SWAP}$ gate behind the first ZZ layer and the one located at the end of circuit. This yields solutions with the minimum number of $\mathrm{CX}$ gates. To accelerate the search process, we can set the calculated optimal $\mathrm{CX}$ gate count as a target and terminate optimization once an initial qubit order achieves this value. It's important to note that the optimal initial order is not unique. Alternatively, we can employ a heuristic approach, where a set of initial qubit order permutations is searched, and the one with the fewest number of $\mathrm{CX}$ gates is selected~\cite{ji2023improving}. For higher depth, we can obtain the solution by utilizing mirror symmetry, which involves alternating between swap layers at depth $p=1$ and their corresponding mirrors, resulting in a circuit repeating every two depths.

\subsection{Variational quantum eigensolver}

In the previous section, we presented optimal routing solutions for QAOA on different types of subtopologies, including linear, T-, and H-shaped. One notable advantage of our approach is its applicability to other VQAs, such as VQE, without requiring additional computational resources. This is because both VQAs involve sequences of parameterized single qubit rotations and fixed two qubit operations. Therefore, optimal qubit routing solutions that we derived for QAOA based on subtopology connectivities can be easily adapted.

VQE~\cite{peruzzo2014variational, tilly2022variational,fedorov2022vqe} is designed to determine the ground state energy or eigenvalue of a Hamiltonian. It has broad applications in various fields such as quantum chemistry~\cite{cao2019quantum}, condensed matter physics~\cite{bauer2020quantum}, and combinatorial optimization~\cite{amaro2022filtering}.
Let ${H}$ be the Hamiltonian of a quantum system, and $\ket{\bm{\psi}}$ a trial wavefunction. The Rayleigh-Ritz quotient is bounded below by the ground state energy $E_0$
\begin{equation}
    E_0 \le \frac{\bra{\bm{\psi}}  {H} \ket{\bm{\psi}}}{\braket{\bm{\psi}}}.
\end{equation}
The objective is to determine a quantum state by examining a parameterized ansatz state, denoted as $\ket{\bm{\psi(\theta)}} = {U}(\bm{\theta})\ket{\bm{0}}$, to minimize expectation value of Hamiltonian. Here, $\ket{\bm{0}}$ represents initial state, and $ {U}(\bm{\theta})$ is the vector of parameters $\bm{\theta}$, also known as variational form or ansatz, which represents a parameterized unitary transformation achievable through quantum circuit. The selection of ansatz circuits plays a critical role in determining the efficacy of VQE. Three prominent categories of ansatz circuits include chemically inspired~\cite{romero2018strategies}, hardware-efficient ansatz (HEA)~\cite{kandala2017hardware}, and Hamiltonian variational~\cite{wecker2015progress}.

In this study, we examine VQE with full entanglement, as demonstrated in previous research (e.g.,~\cite{nannicini2019performance}). The ansatz circuit begins with a layer of parameterized $\mathrm{R_Y}$ gates, followed by controlled-Z ($\mathrm{CZ}$) gates that serve as entangling gates. Unlike $\mathrm{CX}$ gate, which distinguishes between control and target qubits, $\mathrm{CZ}$ gate is undirected. After that, another set of parameterized $\mathrm{R_Y}$ gates is performed, succeeded by measurement. For higher $p$, the subcircuit between the first layer and measurement is repeated $p$ times. The circuit with $n$ qubits and depth $p$ contains $(p+1)n$ parameters that require optimization by a classical optimizer. The full entanglement is achieved through $n(n-1)/2$ $\mathrm{CZ}$ gates, each comprising one $\mathrm{CX}$ and two Hadamard gates.

\begin{algorithm}[tb]
    \DontPrintSemicolon
    \caption{AOQMAP for VQE on linear subtopology}
    \label{alg:vqe_linear}
    \SetKwFunction{ApplyCZGate}{ApplyCZGate}
    \SetKwFunction{ApplyCZSWAPGate}{ApplyCZSWAPGate}
    \KwIn{Number of qubits $n$, VQE depth $p$, Vector of parameters $\bm{\theta}$ with dimension $(p+1)\times n$}
    \KwOut{Circuit satisfying connectivity constraints} 
    \SetKwProg{Fn}{Function}{}{end}
    \Fn{\ApplyCZGate{$i$, $j$}}{
        Apply $\rm{H}$ on $j$\;
        Apply $\rm{CX}$ on ($i$, $j$)\;
        Apply $\rm{H}$ on $j$\;
    }
    \Fn{\ApplyCZSWAPGate{$O$, $i$, $j$}}{
        Apply $\rm{H}$ on $j$\;
        Apply $\rm{CX}$ on ($i$, $j$)\;
        Apply $\rm{SWAP}$ on ($i$, $j$)\;
        $O[i] \leftrightarrow O[j]$ \tcp{Exchange qubit order}
        Apply $\rm{H}$ on $i$\;
    }
    $O \gets \{0, 1, ..., n-1\}$\tcp{Initialize the qubit order}
    Prepare the initial state $| 0 \rangle^{\otimes n}$\;
    \For{$i\coloneqq 0$ to $n$}{
        Apply $\rm{R_Y}(\bm{\theta}[i])$ on $i$\;
    }
    $p_{\max} \gets p$\;
    \While{$p>0$}{
        $s \gets 0$\;
        \While{$s < n$}{
            \For{$q\coloneqq 0$ to $n-1$ step 2}{
                \eIf{$s==0$ or $s==n-1$}{
                    \ApplyCZGate{$q$, $q+1$}
                }{
                    \ApplyCZSWAPGate{$O$, $q$, $q+1$}
                }
            }
            $s \gets s+1$\;
            \uIf{$s<n$}{
                \For{$q\coloneqq 1$ to $n-1$ step 2}{
                    \eIf{$s==0$ or $s==n-1$}{
                        \ApplyCZGate{$q$, $q+1$}
                    }{
                        \ApplyCZSWAPGate{$O$, $q$, $q+1$}
                    }
                }
            }
            $s \gets s+1$\;
        }
        \For{$i\coloneqq 0$ to $n$}{
            Apply $\rm{R_Y}(\bm{\theta}[i+n(p_{\max}-p+1)])$ on $O[i]$\;
        }
        $p \gets p-1$\;
    }
    Measure the qubits ($[0, ..., n-1]$ $\rightarrow$ $[O[0], ..., O[n-1]]$)\;
\end{algorithm}

\begin{figure*}[tb]
  \includegraphics[width=\linewidth]{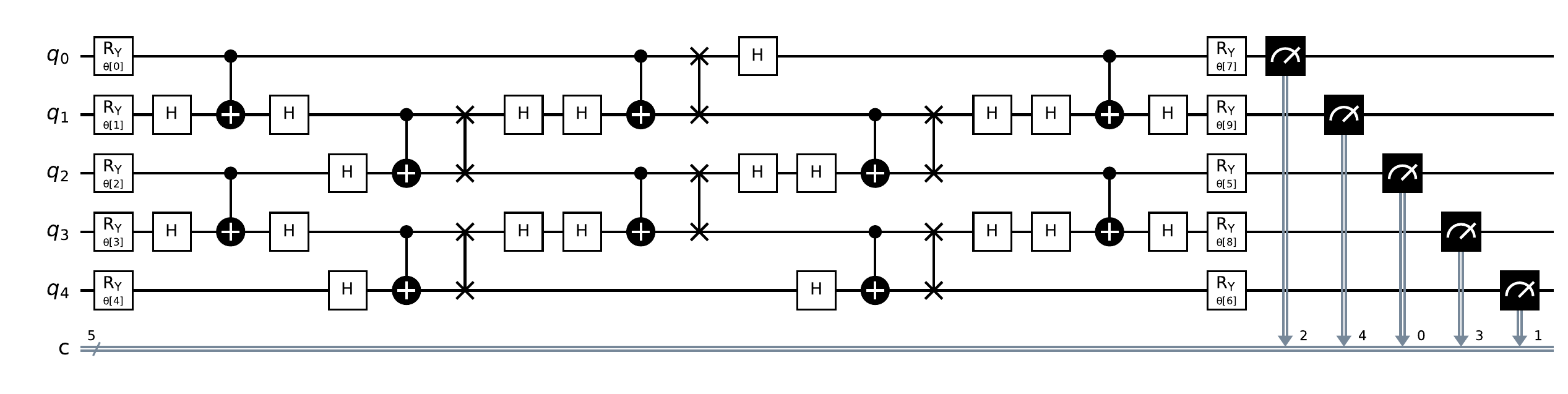}
  \caption{Resulting circuit of a five-qubit VQE at depth $p=1$ with full entanglement on a linear subtopology using AOQMAP. The circuit can be further optimized by canceling two Hadamard gates and performing CX gate cancellation for CX-SWAP gates.}
  \label{fig:5q_vqe_solution}
\end{figure*}

Algorithm~\ref{alg:vqe_linear} describes procedure of AOQMAP for VQE with full entanglement on linear subtopology. It differs from Algorithm~\ref{alg:aoqmap_qaoa_linear} in the gates used. We initiate the implementation of $\mathrm{CZ}$ gate with one $\mathrm{CX}$ and two Hadamard gates that act on the physical qubit representing target qubit of $\mathrm{CX}$ gate. Subsequently, we perform $\mathrm{CZ}$-$\mathrm{SWAP}$ gate by inserting a $\mathrm{SWAP}$ gate after $\mathrm{CX}$ gate and the second Hadamard gate on the physical qubit representing control qubit of $\mathrm{CX}$ instead of target qubit, since the introduced $\mathrm{SWAP}$ gate alters qubit order. Analogously, we construct $n-2$ swap layers for $n$ qubits on linear subtopology, meaning that $\mathrm{CZ}$ gates are performed on the first and last layers, while the constructed  $\mathrm{CZ}$-$\mathrm{SWAP}$ gates are implemented on remaining layers. Finally, $\mathrm{R_Y}$ gates are assigned accordingly, followed by measurement. The qubit mapping solution for VQE circuits on T- and H-shaped subtopologies can be attained similarly. Additionally, Algorithm~\ref{alg:aoqmap_qaoa_maxcut_non} can be employed to obtain solutions for non-fully entangled VQE.
The mapped circuit with depth $p=1$ on five qubits is depicted in {Fig.~\ref{fig:5q_vqe_solution}}. Each $\mathrm{SWAP}$ gate introduces a new qubit order and adds one additional $\mathrm{CX}$ gate, resulting in a total of $p(n-1)^2$ $\mathrm{CX}$ gates. The incorporated $\mathrm{SWAP}$ gates guarantee that all requisite $\mathrm{CZ}$ gates can be executed. $\mathrm{R_Y}$ gates and measurement operators are then assigned according to the current qubit order.

The AOQMAP approach offers several advantages. First, solutions for different subtopologies can be easily adapted between VQAs. This means that once solutions are found for a specific algorithm on target subtopology, they can be applied to other VQAs. Second, AOQMAP facilitates individual block optimization. For instance, the optimization of $\mathrm{CZ}$-$\mathrm{SWAP}$ gate can be achieved by rearranging $\mathrm{SWAP}$ gate before the second Hadamard gate, while altering the qubit that Hadamard gate acts upon. Moreover, this structure enables efficient pulse optimization~\cite{ji2023optimizing} and application of error mitigation strategies~\cite{ji2023improving, ji2024synergistic}.

\subsection{NISQ devices and algorithms beyond VQAs}

AOQMAP facilitates the transfer of solutions among diverse quantum devices, accounting for their specific architectures and noise characteristics. The routing solutions for VQAs on linear, T-, and H-shaped configurations can be adapted to other superconducting quantum devices. For instance, Google's Sycamore processor~\cite{arute2019quantum} employs square lattice qubit connectivity providing more opportunities for the three subtopologies. An H-shaped subtopology can be readily implemented in the processor by identifying two neighboring square faces in the grid and their linear connection. Similarly, Rigetti's processor~\cite{otterbach2017unsupervised} also exhibits such basic subtopologies. The adaptability is then achieved through a decomposition-optimization-remapping workflow, where the routed circuit on a subtopology is decomposed into basis gates of target device, optimized, and mapped onto hardware, taking into account the latest device calibration data. As novel architectures proliferate, AOQMAP's customized adaptability enables algorithms to be easily adjusted, making it a crucial advantage in the NISQ era.

The proposed method is also adaptable to other gate-based quantum architectures, such as trapped ions \cite{monroe2013scaling,pogorelov2021compact,ramette2022any}, where qubit connectivity and gate fidelities may differ significantly. Trapped ion architectures typically support full connectivity on a one-dimensional chain. While this eliminates the need for SWAP layers, first decomposing the circuit into native gates and then selecting high-quality qubits remains essential for improving algorithm performance. Moreover, gate sequences introduced by SWAP layers provide guidance for implementing algorithms. Additionally, research has shown that gate fidelities decrease as the chain size increases \cite{pogorelov2021compact, cetina2022control}, limiting scalability to a large number of qubits. An open question is whether focusing on neighboring qubit interactions can maintain high gate fidelities with increasing system size, thereby achieving scalability. Future research can compare the performance of fully connected two-qubit interactions versus neighboring qubit interactions using our routing solutions.

For photonic architectures \cite{politi2008silica,o2009photonic,wang2020integrated,luo2023recent}, the universal quantum computing models, such as one-way or measurement-based \cite{knill2001scheme, raussendorf2001one,walther2005experimental,menicucci2008one}, differ fundamentally from the standard quantum circuit model, requiring tailored compilation strategies for mapping circuits to photonic hardware \cite{zilk2022compiler,zhang2023oneq,zhang2024oneperc}. While the proposed mapping strategy is not directly applicable to such architectures, this study provides valuable insights into optimizing algorithm performance. For instance, recent work has demonstrated VQA with hardware-efficient ansatzes (HEA) on a quantum photonic platform \cite{agresti2024demonstration}. Our routing solutions with neighboring interactions can be adapted to build HEA for VQAs, including integrating symmetric structures to construct high VQA layers, as employed in AOQMAP-L.

\begin{table*}[tb]
\begin{ruledtabular}
\centering
 \caption{Characterization of IBM QPUs.}
  \label{tab:qpu_info}
  \begin{tabular}{lllllllll}
    Property & ibm\_perth & ibm\_nairobi & ibmq\_ehningen & ibmq\_kaolkata & ibm\_cusco & ibm\_nazca & ibm\_brisbane\\
    \hline
    Qubit number & 7 & 7 & 27 & 27 & 127 & 127 & 127\\
    Single qubit gate error (\%) & 0.049 & 0.030 & 0.033 & 3.732 & 0.270 & 0.086 & 0.024\\
    Two qubit gate error (\%) & 1.14 & 0.87 & 1.03 & 18.55 & 8.86 & 5.49 & 0.75\\
    Readout error (\%) & 2.44 & 3.01 & 1.23 & 2.60 & 5.51 & 5.18 & 14.08\\
    Single qubit gate length (ns) & 35.56 & 35.56 & 32.00 & 35.56 & 44.00 & 60.00 & 60.00\\
    Two qubit gate length (ns) & 485.93 & 306.96 & 346.98 & 450.29 &  487.22 & 658.17 & 660.0\\
    Readout length (ns) & 721.78 & 5560.89 & 846.22 & 640.0 & 4000.0 & 4000.0 & 4000.0\\
    $T_1$ ($\rm{\mu s}$) & 162.06  & 101.06 & 143.82 & 103.18 &  131.51 & 197.66 & 229.71\\
    $T_2$ ($\rm{\mu s}$) & 123.82 & 71.04 & 166.11 & 82.98 & 109.17 & 128.56 & 146.32\\
  \end{tabular}
  \end{ruledtabular}
\end{table*}

The AOQMAP flow presented in Fig.~\ref{fig:aoqmap_flow} can be generalized to optimize various quantum algorithms. Specifically, the division of the process into subtopology adaptation and mapping of subtopology-adapted circuits is particularly promising for improving the performance of algorithms on real quantum devices, as it facilitates the efficient utilization of classical and quantum resources while enabling high-quality implementations. One of the most significant advantages of AOQMAP is its ability to find optimal and scalable solutions for VQAs with fully connected two-qubit interactions on linear, T-, and H-shaped subtopologies. These solutions can be directly adapted to variational-based algorithms, including machine learning models leveraging variational circuits \cite{mitarai2018quantum, benedetti2019parameterized, schuld2020circuit, meyer2023exploiting}. While these solutions cannot be straightforwardly transformed to algorithms beyond VQAs, such as Grover's algorithm \cite{grover1996fast}, our approach to finding solutions by bridging exact and scalable methods offers valuable guidance:
(1) Carefully analyze the specific structure of target algorithm from Hamiltonian to circuit level to identify properties such as symmetry and the most advantageous subtopologies. In the case where the optimal subtopology is not immediately apparent, start with linear, T-, and H-shaped topologies, as they frequently appear in most NISQ devices;
(2) Focus on two-qubit interactions in the algorithm and insert SWAP gates to satisfy connectivity constraints. This simplifies the routing problem, as any single qubit gate can be assigned at the algorithmic level, which we have demonstrated with QAOA by first finding routing solutions on subtopologies and then assigning single qubit gates;
(3) Prioritize placing SWAP gates behind or before two-qubit gates. This allows for optimization of the two-qubit gate following a SWAP gate by leveraging CX gate cancellation (see Fig.~\ref{fig:5q_vqe_solution} for CZ-SWAP and Ref.~\cite{ji2023improving} for ZY-SWAP), thereby reducing the impact of noise due to the insertion of SWAP gates;
(4) Remove any possible initial and final SWAP gates by adjusting the initial and measurement orders.
Integrating these guidelines into the design process of heuristic and exact methods is promising to improve the quality and efficiency of solving qubit mapping problems.

\section{Benchmarking experiments}
\label{sec:bench}

This section presents benchmarking results of QAOA for portfolio optimization on several IBM QPUs. We evaluate the efficiency of our method against three other approaches: Qiskit~\cite{qiskit}, Tket~\cite{sivarajah2020tket}, and SWAP network (SWAPNK)~\cite{harrigan2021quantum}.

Table~\ref{tab:qpu_info} summarizes the characteristics of IBM QPUs employed in our benchmarking experiments. It is worth noting that these characteristics fluctuate over time. The processors used in our study range from 7 to 127 qubits. The native gate set on 7- and 27-qubit QPUs is \{$\mathrm{CX}$, $\mathrm{ID}$, $\mathrm{R_Z}$, $\mathrm{SX}$, $\mathrm{X}$\}, comprising CX, identity gate, single-qubit Z-rotation, $\pi/2$ X-rotation, and Pauli-X. The 127-qubit QPUs implement a basis set of \{$\mathrm{ECR}$, $\mathrm{ID}$, $\mathrm{R_Z}$, $\mathrm{SX}$, $\mathrm{X}$\}, where $\mathrm{ECR}$ is echoed cross-resonance two-qubit gate. The average error rate for single qubit gates varies from $10^{-4}$ to $10^{-2}$, while for two-qubit gates, it is typically an order of magnitude higher, around $10^{-2}$. Additionally, average readout error rates are approximately $10^{-2}$. The average gate lengths for single-qubit gates range from 32 ns to 60 ns, whereas for two-qubit gates, they range from 306.96 ns to 660.00 ns. The readout lengths vary between 640.0 ns and 5560.89 ns. Furthermore, the mean energy relaxation time $T_1$ ranges from $101.06~\rm{\mu s}$ to 229.71 $\rm{\mu s}$ across different processors, and the average dephasing time $T_2$ ranges from $71.04~\rm{\mu s}$ to $166.11~\rm{\mu s}$. These comprehensive devices allow us to validate our optimization techniques across various qubit numbers, connectivity options, and noise properties.

The portfolio optimization instances utilized in this study are obtained from the supplementary information of Ref.~\cite{brandhofer2023benchmarking}. Six different problem sizes are examined, involving the selection of the first 3, 4, 5, 6, 7, and 10 assets from the portfolio optimization dataset.
In QAOA, the key parameters ($q$, $B$, $A$, $\lambda$), as defined in Sec.~\ref{subsec:subtop_circ_adapt}, are set based on the problem size as follows: (0.33, 2, 0, 20.97) for three-qubit, (0.33, 2, 0.13, 17.99) for four-qubit, (0.33, 3, 0.07, 17.51) for five-qubit, (0.33, 3, 0.12, 17.73) for six-qubit, (0.33, 3, 0.13, 14.70) for seven-qubit, and (0.33, 5, 0.05, 6) for ten-qubit scenario. To determine the parameters ($\bm{\beta}, \bm{\gamma}$), we employ the constrained optimization by linear approximation (COBYLA) \cite{powell1994direct} optimizer along with Qiskit's QASM simulator.

\subsection{Evaluation metrics}

We utilize circuit metrics including two-qubit gate count and circuit depth to evaluate various qubit mapping methods. Additionally, we employ approximation ratio (AR) and success probability (SP) to assess the performance of QAOA.

Portfolio optimization refers to the process of maximizing returns while minimizing risk by selecting a subset of $n$ assets ($n$ qubits). Generally, a defined number of assets needs to be selected, which is called the budget constraint. If the solution violates the budget constraint, the value of AR is defined as 0, otherwise it is defined as
\begin{align}
    r = \frac{F-F_{\max}}{F_{\text{opt}}-F_{\max}},
\end{align}
where $F$, $F_{\text{opt}}$, and $F_{\max}$ are the average value found by QAOA, the optimal/minimum value, and the worst-case/maximum value, respectively. The SP is defined as the probability of obtaining an optimal solution.
A higher value of AR or SP implies a more effective algorithm performance, while a lower value of two-qubit gate count or circuit depth indicates a high quality of the qubit mapping solution.

\subsection{Comparison of qubit mapping strategies\label{subsec:compar_of_qubit_mappi}}

To ensure fair benchmarking, we employ various qubit mapping techniques to map circuits onto target quantum hardware, followed by decomposing and optimizing these circuits using Qiskit~\cite{qiskit} with default settings (optimization level 1).
For Tket~\cite{sivarajah2020tket}, we employ NoiseAwarePlacement to select the lowest-noise qubits based on target QPUs' noise properties. The default Tket routing method \textit{RoutingPass} is employed to insert $\mathrm{SWAP}$ gates. For Qiskit~\cite{qiskit}, we transpile with default settings. Finally, to map SWAPNK~\cite{harrigan2021quantum}, we apply the same mapping strategy as AOQMAP, as detailed in Sec.~\ref{subsec:mapping_sub_circ}.

\subsubsection{Circuit properties}

\begin{figure}[tb]
    \centering
    \includegraphics[width=\linewidth]{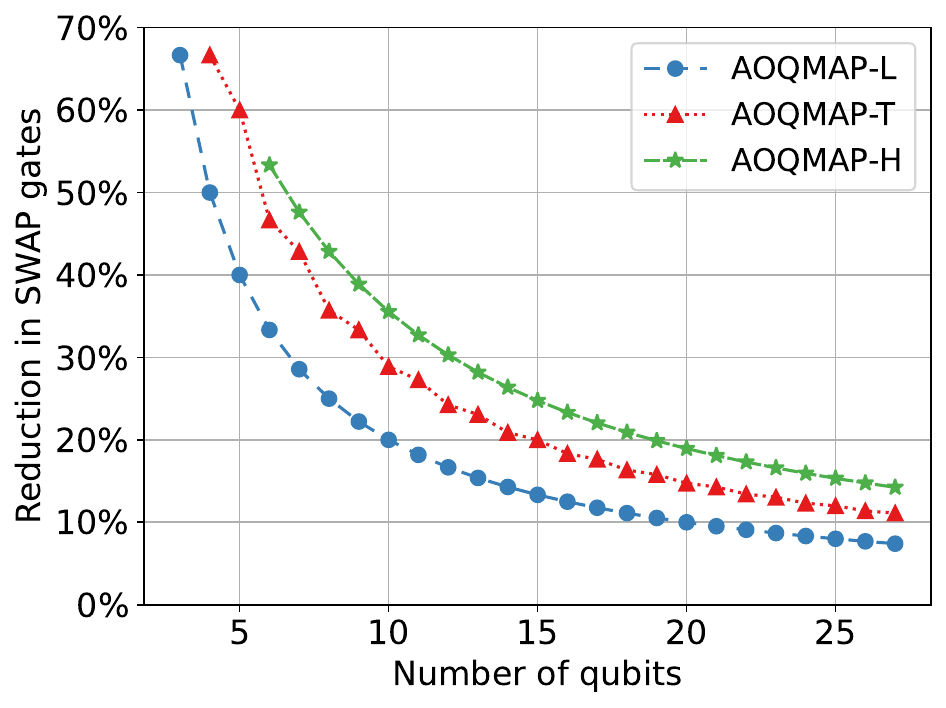}
    \caption{Reduction in the number of $\mathrm{SWAP}$ gates for QAOA with depth $p=1$ using AOQMAP-L, AOQMAP-T, and AOQMAP-H compared to SWAPNK.}\label{fig:circ_prop_comp}
\end{figure}

We first compare our approach with SWAPNK~\cite{harrigan2021quantum}.
Figure~\ref{fig:circ_prop_comp} presents the reduction in $\mathrm{SWAP}$ gates for QAOA at depth $p=1$ using AOQMAP on linear (AOQMAP-L), T- (AOQMAP-T), and H-shaped (AOQMAP-H) subtopologies compared to SWAPNK. Specifically, AOQMAP-L leads to a decrease of $67\%$ in $\mathrm{SWAP}$ gates for three-qubit and $20\%$ for ten-qubit. Moreover, AOQMAP-T reduces $\mathrm{SWAP}$ gates by $67\%$ for four-qubit and $29\%$ for ten-qubit, whereas AOQMAP-H reduces $\mathrm{SWAP}$ gates by $53\%$ for six-qubit and $36\%$ for ten-qubit. For QAOA with depth $p$, the reduction in $\mathrm{SWAP}$ gate count achieved is $p$ multiplied by the reduction attained at depth $p=1$.

\begin{table*}[tb]\footnotesize
\begin{ruledtabular}
\centering
 \caption{Reduction in $\mathrm{CX}$ or $\mathrm{ECR}$ gate counts of AOQMAP compared to other qubit mapping methods using QAOA with $n$ qubits ($n$Q) and depths from 1 to 7 on various IBM QPUs. Higher values indicate better performance of AOQMAP.}
 \label{tab:circ_prop_cx}
  \begin{tabular}{llll}
    Benchmark & Tket & Qiskit & SWAPNK\\
    \hline
    3Q-Perth  & (22\%, 22\%, 30\%, 28\%, 31\%, 30\%, 32\%) & (22\%, 22\%, 30\%, 28\%, 31\%, 30\%, 32\%)   & (22\%, 22\%, 22\%, 22\%, 22\%, 22\%, 22\%) \\
    3Q-Kolkata  & (22\%, 22\%, 30\%, 28\%, 31\%, 30\%, 32\%) & (22\%, 22\%, 30\%, 28\%, 31\%, 30\%, 32\%)   & (22\%, 22\%, 22\%, 22\%, 22\%, 22\%, 22\%) \\
    3Q-Ehningen  & (22\%, 22\%, 30\%, 28\%, 31\%, 30\%, 32\%) & (30\%, 48\%, 40\%, 45\%, 35\%, 38\%, 42\%) & (22\%, 22\%, 22\%, 22\%, 22\%, 22\%, 22\%) \\
    3Q-Nairobi & (22\%, 22\%, 30\%, 28\%, 31\%, 30\%, 32\%) & (22\%, 22\%, 30\%, 28\%, 31\%, 30\%, 32\%)   & (22\%, 22\%, 22\%, 22\%, 22\%, 22\%, 22\%) \\
    5Q-Perth  & (10\%, 28\%, 28\%, 30\%, 30\%, 31\%, 31\%) &  (19\%, 34\%, 35\%, 37\%, 35\%, 39\%, 36\%) & (13\%, 20\%, 20\%, 20\%, 20\%, 20\%, 20\%)\\
    5Q-Kolkata  &  (10\%, 22\%, 22\%, 25\%, 26\%, 25\%, 26\%) &  (21\%, 29\%, 30\%, 32\%, 32\%, 34\%, 34\%)  &  (13\%, 13\%, 13\%, 13\%, 13\%, 13\%, 13\%)\\
    5Q-Cusco & (28\%, 25\%, 27\%, 26\%, 27\%, 26\%, 27\%) & (32\%, 34\%, 38\%, 38\%, 38\%, 39\%, 39\%) & (13\%, 13\%, 13\%, 13\%, 13\%, 13\%, 13\%)\\
    5Q-Ehningen & (7\%, 19\%, 22\%, 24\%, 24\%, 25\%, 25\%) &  (48\%, 50\%, 54\%, 52\%, 53\%, 56\%, 53\%) &  (13\%, 13\%, 13\%, 13\%, 13\%, 13\%, 13\%)\\
    6Q-Ehningen  &  (41\%, 40\%, 41\%, 42\%, 42\%, 42\%, 42\%) &  (39\%, 43\%, 45\%, 43\%, 44\%, 44\%, 45\%) &  (11\%, 11\%, 11\%, 11\%, 11\%, 11\%, 11\%)\\
    10Q-Kolkata  &  (52\%, 50\%, 48\%, 52\%, 53\%, 48\%, 54\%) & (44\%, 43\%, 45\%, 47\%, 48\%, 47\%, 44\%) & (10\%, 7\%, 7\%, 7\%, 7\%, 7\%, 10\%)\\
    10Q-Nazca  &  (48\%, 53\%, 55\%, 55\%, 56\%, 56\%, 56\%) & (46\%, 46\%, 48\%, 49\%, 48\%, 49\%, 49\%) & (7\%, 7\%, 7\%, 7\%, 7\%, 7\%, 7\%)\\
    10Q-Cusco  & (51\%, 53\%, 55\%, 55\%, 56\%, 56\%, 56\%) & (46\%, 46\%, 48\%, 49\%, 48\%, 49\%, 49\%) & (7\%, 7\%, 7\%, 7\%, 7\%, 7\%, 7\%)\\
  \end{tabular}
  \end{ruledtabular}

\centering
 \caption{Reduction in circuit depth. Same benchmarks as in Table~\ref{tab:circ_prop_cx}.}
 \label{tab:circ_prop_depth}
 \begin{ruledtabular}
  \begin{tabular}{llll}
    Benchmark & Tket & Qiskit & SWAPNK\\
    \hline
    3Q-Perth  & (30\%, 21\%, 29\%, 25\%, 29\%, 26\%, 29\%) & (27\%, 21\%, 22\%, 25\%, 25\%, 26\%, 26\%) & (10\%, 11\%, 11\%, 11\%, 11\%, 11\%, 11\%) \\
    3Q-Kolkata  & (30\%, 21\%, 29\%, 25\%, 29\%, 26\%, 29\%)  &  (27\%, 21\%, 22\%, 25\%, 25\%, 26\%, 26\%)  &  (10\%, 11\%, 11\%, 11\%, 11\%, 11\%, 11\%)\\
    3Q-Ehningen  & (27\%, 21\%, 28\%, 25\%, 28\%, 26\%, 28\%) &  (14\%, 28\%, 22\%, 26\%, 19\%, 22\%, 25\%) &  (10\%, 11\%, 11\%, 11\%, 11\%, 11\%, 11\%)\\
    3Q-Nairobi &  (30\%, 21\%, 29\%, 25\%, 29\%, 26\%, 29\%) &  (27\%, 21\%, 22\%, 25\%, 25\%, 26\%, 26\%) &  (10\%, 11\%, 11\%, 11\%, 11\%, 11\%, 11\%)\\
    5Q-Perth  & (29\%, 13\%, 18\%, 19\%, 20\%, 20\%, 21\%)  & (43\%, 23\%, 27\%, 34\%, 28\%, 36\%, 28\%)  & (7\%, -26\%, -27\%, -27\%, -27\%, -27\%, -27\%)\\
    5Q-Kolkata  & (29\%, 36\%, 40\%, 46\%, 47\%, 46\%, 47\%) &  (47\%, 43\%, 51\%, 50\%, 48\%, 54\%, 53\%) &  (7\%, 7\%, 8\%, 8\%, 8\%, 8\%, 8\%)\\
    5Q-Cusco & (53\%, 50\%, 50\%, 49\%, 50\%, 49\%, 49\%) &   (48\%, 47\%, 48\%, 52\%, 50\%, 52\%, 52\%) &   (13\%, 13\%, 12\%, 12\%, 11\%, 12\%, 11\%)\\
    5Q-Ehningen &  (21\%, 26\%, 28\%, 29\%, 30\%, 30\%, 31\%) & (51\%, 55\%, 57\%, 57\%, 57\%, 59\%, 57\%) & (7\%, 7\%, 8\%, 8\%, 8\%, 8\%, 8\%)\\
    6Q-Ehningen  & (48\%, 43\%, 44\%, 48\%, 45\%, 46\%, 48\%) & (44\%, 45\%, 50\%, 50\%, 47\%, 49\%, 49\%) & (6\%, 6\%, 7\%, 7\%, 7\%, 7\%, 7\%)\\
    10Q-Kolkata  & (57\%, 69\%, 65\%, 71\%, 71\%, 66\%, 59\%) & (54\%, 66\%, 67\%, 69\%, 72\%, 71\%, 27\%) & (-33\%, 4\%, 4\%, 4\%, 4\%, 4\%, -35\%)\\
    10Q-Nazca  & (74\%, 75\%, 79\%, 80\%, 81\%, 82\%, 82\%) & (71\%, 71\%, 52\%, 72\%, 71\%, 76\%, 74\%) & (6\%, 6\%, 6\%, 5\%, 5\%, 5\%, 5\%)\\
    10Q-Cusco  & (75\%, 75\%, 79\%, 80\%, 81\%, 81\%, 81\%) & (71\%, 71\%, 52\%, 71\%, 71\%, 76\%, 74\%) & (7\%, 7\%, 6\%, 6\%, 6\%, 7\%, 6\%)\\
  \end{tabular}
  \end{ruledtabular}
\end{table*}

We then compare our approach with Tket, Qiskit, and SWAPNK on IBM QPUs. The examined QAOA has depths ranging from 1 to 7 and qubit numbers 3, 5, 6, and 10. Tables~\ref{tab:circ_prop_cx} and~\ref{tab:circ_prop_depth} present the reduction in two-qubit gate counts ($\mathrm{CX}$ or $\mathrm{ECR}$) and circuit depth, respectively, of mapped circuits generated by AOQMAP compared to other approaches across various QPUs. We note that the 127-qubit QPUs natively implement $\mathrm{ECR}$ gates instead of $\mathrm{CX}$ gates.
The results demonstrate that AOQMAP leads to reductions in $\mathrm{CX}$ or $\mathrm{ECR}$ gates and/or circuit depth compared to other approaches. For three-qubit QAOA, the number of $\mathrm{CX}$ gates remains the same reduction for AOQMAP compared to Tket and SWAPNK since three qubits can only form a linear subtopology. The difference in reduction in Qiskit is due to the randomness of Qiskit's transpiler.
However, for five-qubit QAOA with depths ranging from 2 to 7, AOQMAP exhibits a higher $\mathrm{CX}$ gate reduction but lower circuit depth reduction on ibm\_perth (5Q-Perth) than ibmq\_kolkata (5Q-Kolkata). The reason is that a circuit adapted to T-shaped subtopology yields a lower expectation value of problem Hamiltonian and is selected. This selection of T-shaped topology further reduces the number of CX gates, but also increases circuit depth.
For six-qubit QAOA on ibmq\_ehningen, AOQMAP uses only linear subtopology, which has the least number of $\mathrm{CX}$ gates and the shortest depths compared to others. A T-shaped subtopology is also selected in AOQMAP for ten-qubit QAOA on ibmq\_kolkata at depths 1 and 7.
The solutions provided by AOQMAP achieve the shortest circuit depth and/or the fewest number of $\mathrm{CX}$ or $\mathrm{ECR}$ gates compared to other approaches. Specifically, AOQMAP reduces two-qubit gate count by 29\% (up to 56\%) and circuit depth by 31\% (up to 82\%) on average compared to Qiskit, Tket, and SWAPNK.

It is important to note that the solution quality in Qiskit is sensitive to the optimization level settings. As shown in Ref.~\cite{ji2023improving}, where AOQMAP is applied to improve digitized counterdiabatic QAOA, for QAOA circuits with depth $p=1$, Qiskit’s optimization levels 1 and 2 produce comparable results, while level 3 shows improved performance. However, AOQMAP consistently outperforms all Qiskit optimization levels tested, highlighting its superior efficiency and effectiveness.

\subsubsection{Computational resources}

This section examines the computational resources needed for AOQMAP and compares them with the requirements of Qiskit and Tket. We show that the performance enhancements achieved by AOQMAP do not come at the expense of increased computational complexity, making it a practical approach.
Specifically, we first divide the runtime of AOQMAP into three parts: adaptation, verification, and qubit selection. As illustrated in Fig.~\ref{fig:aoqmap_flow}, VQAs are routed, decomposed, and optimized on subtopologies in adaptation process, ensuring that the circuit can be executed on QPU's subtopologies. Then, the circuit correctness is verified by computing the Hellinger distance of adapted and original circuits using QASM simulator. Finally, mapomatic \cite{nation2023suppressing} is employed to select optimal qubits for execution.

\begin{table*}[tb]
\centering
 \caption{Runtime [$\rm{s}$] of three components in AOQMAP on T-shaped subtopology using QAOA with the number of qubits $n$ ranging from 4 to 10 and depth $p$ of 1, 3, 5, and 7. The runtime is divided into adaptation, verification, and qubit selection. Avg denotes the average runtime for different qubit numbers.}
 \label{tab:runtime_aoqmap}
 \begin{ruledtabular}
  \begin{tabular}{llllllllll}
    $p$ & $n$ & 4 & 5 & 6 & 7 & 8 & 9 & 10 & Avg\\
    \hline
    \multirow{3}{*}{1} & Adaptation & 0.04 &  0.05 & 0.08 & 0.06 & 0.12 & 0.09 & 0.08 & 0.07 \\
    & Verification & 0.13 & 0.12 & 0.16 & 0.16 & 0.19 & 0.25 & 0.21 & 0.17 \\
    & Qubit selection & 0.03 & 0.05 & 0.11 & 0.07 & 0.09 & 0.09 & 0.14 & 0.08 \\
    \hline
    \multirow{3}{*}{3} & Adaptation & 0.02 & 0.02 & 0.03 & 0.03 & 0.05 & 0.06 & 0.08 & 0.04 \\
    & Verification & 0.12 & 0.14 & 0.14 & 0.24 & 0.19 & 0.2 & 0.23 & 0.18 \\
    & Qubit selection & 0.11 & 0.06 & 0.09 & 0.09 & 0.16 & 0.14 & 0.23 & 0.13 \\
    \hline
    \multirow{3}{*}{5} & Adaptation & 0.03 & 0.03 & 0.05 & 0.05 & 0.06 & 0.08 & 0.1 & 0.06 \\
    & Verification & 0.12 & 0.2 & 0.16 & 0.31 & 0.2 & 0.23 & 0.28 & 0.21 \\
    & Qubit selection & 0.05 & 0.06 & 0.11 & 0.13 & 0.2 & 0.2 & 0.33 & 0.15 \\
    \hline
    \multirow{3}{*}{7} & Adaptation & 0.03 & 0.03 & 0.06 & 0.06 & 0.08 & 0.11 & 0.2 & 0.08 \\
    & Verification & 0.13 & 0.14 & 0.16 & 0.19 & 0.22 & 0.25 & 0.28 & 0.2 \\
    & Qubit selection & 0.08 & 0.08 & 0.12 & 0.16 & 0.27 & 0.25 & 0.44 & 0.2 \\
  \end{tabular}
  \end{ruledtabular}
\end{table*}

\begin{table*}[tb]
\centering
 \caption{Runtime comparison of different qubit mapping methods using QAOA, with the number of qubits $n$ ranging from 4 to 10 and depth $p$ of 1, 3, 5, and 7. Avg denotes the average runtime for different qubit numbers.}
 \label{tab:runtime_comp}
 \begin{ruledtabular}
  \begin{tabular}{llllllllll}
    $p$ & $n$ & 4 & 5 & 6 & 7 & 8 & 9 & 10 & Avg\\
    \hline
    \multirow{3}{*}{1} & AOQMAP-T & 0.2 & 0.22 & 0.34 & 0.29 & 0.4 & 0.44 & 0.44 & 0.33 \\
    & Qiskit & 0.02 & 0.03 & 0.03 & 0.03 & 0.03 & 0.04 & 0.05 & 0.03 \\
    & Tket & 0.05 & 0.36 & 0.19 & 0.13 & 0.64 & 0.77 & 2.99 & 0.73 \\
    \hline
    \multirow{3}{*}{3} & AOQMAP-T & 0.25 & 0.22 & 0.27 & 0.36 & 0.39 & 0.41 & 0.55 & 0.35\\
    & Qiskit & 0.02 & 0.04 & 0.03 & 0.13 & 0.06 & 0.06 & 0.08 & 0.06\\
    & Tket & 0.07 & 0.42 & 0.27 & 0.23 & 1.42 & 0.92 & 3.16 & 0.93 \\
    \hline
    \multirow{3}{*}{5} & AOQMAP-T & 0.2 & 0.23 & 0.31 & 0.48 & 0.47 & 0.52 & 0.7 & 0.42 \\
    & Qiskit & 0.05 & 0.05 & 0.05 & 0.09 & 0.08 & 0.12 & 0.14 & 0.08 \\
    & Tket & 0.13 & 0.48 & 0.35 & 0.34 & 0.97 & 1.09 & 3.19 & 0.94 \\
    \hline
    \multirow{3}{*}{7} & AOQMAP-T & 0.24 & 0.25 & 0.34 & 0.41 & 0.56 & 0.62 & 0.92 & 0.48 \\
    & Qiskit & 0.04 & 0.06 & 0.06 & 0.09 & 0.19 & 0.13 & 0.17 & 0.11 \\
    & Tket & 0.16 & 0.53 & 0.49 & 0.44 & 1.13 & 1.34 & 3.67 & 1.11 \\
  \end{tabular}
  \end{ruledtabular}
\end{table*}

\begin{figure*}[tb]
\centering
  \includegraphics[width=\linewidth]{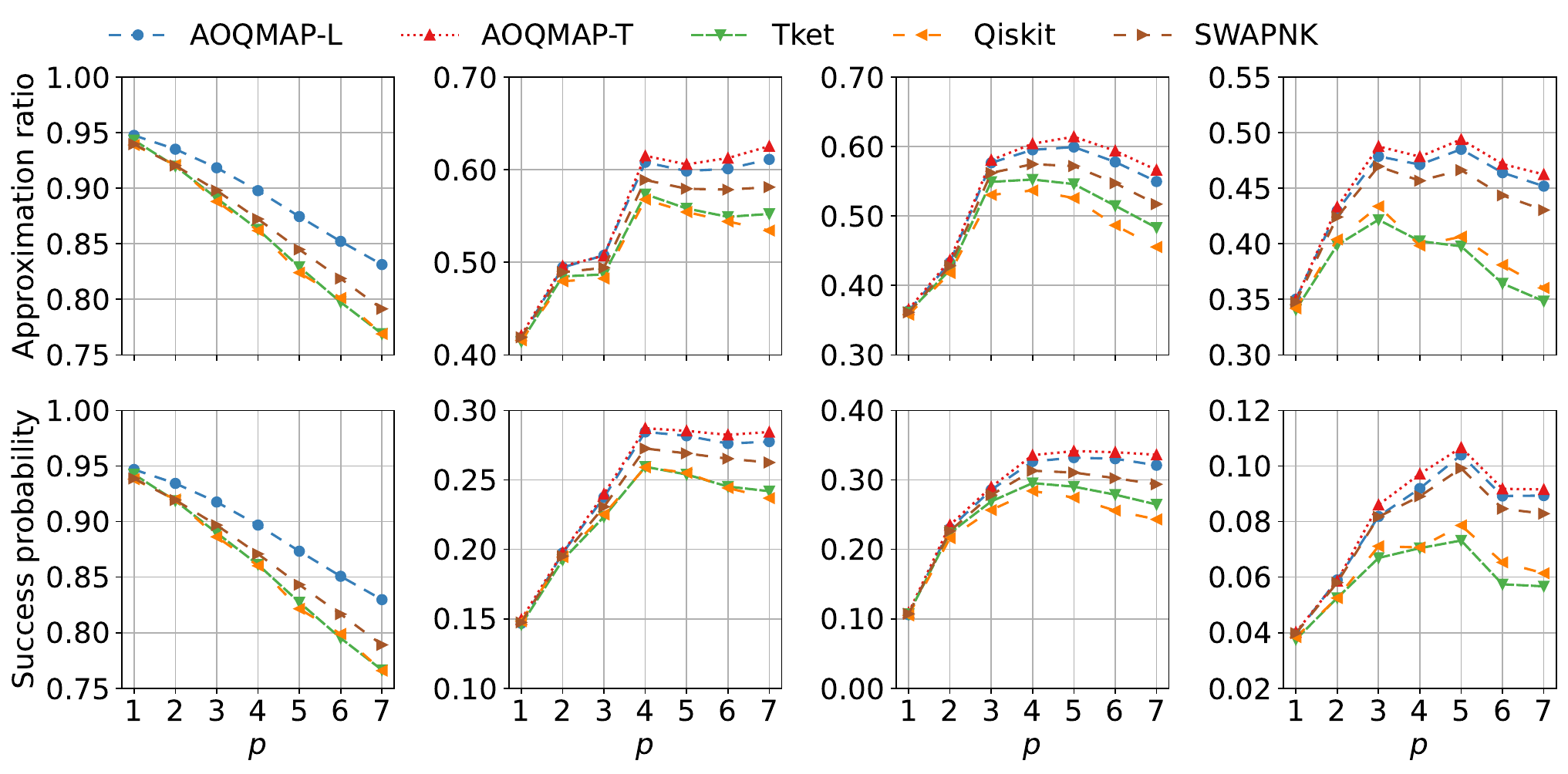}
  \put(-\linewidth,0.455\linewidth){{\textbf{(a)}}}
  \put(-\linewidth,0.24\linewidth){{\textbf{(b)}}}
  \caption{Simulation under depolarizing noise of QAOA for portfolio optimization mapped with AOQMAP-L, AOQMAP-T, Tket, Qiskit, and SWAPNK. (a) Approximation ratio and (b) success probability for QAOA with three, four, five, and six qubits. Three-qubit QAOA has an absence of AOQMAP-T, as the minimum number of qubits required for the T-shaped subtopology is four.}
  \label{fig:noise_sim_dp}
\end{figure*}

AOQMAP, Qiskit, and Tkit employ the same experimental setup described at the beginning of Sec.~\ref{subsec:compar_of_qubit_mappi}.
We benchmark the QAOA for MaxCut on complete graphs with varying qubit numbers and depths, mapping these circuits onto a 27-qubit QPU. We take AOQMAP-T as an example, and its analysis can be extended to AOQMAP-L and AOQMAP-H.
Tables~\ref{tab:runtime_aoqmap} and \ref{tab:runtime_comp} provide a comprehensive overview of runtime for three parts of AOQMAP and an overall comparison, respectively. Our investigation reveals that adaptation in AOQMAP consumes the least time, with its duration increasing minimally as the number of qubits grows. This process is also minimally affected by variations in QAOA depths. In contrast, verification is the most time-intensive, requiring the execution of circuits with QASM simulator. Moreover, verification runtime increases with the number of qubits but is relatively unaffected by QAOA depths. The runtime for qubit selection increases with both the number of qubits and QAOA depths. We also observe that the runtime for deeper circuits might be shorter than that for shallower circuits. This is potentially due to the stochastic nature of Qiskit transpiler employed for decomposition and optimization. As indicated in Table~\ref{tab:runtime_comp}, Qiskit demonstrates the shortest runtime. Additionally, AOQMAP outperforms the heuristic strategy Tket. Tket's runtime increases with QAOA depths and qubit numbers, and exhibits significant fluctuations, potentially due to the interface transforming circuits implemented with Qiskit and Tket.

\subsubsection{Simulation under noise}

To examine the impact of noise on the performance of QAOA with different qubit mapping strategies, we simulate algorithms mapped onto a 27-qubit QPU with the topology shown in Fig.~\ref{fig:topology_qpu}. We use a depolarizing noise model~\cite{georgopoulos2021modeling, zeng2021simulating}, which is a common simple model used to approximate the effects of mixed noise processes in quantum systems. This model captures noise effects by applying random single-qubit bit-flip, phase-flip, and combined bit- and phase-flip errors to each gate, providing a good approximation of the overall noise behavior and impact on algorithm performance. The depolarizing error channel, acting on qubits described by density matrix $\mathrm{\rho}$, is defined as
\begin{equation}
    {E_D}({\rho}) = \frac{\epsilon}{4}\left({X\rho X} + {Y\rho Y} + {Z\rho Z}\right)+(1 - \frac{3\epsilon}{4}) {\rho},
\end{equation}
where ${X}$, ${Y}$, and ${Z}$ are Pauli matrices, and $\epsilon$ is noise strength.

We compare AOQMAP-L and AOQMAP-T with Tket, Qiskit, and SWAPNK. The mapped circuits are simulated under depolarizing noise with strength $0.005$ for two-qubit gates and $\epsilon/10$ for single-qubit gates. Figure~\ref{fig:noise_sim_dp} presents the approximation ratio and success probability of QAOA on 3, 4, 5, and 6 qubits at depths from 1 to 7.
Compared to other quantum mapping strategies, AOQMAP-L and AOQMAP-T demonstrate better performance at higher depths and comparable performance at depth $p=1$. This suggests that increased depth of QAOA leads to a more significant noise effect, emphasizing advantages of AOQMAP. AOQMAP-T requires fewer $\mathrm{CX}$ gates during the mapping stage compared to others, which mitigates the detrimental effects of noise accumulation, leading to improved performance.
For three-qubit QAOA, AR and SP decrease with increased depth, indicating that noise effects outweigh performance improvement at higher depths. In comparison, for QAOA with 4, 5, and 6 qubits, AR and SP increase and then stabilize or slightly decrease, suggesting a balance between performance improvements and noise effects.

\subsubsection{Advantages of postselection on subtopologies}\label{subsec:advan_postsel_subtop}

In this section, we investigate the performance benefits of restricting circuit execution to identified subtopologies and applying postselection, compared to directly mapping circuits onto the full device topology. Specifically, we first compare subtopology-based circuits with those mapped to the entire topology of a 127-qubit QPU ibm\_brisbane in terms of circuit depth and the number of two-qubit gates. We then show that postselection among results obtained on subtopologies, executed using a noise simulator configured to mimic the ibm\_brisbane, yields higher performance than full-topology mapping. Moreover, we evaluate the performance of Qiskit and Tket in comparison to our AOQMAP approach and show that AOQMAP consistently outperforms both, underscoring its potential for large scale quantum computing applications.

\begin{figure}[tb]
    \centering
    \includegraphics[width=\linewidth]{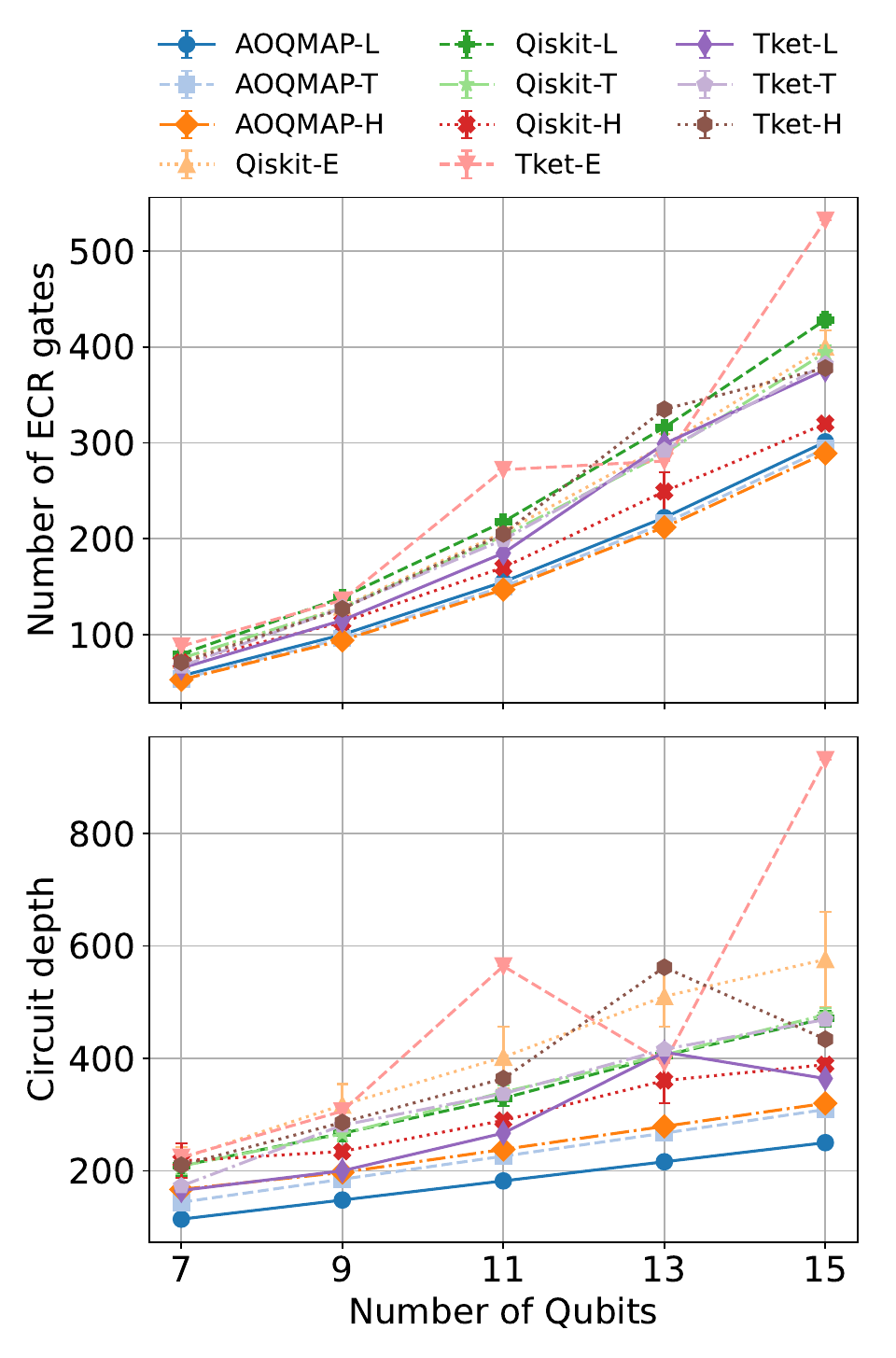}
    \caption{Comparison of ECR gate count and circuit depth in $n$-qubit QAOA with $p=1$, transpiled on the entire topology (denoted by ``-E") and subtopologies (denoted by ``-L", ``-T", and ``-H") of a 127-qubit QPU ibm\_brisbane. Each data point represents the mean of 10 repetitions, with error bars indicating the standard error of the mean.}\label{fig:cx_depth_compar_subtop}
\end{figure}

\begin{table}[tb]
\begin{ruledtabular}
\centering
\caption{Average reductions in ECR gate count and circuit depth with Qiskit and Tket on subtopologies compared to the full topology, derived from the dataset in Fig.~\ref{fig:cx_depth_compar_subtop}.}
\label{tab:cx_depth_compar_subtop}
\begin{tabular}{lllll}
Transpiler & Subtopology & $\phantom{-}$ECR reduction & Depth reduction \\
       & Linear & $-6.87$\% & 15.75\% \\
Qiskit & T-shaped & $\phantom{-}0.90$\%  & 15.04\% \\
       & H-shaped & $\phantom{-}13.92$\% & 23.56\% \\
\hline
       & Linear   & $\phantom{-}19.30$\% & 33.94\% \\
Tket   & T-shaped   & $\phantom{-}16.17$\% & 22.94\% \\
       & H-shaped   & $\phantom{-}12.06$\% & 11.65\% \\
\end{tabular}
\end{ruledtabular}
\end{table}

Here we employ the highest optimization level available in Qiskit, level 3, which is consistently applied for Qiskit and the decomposition process employed in AOQMAP and Tket to ensure a fair comparison. To accurately emulate the noise characteristics of a real quantum device, we configure the Qiskit AerSimulator using \texttt{AerSimulator.from\_backend(device\_backend)}. This method integrates the simulator with the noise model and characteristics of an actual backend, specifically ibm\_brisbane. By replicating the practical noise environment of a quantum processor, this simulation framework enables a robust evaluation of algorithmic performance under realistic conditions.

Figure~\ref{fig:cx_depth_compar_subtop} illustrates the impact of topology selection on ECR gate count and circuit depth for $n$-qubit QAOA circuits with $p=1$ on ibm\_brisbane.
Restricting the layout to subtopologies reduces the number of two-qubit gates and/or circuit depth compared to the full topology. Among Qiskit implementations, the H-shaped subtopology achieves the best performance, while Tket performs optimally on linear subtopologies. AOQMAP consistently outperforms Qiskit and Tket across all subtopologies, with AOQMAP-L yielding the lowest circuit depth and AOQMAP-H achieving the minimal ECR gate count.
These results highlight the advantages of subtopology-based transpilation approaches.
Table~\ref{tab:cx_depth_compar_subtop} provides a quantitative summary of the average reductions in ECR count and circuit depth achieved by Qiskit and Tket on subtopologies relative to the full topology. For Qiskit, the H-shaped subtopology achieves the greatest reductions, with a 13.92\% decrease in ECR gate count and a 23.56\% decrease in circuit depth. In contrast, Tket performs best on the linear subtopology, achieving notable reductions of 19.30\% in ECR gate count and 33.94\% in circuit depth.
While Qiskit exhibits less efficient performance on the linear subtopology due to an increase in ECR gates, it still achieves a 15.75\% reduction in circuit depth. Notably, Qiskit is not limited to linear configurations, as it demonstrates substantial decreases in both ECR gates and circuit depth on the H-shaped subtopology. This underscores Qiskit's adaptability and improved performance when employing our proposed approach, which is designed to leverage multiple subtopology types effectively.

\begin{figure}[tb]
    \centering
    \includegraphics[width=\linewidth]{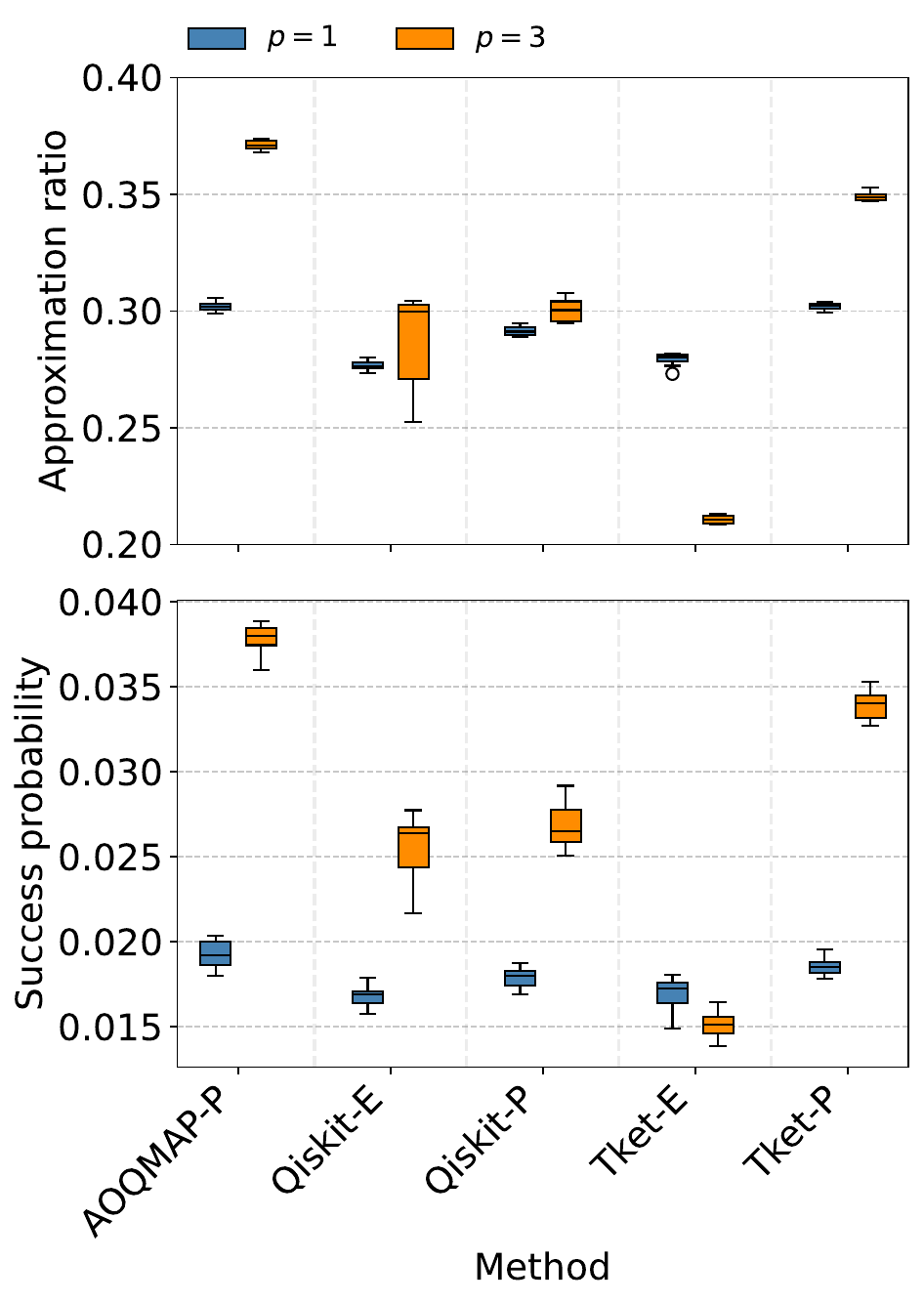}
    \caption{Comparison of mapping strategies for a 7-qubit QAOA with $p=1$ (blue, left) and $p=3$ (orange, right) on a noisy simulator mimicking the 127-qubit QPU ibm\_brisbane. Each box plot represents 10 repetitions of the corresponding mapping strategy. The results contrast entire topology mapping (denoted by ``-E") with subtopology mapping followed by postselection (denoted by ``-P").}\label{fig:compar_entrie_sub_posts}
\end{figure}

Figure~\ref{fig:compar_entrie_sub_posts} compares the performance of mapping strategies for a 7-qubit QAOA circuit with $p=1$ and $p=3$ on a noisy simulator modeled after the 127-qubit ibm\_brisbane. The results indicate that subtopology-based postselection consistently outperforms direct mapping on the full topology. Notably, Tket demonstrates more significant performance improvements from postselection compared to Qiskit, particularly for higher-depth circuits ($p=3$). Among the strategies evaluated, AOQMAP achieves the highest performance. These findings underscore the significant performance benefits of postselection on subtopologies, particularly in enhancing circuit efficiency and reliability across different transpilation frameworks. They also provide valuable insights into optimizing other classes of algorithms for large-scale quantum devices, demonstrating the broader applicability of this approach.

\subsection{Demonstration on IBM quantum devices\label{subsec:demons_on_ibm_quant}}

This section evaluates the performance of QAOA for portfolio optimization across six IBM QPUs, comparing different mapping approaches on 7-qubit ibm\_perth and ibm\_nairobi, 27-qubit ibmq\_kolkata and ibmq\_ehningen, and 127-qubit ibm\_nazca and ibm\_cusco. The characterization of the IBM QPUs at the time of the demonstration is detailed in Appendix~\ref{app:cloud_platf}. The problem sizes QAOA include 3, 4, 5, 6, and 10 qubits, and depths ranging from 1 to 7.

\begin{figure}[tb]
\centering
  \includegraphics[width=\linewidth]{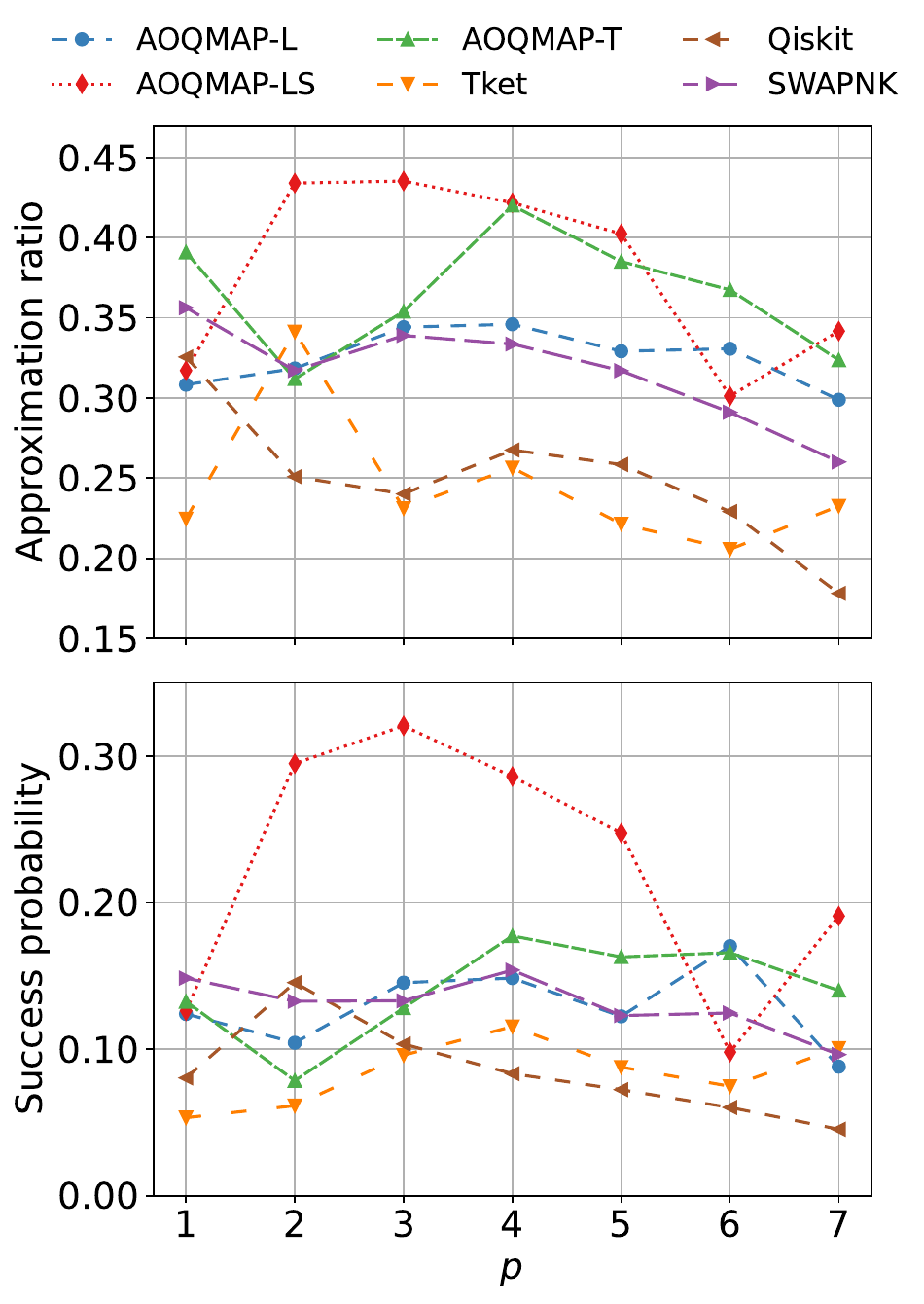}
  \put(-\linewidth,1.25\linewidth){{\textbf{(a)}}}
    \put(-\linewidth,0.65\linewidth){{\textbf{(b)}}}
    \vspace{-6pt}
  \caption{Demonstration of 4-qubit QAOA on a 7-qubit QPU ibm\_nairobi with various qubit mapping methods. (a) Approximation ratio and (b) success probability. QAOA depths range from 1 to 7.}
  \label{fig:arsp_4q_ibm_nairobi}
\end{figure}

As discussed in Sec.~\ref{subsec:subtop_circ_adapt}, two methods exist for constructing higher QAOA depth, including repeating swap layers in depth $p=1$ circuit and leveraging their symmetry.
To examine the influence of symmetry, we conduct a comparative analysis of AOQMAP on linear subtopology with repetitive routing at depth $p=1$ (AOQMAP-L) and with mirror-symmetric swap layers (AOQMAP-LS), as well as AOQMAP on T-shaped subtopology (AOQMAP-T). We consider QAOA instances with four qubits and depths from 1 to 7. Figure~\ref{fig:arsp_4q_ibm_nairobi} compares these variants to Tket, Qiskit, and SWAPNK on the 7-qubit ibm\_nairobi. AOQMAP on linear subtopology utilizing symmetry significantly enhances the performance, increasing success probability by an average of 82\% (up to 1.83$\times$) compared to AOQMAP-L, 76\% (up to 2.76$\times$) compared to AOQMAP-T, 1.72$\times$ (up to 3.80$\times$) compared to Tket, 1.77$\times$ (up to 3.21$\times$) compared to Qiskit, and 73\% (up to 1.41$\times$) compared to SWAPNK. These results demonstrate the effectiveness of symmetry for improving performance on NISQ devices, validating the advantage of AOQMAP's symmetry-incorporated mapping approach.

\begin{figure*}[tb]
\centering
  \includegraphics[width=\linewidth]{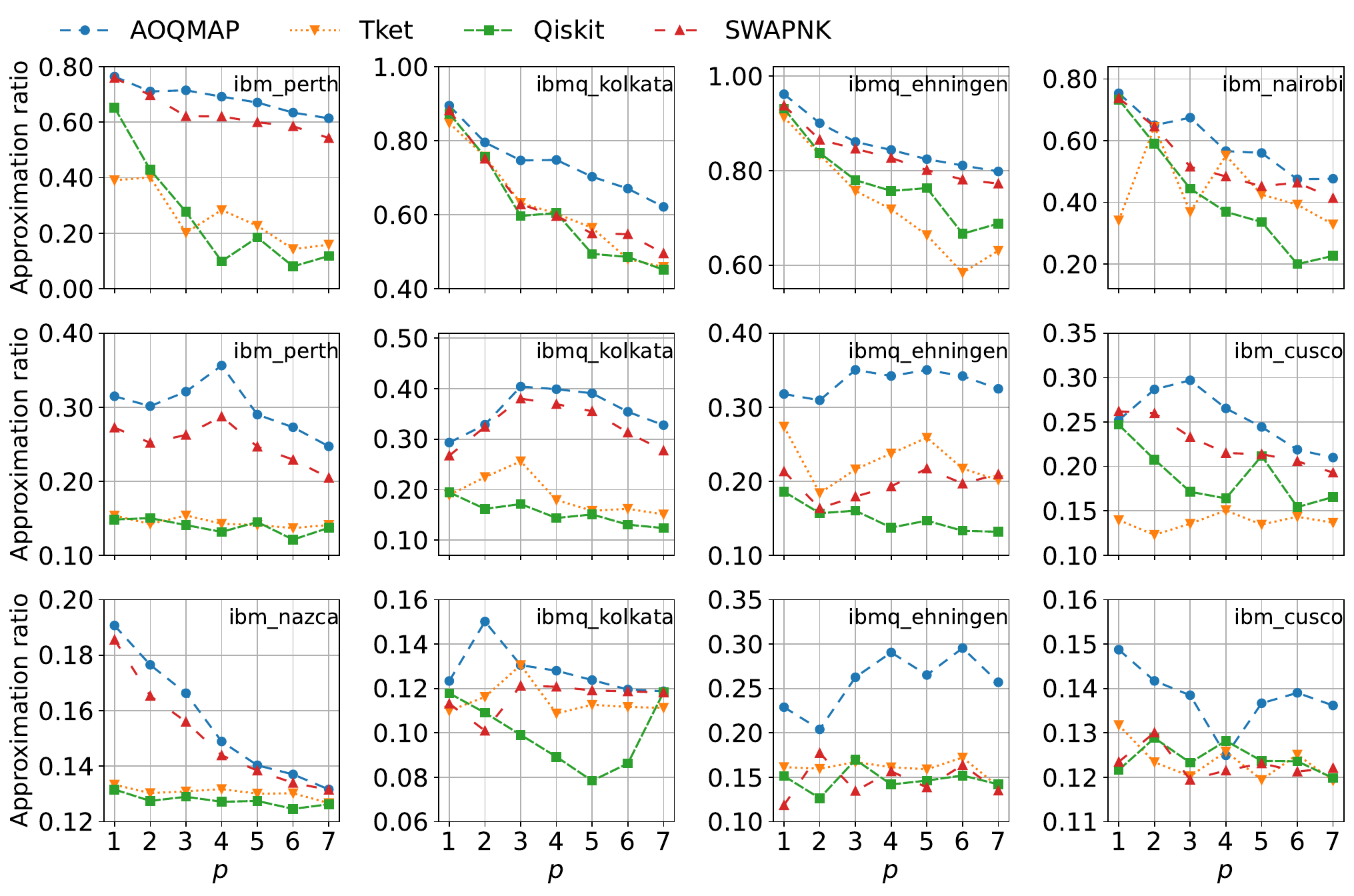}
    \put(-\linewidth,0.621\linewidth){{\textbf{(a)}}}
    \put(-\linewidth,0.428\linewidth){{\textbf{(b)}}}
    \put(-\linewidth,0.23\linewidth){{\textbf{(c)}}}
    \vspace{-6pt}
  \caption{Approximation ratio of QAOA obtained on various IBM QPUs using AOQMAP, Tket, Qiskit, and SWAPNK. The numbers of qubits are (a) three, (b) five, and (c) ten (six for ibmq\_ehningen). AOQMAP demonstrates the highest performance on all QPUs in comparison to other approaches.}
  \label{fig:ibmq_run_ar_all}
\end{figure*}

\begin{figure*}[tb]
\centering
  \includegraphics[width=\linewidth]{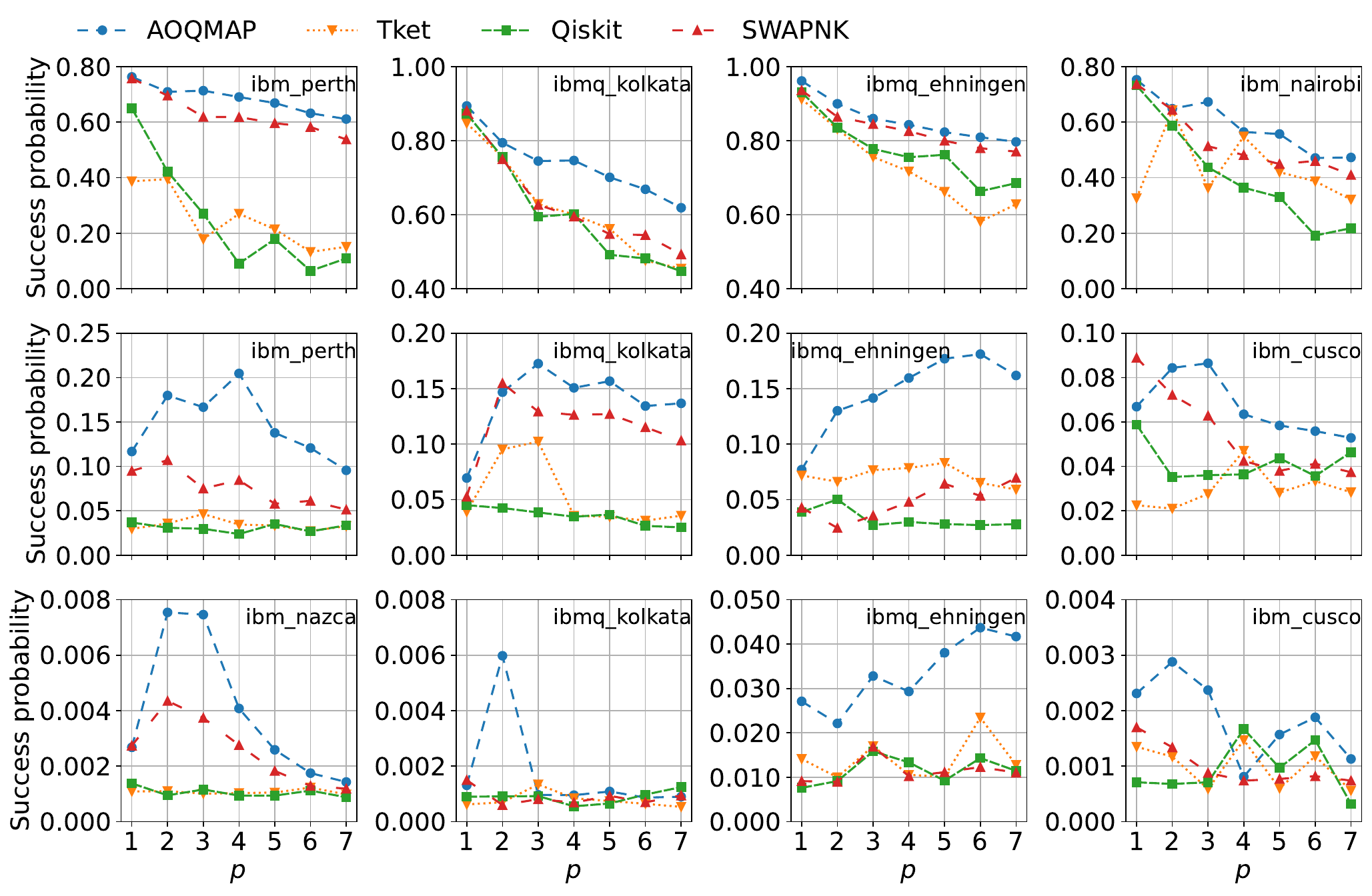}
  \put(-\linewidth,0.62\linewidth){{\textbf{(a)}}}
    \put(-\linewidth,0.422\linewidth){{\textbf{(b)}}}
    \put(-\linewidth,0.228\linewidth){{\textbf{(c)}}}
    \vspace{-6pt}
  \caption{Success probability of QAOA obtained on various QPUs with AOQMAP, Tket, Qiskit, and SWAPNK. The labels are the same as Fig.~\ref{fig:ibmq_run_ar_all}.}
  \label{fig:ibmq_run_sp_all}
\end{figure*}

For the demonstration on ibmq\_ehningen, we directly compare AOQMAP-L with other methods, while for evaluations on other QPUs, a postselection process is utilized to generate the AOQMAP solution. Specifically, we construct the algorithms with AOQMAP-L, AOQMAP-LS, and AOQMAP-T. Each adapted circuit is mapped to the target device using the method described in Sec.~\ref{subsec:mapping_sub_circ}. The demonstration solution corresponding to the minimum expectation value of the problem Hamiltonian is then chosen.
As shown in Fig.~\ref{fig:ibmq_run_ar_all}(a), AOQMAP achieves the highest performance on all QPUs. Tket and Qiskit perform comparably on all QPUs, while SWAPNK performs better on ibm\_perth and ibmq\_ehningen than ibmq\_kolkata and ibm\_nairobi. The approximation ratio decreases with increased depth, indicating that the negative impact of noise dominates the improved performance at higher depths.
For five-qubit QAOA (Fig.~\ref{fig:ibmq_run_ar_all}(b)), AOQMAP maintains the highest AR across all QPUs. Tket outperforms Qiskit on ibmq\_kolkata and ibmq\_ehningen, whereas Qiskit performs better on ibm\_cusco. SWAPNK has a better performance on ibmq\_perth and ibmq\_kolkata, while it performs worse on ibmq\_ehningen.
For six-qubit QAOA on ibmq\_ehningen, AOQMAP has a significantly higher AR value than Tket, Qiskit, and SWAPNK (Fig.~\ref{fig:ibmq_run_ar_all}(c)). For the largest ten-qubit QAOA, AOQMAP consistently achieves the highest AR across all tested QPUs. SWAPNK has a higher AR only on ibm\_nazca, while Qiskit and Tket perform worse on all QPUs. On ibmq\_kolkata, AOQMAP achieves the highest AR value at depth two, outperforming others. With increased depth, the noise in large-scale circuit increases significantly, resulting in a reduced AR value. For ten-qubit QAOA on ibm\_cusco, AOQMAP has the highest AR value at depth $p=1$. The presence of the second-highest AR at depth six demonstrates a trade-off between improved performance and introduced noise of the increased QAOA depth.

The success probability of QAOA with 3, 5, and 10 qubits (6 qubits for ibmq\_ehningen) is displayed in {Figs.~\ref{fig:ibmq_run_sp_all}(a-c)}, respectively, following a similar trend as the AR. AOQMAP demonstrates the highest SP overall with different circuit sizes across all QPUs. For QAOA with five and ten qubits, the SP initially increases with depth and then decreases, peaking at different $p$ depending on device noise. This phenomenon indicates a balance between performance improvement and noise impact in high-depth circuits. For five-qubit QAOA, Qiskit performs better on ibm\_cusco, Tket has higher performance on ibmq\_kolkata and ibmq\_ehningen, and SWAPNK has higher SP values on ibmq\_kolkata and ibm\_cusco. For ten-qubit QAOA, Tket and Qiskit have lower success probabilities on all QPUs, whereas SWAPNK performs better only on ibm\_nazca.

These findings demonstrate that AOQMAP provides a significant improvement in approximation ratio and success probability, surpassing other popular qubit mapping approaches. In particular, AOQMAP improves AR by an average of 54\% and SP by an average of 138\% compared to Qiskit, Tket, and SWAPNK, demonstrating robust high performance on NISQ devices. Furthermore, our experiments on various QPUs highlight the limitations of solely relying on noise model simulations for assessing circuit performance. While such simulations provide valuable insights, they often fail to accurately capture the complex and multifaceted noise characteristics inherent to real quantum hardware. Although different qubit mapping strategies exhibit comparable behavior under simulated noise conditions, their performance diverges significantly when executed on physical devices. This disparity emphasizes the critical role of efficient qubit mapping in achieving high algorithm performance on NISQ devices.

\section{Conclusion}
\label{sec:concl}

We present an efficient qubit mapping methodology designed for VQAs on near term quantum devices. The method involves two essential steps: adapting circuits to subtopologies and mapping adapted circuits onto target QPU, followed by postselecting execution results across different subtopologies. In the adaptation step, optimal routing solutions with the shortest circuit depth are provided for diverse subtopologies, including linear, T-, and H-shaped configurations. These solutions inherently scale to arbitrary depths and numbers of qubits for algorithms characterized by fully connected two qubit interactions. For partially connected interactions, optimizing initial qubit order enables the derivation of optimal depth-one solutions. Leveraging mirror symmetry, these solutions can be extended to higher QAOA depths without additional computational overhead.
In the mapping step, the adapted circuit is mapped onto target device, minimizing circuit error with a cost function. A postselection process is then employed to choose the optimal solution that minimizes the expectation value of problem Hamiltonian. Demonstrations of QAOA for portfolio optimization on six IBM QPUs with 7, 27, and 127 qubits exhibit improved performance of AOQMAP compared to existing methods. We also show that the symmetry used in the circuit structure can improve algorithm performance on QPUs. These results and solutions can be easily applied to other NISQ devices. Our qubit mapping strategy tailors mappings to algorithmic structures, promising to maximize the capabilities of near term devices for VQAs.

Future research directions include extending our method to handle more complex topologies, such as T- and H-shaped variants, and heavy-hex. Furthermore, exploring strategies to adapt Hamiltonian with partially connected two-qubit gate interactions presents a promising avenue of exploration. While mirror symmetry of swap layers enables scaling solutions to high depths, the combinatorial complexity associated with the depth-one circuits remains a significant challenge. Developing heuristic optimization approaches to refine initial qubit mapping can accelerate the search process. The trade-off between computational cost and solution quality is essential for achieving efficient and effective optimization.
Additionally, a comprehensive cost function and an accurate noise model, including crosstalk effects, are essential to determine the optimal subtopology for a given circuit. While linear subtopologies minimize circuit depth, T- and H-shaped subtopologies with increasing connectivity reduce SWAP gates at the expense of increased depth. The cost function should carefully balance circuit fidelity and schedule duration. The well-crafted error model, such as by integrating experimentally characterized information on real hardware \cite{dahlhauser2021modeling, dahlhauser2024benchmarking}, can assist in identifying the most suitable qubits. Designing cost functions and accurately modeling noise in QPUs, combined with a postselection process to choose between different subtopologies, represents a promising avenue for future research to significantly enhance algorithm performance on near-term quantum devices.
Finally, extending AOQMAP to optimize circuits with many-body interactions, such as VQE with the unitary coupled cluster ansatz \cite{romero2018strategies}, by addressing increased connectivity requirements and operator non-commutativity represents an exciting avenue for broadening its applicability to more complex quantum algorithms.

The implementation of AOQMAP is publicly available on Github \cite{aoqmap2024}.

\begin{acknowledgments}

The authors would like to thank Kathrin F. Koenig, Thomas Wellens, Joris Kattemölle, Sebastian Brandhofer, Andreas Ketterer, Koushik Paul, Pranav Chandarana, Juli\'{a}n Ferreiro V\'{e}lez, and Pengcheng Zhu for their useful discussions and exchanges.
We acknowledge the use of IBM Quantum services for this work and to advanced services provided by the IBM Quantum Researchers Program. The views expressed are those of the authors, and do not reflect the official policy or position of IBM or the IBM Quantum team.
The authors acknowledge the financial support of the project QORA by Ministerium f\"{u}r Wirtschaft, Arbeit und Tourismus Baden-W\"{u}rttemberg in the frame of the Competence Center Quantum Computing Baden-Württemberg, the Basque Government through Grant No. IT1470-22, the Project Grant No. PID2021-126273NB-I00 funded by MCIN/AEI/10.13039/501100011033 and by “ERDF A way of making Europe” and “ERDF Invest in your Future,” Nanoscale NMR and complex systems (Grant No. PID2021-
126694NB-C21), EU FET Open Grant EPIQUS (No. 899368), the ELKARTEK Program by the Basque
Government under Grant KK-2022/00041, BRTA QUANTUM Hacia una especializaci\'{o}n armonizada en tecnologías cu\'{a}nticas en BRTA.  X.C. acknowledges ayudas para contratos Ramón y Cajal–2015-2020
(Grant No. RYC-2017-22482) and open access funding provided by UPV/EHU.
Y.B. acknowledges the Ramón y Cajal (RYC2023-042699-I) research fellowship.

\end{acknowledgments}

\appendix

\section{Circuit verification in AOQMAP \label{appen:verif}}

Here we detail the circuit verification process in the AOQMAP. To maintain effectiveness, it is necessary to enforce the same or reduced number of $\mathrm{CX}$ gates during decomposition and optimization. Moreover, we verify the accuracy of adapted circuit by comparing output probability distributions of adapted and original circuits with Hellinger distance~\cite{hellinger1909neue}, which ranges from 0 to 1, with 0 indicating identical distributions. The Hellinger distance of two probability distributions is given by
\begin{equation}
    H(P, Q) = (1-\sum_j \sqrt{p_j q_j})^{1/2},
\end{equation}
where $p_j$ and $q_j$ are probabilities of outcome $j$ in distributions $P$ and $Q$, respectively.
The verification process begins by simulating each circuit using a simulator on a classical computer to obtain the expected output state in the absence of noise. In this study, we employ the QASM simulator provided by Qiskit to execute original and mapped circuits. We then calculate the Hellinger distance between probability distributions of simulated output states.

Table~\ref{tab:hellinger_distance} reports the Hellinger distance between original and mapped circuits generated using AOQMAP on linear subtopology (AOQMAP-L). The benchmark circuits are QAOA for portfolio optimization with 3, 4, 5, 6, and 10 qubits and depths ranging from 1 to 7. Each data point is averaged over 50 circuit repetitions, with a standard error of the means that is more than 15 orders of magnitude smaller than the mean. The Hellinger distance increases with qubit number but remains two orders of magnitude smaller than one. Values greater than 0 are due to measurement noise or shot noise generated when executing circuits using the QASM simulator. This observation confirms that the mapped circuits are consistent with the originals across the problem instances studied.

\begin{table}[tb]
\footnotesize
  \begin{ruledtabular}
  
\centering
 \caption{Hellinger distance between original and mapped circuits with AOQMAP-L using QAOA at a depth ranging from 1 to 7.}
 \label{tab:hellinger_distance}
  \resizebox{\columnwidth}{!}{
  \begin{tabular}{llllllll}
    $p$ & 1 & 2 & 3 & 4 & 5 & 6 & 7\\
    \hline
    3-qubit & 0.0045 & 0.0036 & 0.0033 & 0.005 & 0.0036 & 0.0048 & 0.0051 \\
    4-qubit & 0.0044 & 0.0064 & 0.0057 & 0.0051 & 0.005 & 0.006 & 0.0044 \\
    5-qubit & 0.0087 & 0.0067 & 0.0086 & 0.0103 & 0.0072 & 0.0095 & 0.0071\\
    6-qubit & 0.0138 & 0.0117 & 0.0138 & 0.0102 & 0.0123 & 0.0131 & 0.0124\\
    10-qubit & 0.0120 & 0.0127 & 0.0137 & 0.0127 & 0.0106 & 0.0158 & 0.0123\\
  \end{tabular}}
\end{ruledtabular}
\end{table}

While we directly execute adapted and original circuits in the ideal case, without considering noise, and employ Hellinger distance to verify correctness, incorporating noise models is crucial for a more accurate assessment of AOQMAP's performance in real-world scenarios. However, verifying quantum circuits under noise is challenging due to the exponential scaling of traditional strategies such as quantum state tomography \cite{leibfried1996experimental,cramer2010efficient, gross2010quantum,christandl2012reliable}, quantum process tomography \cite{poyatos1997complete,mohseni2008quantum}, and randomized benchmarking \cite{magesan2011scalable,knill2008randomized,helsen2022general} with the size of qubits to be characterized.
Moreover, for large numbers of qubits, directly executing reference and original circuits with a QASM simulator and comparing their probability distributions becomes inaccurate and impractical.
Current research on verifying quantum circuits typically involves decomposing the circuit into subcircuits and quantifying the difference between noisy and ideal subcircuit outputs using metrics such as total variation distance \cite{maciejewski2020mitigation, dahlhauser2021modeling} and fidelity \cite{takeuchi2022divide}. The resulting circuits produced by AOQMAP for VQAs on linear, T-, and H-shaped topologies maintain their structure with increasing qubit numbers and VQA depths, enabling efficient identification of subcircuits. These specifically structured circuits can also be employed to test the performance of NISQ devices as they minimize the impact of compilation quality on their performance and contain different types of topologies to capture more noise information such as crosstalk, providing valuable insights into the scalability of VQAs on real devices. Future research can verify subtopology-adapted VQAs on noisy quantum devices and benchmark the scalability of VQAs on real devices.

\section{Cloud platform details \label{app:cloud_platf}}

Here we present calibration data for IBM QPUs used in the demonstration described in Sec.~\ref{subsec:demons_on_ibm_quant}.
Tables~\ref{tab:qubits_used_fig12} and \ref{tab:properties_fig12} summarize the qubits used for the QAOA implementation on ibm\_nairobi (Fig.~\ref{fig:arsp_4q_ibm_nairobi}) along with their respective properties.
In particular, $T_1$ and $T_2$ correspond to the relaxation and dephasing times, respectively. ``Freq." and ``Anh." denote the qubit frequency and anharmonicity, respectively. $P_{01}$ and $P_{10}$ represent the probability of measuring $\ket{0}$ after preparing $\ket{1}$ and measuring $\ket{1}$ after preparing $\ket{0}$, respectively. ``RO err." indicates the readout error, while ``Gate err." refers to the single-qubit gate error rate. ``CX gate" and ``ECR gate" specify interactions between qubits via the CX and ECR gates, respectively, with ``CX err." and ``ECR err." representing their associated error rates.
Similarly, Table~\ref{tab:qubits_used_fig13} lists the qubits used for the QAOA implementation on IBM QPUs in Figs.~\ref{fig:ibmq_run_ar_all} and \ref{fig:ibmq_run_sp_all}. The calibration data of ibm\_perth, ibm\_nazca, ibmq\_kolkata, ibmq\_ehningen, ibm\_nairobi, and ibm\_cusco are detailed in Tables~\ref{tab:properties_fig13_perth}, \ref{tab:properties_fig13_nazca}, \ref{tab:properties_fig13_kolkata}, \ref{tab:properties_fig13_ehningen}, \ref{tab:properties_fig13_nairobi}, and \ref{tab:properties_fig13_cusco}. We note that for the demonstration of QAOA with the same number of qubits on ibmq\_ehningen, different transpilation methods are applied at different times, potentially leading to variations in the properties of the same qubits or CX gates. In contrast, for QAOA demonstrations on other QPUs, all transpilation methods are applied simultaneously, ensuring that the properties of the same qubits and two-qubit gates remain unchanged.

\begin{table*}[tb]
\caption{\label{tab:qubits_used_fig12}%
Qubits used in the demonstration of $4$-qubit QAOA with depth $p$ on ibm\_nairobi in Fig.~\ref{fig:arsp_4q_ibm_nairobi}.}
\begin{ruledtabular}
\begin{tabular}{lllllll}
$p$ &
\multicolumn{1}{l}{\textrm{AOQMAP-L}}&
\multicolumn{1}{l}{\textrm{AOQMAP-LS}}&
\multicolumn{1}{l}{\textrm{AOQMAP-T}}&
\multicolumn{1}{l}{\textrm{Tket}}&
\multicolumn{1}{l}{\textrm{Qiskit}}&
\multicolumn{1}{l}{\textrm{SWAPNK}}\\
\colrule
$1$ & $[1, 3, 4, 5]$ & $[1, 3, 4, 5]$ & \multirow{2}{*}{$[0,1,2,3]$} & \multirow{2}{*}{$[0,1,2,3]$} & $[1, 2, 3, 5]$ & $[1, 3, 4, 5]$\\
$2,\dots,7$ & $[1, 2, 3, 5]$ & $[1, 2, 3, 5]$ & & & $[0, 1, 2, 3]$ & $[1, 2, 3, 5]$\\
\end{tabular}
\end{ruledtabular}
\end{table*}

\begin{table*}[tb]
\caption{\label{tab:properties_fig12}%
Calibration data for the $4$-qubit QAOA demonstration on ibm\_nairobi in Fig.~\ref{fig:arsp_4q_ibm_nairobi}.}
\begin{ruledtabular}
\begin{tabular}{lllllllllll}
Qubit & $T_1~(\rm{\mu s})$ & $T_2~(\rm{\mu s})$ & Freq. (GHz) & Anh. (GHz) & $P_{01}$ (\%) & $P_{10}$ (\%) & RO err. (\%) & Gate err. (\%) & CX gate & CX err. (\%)\\
\colrule
0 & 118.2 & 30.97 & 5.26 & $-0.3398$ & 3.1  & 1.56 & 2.33 & 0.028 & (0,1) & 1.03\\
1 & 115.27 & 108.96 & 5.17 & $-0.3406$ & 3.42 & 1.18 & 2.3  & 0.03  & (1,2) & 0.8\\
2 & 109.8  & 131.81 & 5.274 & $-0.3389$ & 4.8  & 0.72 & 2.76 & 0.053 & (1,3) & 0.6\\
3 & 100.25 & 66.01  & 5.027 & $-0.3425$ & 4.02 & 0.82 & 2.42 & 0.042 & (3,5) & 1.62\\
4 & 83.88  & 86.5   & 5.177 & $-0.3406$ & 3.04 & 1.08 & 2.06 & 0.025 & (4,5) & 0.92\\
5 & 70.59  & 18.87  & 5.293 & $-0.3405$ & 14.94 & 6.52 & 10.73& 0.074 &  &  \\
\end{tabular}
\end{ruledtabular}
\end{table*}

\begin{table*}[tb]
\caption{\label{tab:qubits_used_fig13}%
Qubits used in the demonstration of $n$-qubit QAOA with depth $p$, represented as $D_n^p$, in Figs.~\ref{fig:ibmq_run_ar_all} and \ref{fig:ibmq_run_sp_all}.}
\begin{ruledtabular}
\begin{tabular}{llllll}
IBM QPU & QAOA &
\multicolumn{1}{l}{\textrm{AOQMAP}}&
\multicolumn{1}{l}{\textrm{Tket}}&
\multicolumn{1}{l}{\textrm{Qiskit}}&
\multicolumn{1}{l}{\textrm{SWAPNK}}\\
\colrule
\multirow{3}{*}{ibm\_perth} & $D_3^1, \dots, D_3^7$ & $[0, 1, 3]$ & $[3,5,6]$ & $[1, 3, 5]$ & $[0, 1, 3]$\\
\cline{2-6}
& $D_5^1, D_5^3, \dots, D_5^7$ & \multirow{2}{*}{$[0,1,3,4,5]$} & \multirow{2}{*}{$[1,3,4,5,6]$} & $[0,1,2,3,5]$ & \multirow{2}{*}{$[0,1,3,4,5]$}\\
& $D_5^2$ & & & $[1,3,4,5,6]$ \\
\hline
\multirow{4}{*}[-12pt]{ibm\_nazca} & $D_{10}^1$ & \multirow{4}{*}[-12pt]{\makecell[l]{$[0, 1, 2, 3, 4,$ \\ $5, 6, 7, 8, 9]$}} & \makecell[l]{$[60, 61, 62, 63, 64,$ \\ $72, 80, 81, 82, 83]$} & \makecell[l]{$[22, 23, 24, 34, 40,$ \\ $41, 42, 43, 44, 45]$} & \multirow{4}{*}[-12pt]{\makecell[l]{$[0, 1, 2, 3, 4,$ \\ $5, 6, 7, 8, 9]$}}\\
\cline{4-4}
& $D_{10}^2$ & & \multirow{3}{*}[-9pt]{\makecell[l]{$[60, 61, 62, 63, 64,$ \\ $72, 79, 80, 81, 82]$}} & \makecell[l]{$[1, 2, 3, 4, 5,$ \\ $15, 20, 21, 22, 23]$} &\\
& $D_{10}^3$ & & & \makecell[l]{$[20, 21, 22, 23, 24,$ \\ $34, 39, 40, 42, 43]$} & \\
& $D_{10}^4, \dots, D_{10}^7$ & & & \makecell[l]{$[4, 15, 21, 22, 23,$ \\ $24, 34, 42, 43, 44]$} &\\
\hline
\multirow{12}{*}[-24pt]{ibmq\_kolkata} & $D_3^1, \dots, D_3^7$ & $[21, 23, 24]$ & $[21, 23, 24]$ & $[21, 23, 24]$ & $[21, 23, 24]$\\
\cline{2-6}
& $D_5^1$ & \multirow{2}{*}{$[12,13,15,17,18]$} & \multirow{3}{*}{$[7, 10, 12, 13, 15]$} & \multirow{4}{*}{$[10, 12, 13, 15, 18]$} & $[12, 13, 15, 17, 18]$\\
\cline{6-6}
& $D_5^2$ & & & & \multirow{3}{*}{$[12, 13, 14, 15, 16]$}\\
\cline{3-3}
& $D_5^3$ & \multirow{2}{*}{$[12, 13, 14, 15, 16]$} & & &\\
\cline{4-4}
& $D_5^4, \dots D_5^7$ &  & $[16, 19, 20, 22, 25]$ & & \\
\cline{2-6}
& $D_{10}^1$ & \makecell[l]{$[1, 4, 7, 10, 12,$ \\ $15, 18, 21, 23, 24]$} & \makecell[l]{$[7, 8, 10, 11, 12,$ \\ $13, 14, 15, 16, 19]$} & \makecell[l]{$[7, 10, 12, 13, 14,$ \\ $15, 17, 18, 21, 23]$} & \makecell[l]{$[1, 4, 7, 10, 12,$ \\ $15, 18, 21, 23, 24]$} \\
\cline{6-6}
& $D_{10}^2$ & \makecell[l]{$[12, 13, 14, 15, 16,$ \\ $18, 19, 21, 23, 24]$} & \makecell[l]{$[7, 8, 10, 11, 12,$ \\ $13, 14, 15, 16, 18]$} & \makecell[l]{$[7, 10, 12, 13, 15,$ \\ $17, 18, 21, 23, 24]$} & \multirow{5}{*}[-15pt]{\makecell[l]{$[0, 1, 4, 7, 10,$ \\ $12, 13, 14, 16, 19]$}} \\
\cline{3-3}
& $D_{10}^3$ & \multirow{4}{*}[-8pt]{\makecell[l]{$[0, 1, 4, 7, 10,$ \\ $12, 13, 14, 16, 19]$}} & \makecell[l]{$[8, 10, 11, 12, 13,$ \\ $14, 15, 16, 19, 20]$} & \makecell[l]{$[10, 12, 13, 14, 15,$ \\ $17, 18, 21, 23, 24]$} & \\
\cline{5-5}
& $D_{10}^4, D_{10}^5$ &  & \makecell[l]{$[7, 8, 10, 11, 12,$ \\ $13, 14, 15, 16, 18]$} & \multirow{2}{*}[-6pt]{\makecell[l]{$[7, 10, 12, 13, 14,$ \\ $15, 17, 18, 21, 23]$}} & \\
& $D_{10}^6$ &  & \makecell[l]{$[8, 10, 11, 12, 13,$ \\ $14, 15, 16, 19, 20]$} &  & \\
\cline{5-5}
& $D_{10}^7$ &  & \makecell[l]{$[7, 8, 10, 11, 12,$ \\ $13, 14, 15, 16, 18]$} & \makecell[l]{$[12, 13, 14, 15, 16,$ \\ $19, 22, 23, 24, 25]$} & \\
\hline
\multirow{3}{*}{ibmq\_ehningen} & $D_3^1, \dots, D_3^7$ & $[1, 4, 7]$ & $[1, 4, 7]$ & $[1, 4, 7]$ & $[1, 4, 7]$\\
& $D_5^1, \dots, D_5^7$ & $[10, 11, 12, 13, 14]$ & $[13, 14, 16, 19, 20]$ & $[16, 19, 20, 22, 25]$ & $[19, 20, 22, 24, 25]$\\
& $D_6^1, \dots, D_6^7$ & $[1, 4, 7, 10, 12, 13]$ & $[12, 13, 14, 16, 19, 20]$ & $[14, 16, 19, 20, 22, 25]$ & $[19, 20, 22, 23, 24, 25]$\\
\hline
\multirow{1}{*}{ibm\_nairobi} & $D_3^1, \dots, D_3^7$ & $[1, 2, 3]$ & $[0, 1, 3]$ & $[1, 2, 3]$ & $[1, 2, 3]$\\
\hline
\multirow{2}{*}{ibm\_cusco} & $D_5^1, \dots, D_5^7$ & $[0, 1, 2, 3, 4]$ & $[44, 45, 46, 54, 64]$ & $[4, 5, 6, 7, 15]$ & $[0, 1, 2, 3, 4]$\\
\cline{2-6}
& $D_{10}^1$ & \multirow{6}{*}[-21pt]{\makecell[l]{$[0, 1, 2, 3, 4,$ \\ $5, 6, 7, 8, 9]$}} & \makecell[l]{$[60, 61, 62, 63, 72,$ \\ $79, 80, 81, 82, 83]$} & \makecell[l]{$[1, 2, 3, 4, 5,$ \\ $6, 7, 8, 15, 22]$} & \multirow{6}{*}[-21pt]{\makecell[l]{$[0, 1, 2, 3, 4,$ \\ $5, 6, 7, 8, 9]$}}\\
\cline{4-4}
& $D_{10}^2$ & & \multirow{6}{*}[-15pt]{\makecell[l]{$[60, 61, 62, 63, 64,$ \\ $72, 79, 80, 81, 82]$}} & \makecell[l]{$[3, 4, 5, 6, 7,$ \\ $15, 21, 22, 23, 24]$} & \\
& $D_{10}^3$ & & & \makecell[l]{$[0, 1, 2, 3, 4,$ \\ $14, 15, 19, 20, 22]$} & \\
& $D_{10}^4, D_{10}^5$ & & & \makecell[l]{$[3, 4, 5, 6, 7,$ \\ $8, 9, 15, 16, 22]$} & \\
& $D_{10}^6$ & & & \makecell[l]{$[3, 4, 5, 6, 7,$ \\ $15, 21, 22, 23, 24]$} & \\
& $D_{10}^7$ & & & \makecell[l]{$[3, 4, 5, 6, 15,$ \\ $21, 22, 23, 24, 25]$} & \\
\end{tabular}
\end{ruledtabular}
\end{table*}

\begin{table*}[tb]
\caption{\label{tab:properties_fig13_perth}%
Calibration data for the $n$-qubit QAOA demonstration on ibm\_perth in Figs.~\ref{fig:ibmq_run_ar_all} and \ref{fig:ibmq_run_sp_all}.}
\begin{ruledtabular}
\begin{tabular}{llllllllllll}
$n$ &Qubit & $T_1~(\rm{\mu s})$ & $T_2~(\rm{\mu s})$ & Freq. (GHz) & Anh. (GHz) & $P_{01}$ (\%) & $P_{10}$ (\%) & RO err. (\%) & Gate err. (\%) & CX gate & CX err. (\%)\\
\colrule
\multirow{5}{*}{3} & 0 & 147.9  & 82.13  & 5.158 & $-0.3415$ & 3.26 & 2.48 & 2.87 & 0.073 & (0,1) & 0.6\\
& 1 & 166.58 & 51.27  & 5.034 & $-0.3444$ & 2.82 & 2.84 & 2.83 & 0.031 & (1,3) & 0.46\\
& 3 & 113.38 & 95.99  & 5.125 & $-0.3404$ & 1.52 & 1.72 & 1.62 & 0.021 & (3,5) & 0.65\\
& 5 & 144.73 & 108.09 & 4.979 & $-0.3460$ & 3.08 & 2.74 & 2.91 & 0.026 & (5,6) & 1.14\\
& 6 & 145.99 & 298.35 & 5.157 & $-0.3405$ & 1.10 & 1.14 & 1.12 & 0.032 &    & \\
\hline
\multirow{7}{*}{5} & 0 & 145.05 & 89.48  & 5.158 & $-0.3415$ & 2.80 & 2.44 & 2.62 & 0.035 & (0,1) & 0.73\\
& 1 & 129.60 & 49.00  & 5.034 & $-0.3444$ & 2.70 & 2.68 & 2.69 & 0.032 & (1,2) & 0.81\\
& 2 & 132.88 & 76.48  & 4.863 & $-0.3473$ & 2.84 & 11.90 & 7.37 & 0.036 & (1,3) & 0.48\\
& 3 & 174.95 & 185.66 & 5.125 & $-0.3404$ & 2.48 & 1.80 & 2.14 & 0.026 & (3,5) & 0.72\\
& 4 & 138.13 & 98.61  & 5.159 & $-0.3334$ & 3.00 & 2.46 & 2.73 & 0.036 & (4,5) & 0.83\\
& 5 & 134.09 & 135.52 & 4.979 & $-0.3460$ & 3.46 & 3.42 & 3.44 & 0.028 & (5,6) & 1.18\\
& 6 & 158.21 & 237.22 & 5.157 & $-0.3405$ & 1.56 & 0.98 & 1.27 & 0.030 &  & \\
\end{tabular}
\end{ruledtabular}

\caption{\label{tab:properties_fig13_nazca}%
Calibration data for the 10-qubit QAOA demonstration on ibm\_nazca in Figs.~\ref{fig:ibmq_run_ar_all} and \ref{fig:ibmq_run_sp_all}.}
\begin{ruledtabular}
\begin{tabular}{lllllllllll}
Qubit & $T_1~(\rm{\mu s})$ & $T_2~(\rm{\mu s})$ & Freq. (GHz) & Anh. (GHz) & $P_{01}$ (\%) & $P_{10}$ (\%) & RO err. (\%) & Gate err. (\%) & ECR gate & ECR err. (\%)\\
\colrule
0  & 251.62 & 65.56  & 5.092 & $-0.3064$ & 1.28 & 1.14 & 1.21 & 0.044 & (0, 1) & 1.10\\
1  & 97.45  & 133.83 & 4.971 & $-0.3076$ & 3.22 & 2.74 & 2.98 & 0.017 & (1, 2) & 0.43\\
2  & 289.35 & 224.95 & 4.891 & $-0.3089$ & 2.80 & 2.90 & 2.85 & 0.015 & (3, 2) & 0.69\\
3  & 213.89 & 142.91 & 5.053 & $-0.3067$ & 2.06 & 2.68 & 2.37 & 0.041 & (3, 4) & 0.63\\
4  & 293.74 & 267.30 & 4.978 & $-0.3078$ & 3.82 & 3.72 & 3.77 & 0.015 & (4, 15) & 0.57\\
5  & 142.85 & 102.58 & 5.041 & $-0.3067$ & 2.60 & 2.22 & 2.41 & 0.037 & (5, 4) & 0.67\\
6  & 155.86 & 63.35  & 5.170 & $-0.3038$ & 3.78 & 2.50 & 3.14 & 0.149 & (6, 5) & 1.13\\
7  & 172.99 & 68.09  & 5.081 & $-0.3059$ & 0.56 & 0.40 & 0.48 & 0.044 & (6, 7) & 1.04\\
8  & 228.30 & 64.16  & 5.003 & $-0.3076$ & 0.56 & 0.46 & 0.51 & 0.039 & (7, 8) & 0.99\\
9  & 212.86 & 23.48  & 5.102 & $-0.3062$ & 2.30 & 2.48 & 2.39 & 0.067 & (9, 8) & 1.61\\
15 & 201.55 & 112.02 & 5.153 & $-0.3054$ & 1.00 & 0.68 & 0.84 & 0.021 & (15, 22) & 0.49\\
20 & 228.55 & 87.75  & 5.029 & $-0.3062$ & 5.88 & 5.68 & 5.78 & 0.039 & (20, 21) & 0.63\\
21 & 213.32 & 236.54 & 5.169 & $-0.3041$ & 4.98 & 5.62 & 5.30 & 0.021 & (21, 22) & 0.52\\
22 & 270.45 & 269.29 & 5.061 & $-0.3063$ & 1.32 & 1.16 & 1.24 & 0.014 & (23, 22) & 0.53\\
23 & 227.09 & 21.28  & 5.005 & $-0.3072$ & 3.92 & 3.38 & 3.65 & 0.021 & (24, 23) & 0.78\\
24 & 178.20 & 198.23 & 5.074 & $-0.3059$ & 3.18 & 2.54 & 2.86 & 0.022 & (25, 24) & 1.14\\
25 & 145.51 & 135.14 & 5.164 & $-0.3054$ & 2.48 & 2.38 & 2.43 & 0.049 & (34, 24) & 0.63\\
34 & 257.88 & 157.67 & 4.935 & $-0.3081$ & 1.66 & 1.32 & 1.49 & 0.016 & (34, 43) & 0.59\\
39 & 217.16 & 147.13 & 5.019 & $-0.3071$ & 3.66 & 3.26 & 3.46 & 0.027 & (40, 39) & 0.74\\
40 & 281.53 & 344.42 & 5.085 & $-0.3062$ & 2.34 & 2.82 & 2.58 & 0.023 & (40, 41) & 0.55\\
41 & 229.28 & 265.23 & 4.953 & $-0.3083$ & 1.66 & 2.38 & 2.02 & 0.016 & (41, 42) & 0.83\\
42 & 177.87 & 220.57 & 5.092 & $-0.3062$ & 3.58 & 2.86 & 3.22 & 0.033 & (42, 43) & 0.85\\
43 & 372.59 & 244.69 & 4.870 & $-0.3094$ & 2.26 & 1.92 & 2.09 & 0.015 & (43, 44) & 0.49\\
44 & 150.64 & 20.88  & 5.024 & $-0.3073$ & 1.98 & 1.40 & 1.69 & 0.016 & (45, 44) & 0.67\\
45 & 165.29 & 91.74  & 5.112 & $-0.3051$ & 5.76 & 6.40 & 6.08 & 0.036 &  & \\
60 & 233.08 & 164.62 & 5.113 & $-0.3055$ & 1.44 & 1.24 & 1.34 & 0.026 & (61, 60) & 0.73\\
61 & 293.83 & 194.05 & 4.912 & $-0.3085$ & 6.72 & 7.00 & 6.86 & 0.026 & (62, 61) & 0.58\\
62 & 245.28 & 147.00 & 5.015 & $-0.3067$ & 6.04 & 10.26 & 8.15 & 0.021 & (63, 62) & 1.40\\
63 & 129.04 & 179.81 & 5.080 & $-0.3068$ & 1.16 & 0.66 & 0.91 & 0.059 & (64, 63) & 2.01\\
64 & 89.95  & 20.21  & 5.281 & $-0.3038$ & 2.16 & 2.58 & 2.37 & 0.061 & (72, 62) & 0.81\\
72 & 214.77 & 118.50 & 5.078 & $-0.3049$ & 1.44 & 1.28 & 1.36 & 0.043 & (72, 81) & 0.76\\
79 & 190.09 & 17.58  & 4.927 & $-0.3086$ & 3.26 & 2.44 & 2.85 & 0.025 & (80, 79) & 0.57\\
80 & 185.49 & 160.87 & 4.993 & $-0.3077$ & 2.40 & 2.42 & 2.41 & 0.019 & (80, 81) & 0.86\\
81 & 198.15 & 84.38  & 5.153 & $-0.3051$ & 3.38 & 5.00 & 4.19 & 0.027 & (81, 82) & 0.76\\
82 & 212.53 & 235.84 & 5.205 & $-0.3046$ & 0.86 & 1.28 & 1.07 & 0.019 & (82, 83) & 2.36\\
83 & 230.82 & 184.81 & 5.103 & $-0.3063$ & 1.26 & 1.04 & 1.15 & 0.042 &  & \\
\end{tabular}
\end{ruledtabular}
\end{table*}

\begin{table*}[tb]
\caption{\label{tab:properties_fig13_kolkata}%
Calibration data for the $n$-qubit QAOA demonstration on ibmq\_kolkata in Figs.~\ref{fig:ibmq_run_ar_all} and \ref{fig:ibmq_run_sp_all}.}
\begin{ruledtabular}
\begin{tabular}{llllllllllll}
$n$ & Qubit & $T_1~(\rm{\mu s})$ & $T_2~(\rm{\mu s})$ & Freq. (GHz) & Anh. (GHz) & $P_{01}$ (\%) & $P_{10}$ (\%) & RO err. (\%) & Gate err. (\%) & CX gate & CX err. (\%)\\
\colrule
\multirow{3}{*}{3} & 21 & 79.08  & 19.49  & 5.274 & $-0.3408$ & 0.38 & 0.54 & 0.46 & 0.022 & (21, 23) & 0.47\\
& 23 & 108.35 & 49.07  & 5.138 & $-0.3431$ & 0.80 & 0.34 & 0.57 & 0.016 & (23, 24) & 0.39\\
& 24 & 134.94 & 83.96  & 5.005 & $-0.3459$ & 1.04 & 0.68 & 0.86 & 0.015 & & \\
\hline
\multirow{13}{*}{5} & 7  & 124.68 & 41.33  & 5.031 & $-0.3457$ & 2.68 & 2.14 & 2.41 & 0.021 & (7, 10) & 2.25\\
& 10 & 76.02  & 46.92  & 5.178 & $-0.3416$ & 1.06 & 0.88 & 0.97 & 0.049 & (10, 12) & 2.10\\
& 12 & 118.60 & 178.39 & 4.961 & $-0.3465$ & 0.78 & 0.58 & 0.68 & 0.023 & (12, 13) & 0.70\\
& 13 & 131.15 & 218.25 & 5.018 & $-0.3459$ & 0.92 & 0.76 & 0.84 & 0.029 & (12, 15) & 0.53\\
& 14 & 126.66 & 163.69 & 5.118 & $-0.3428$ & 3.96 & 17.70 & 10.83 & 0.024 & (13, 14) & 0.65\\
& 15 & 143.99 & 145.95 & 5.041 & $-0.3439$ & 0.70 & 0.64 & 0.67 & 0.031 & (14, 16) & 0.65\\
& 16 & 112.30 & 90.13  & 5.222 & $-0.3402$ & 1.02 & 2.04 & 1.53 & 0.016 & (15, 18) & 0.76\\
& 17 & 50.03  & 28.86  & 5.236 & $-0.3400$ & 4.26 & 0.48 & 2.37 & 0.033 & (16, 19) & 0.76\\
& 18 & 135.08 & 136.00 & 5.097 & $-0.3444$ & 1.16 & 1.10 & 1.13 & 0.048 & (17, 18) & 1.20\\
& 19 & 77.11  & 23.42  & 5.002 & $-0.3449$ & 3.20 & 11.86 & 7.53 & 0.041 & (19, 20) & 1.19\\
& 20 & 78.02  & 20.40  & 5.187 & $-0.3408$ & 0.98 & 1.18 & 1.08 & 0.023 & (19, 22) & 0.97\\
& 22 & 93.15  & 35.55  & 5.127 & $-0.3433$ & 3.74 & 3.84 & 3.79 & 0.026 & (22, 25) & 0.60\\
& 25 & 101.55 & 117.17 & 4.921 & $-0.3473$ & 0.68 & 0.68 & 0.68 & 0.018 & & \\
\hline 
\multirow{21}{*}{10} & 0  & 90.1  & 92.67  & 5.204 & $-0.3415$ & 1.04 & 0.74 & 0.89 & 0.021 & (0, 1) & 0.42\\
& 1  & 128.73 & 221.16 & 4.991 & $-0.3453$ & 1.28 & 0.86 & 1.07 & 0.018 & (1, 4) & 0.54\\
& 4  & 89.1   & 134.19 & 5.225 & $-0.3410$ & 2.58 & 1.92 & 2.25 & 0.026 & (4, 7) & 1.06\\
& 7  & 88.36  & 40.82  & 5.031 & $-0.3457$ & 2.42 & 2.36 & 2.39 & 0.033 & (7, 10) & 0.77\\
& 8  & 119.69 & 55.53  & 4.928 & $-0.3454$ & 5.26 & 2.80 & 4.03 & 0.045 & (8, 11) & 1.30\\
& 10 & 116.95 & 45.47  & 5.178 & $-0.3416$ & 1.00 & 0.88 & 0.94 & 0.020 & (10, 12) & 0.88\\
& 11 & 44.99  & 29.61  & 4.868 & $-0.3734$ & 13.96 & 14.20 & 14.08 & 0.025 & (11, 14) & 100\\
& 12 & 84.02  & 156.07 & 4.961 & $-0.3465$ & 0.80 & 0.72 & 0.76 & 0.016 & (12, 13) & 0.66\\
& 13 & 84.04  & 202.42 & 5.018 & $-0.3459$ & 0.92 & 0.66 & 0.79 & 0.024 & (12, 15) & 0.92\\
& 14 & 134.37 & 279.81 & 5.118 & $-0.3428$ & 3.76 & 12.62 & 8.19 & 0.017 & (13, 14) & 0.52\\
& 15 & 121.91 & 287.98 & 5.041 & $-0.3439$ & 0.64 & 0.66 & 0.65 & 0.027 & (14, 16) & 0.62\\
& 16 & 116.7  & 91.08  & 5.222 & $-0.3402$ & 0.78 & 1.32 & 1.05 & 0.019 & (15, 18) & 0.65\\
& 17 & 78.53  & 31.69  & 5.236 & $-0.3400$ & 0.80 & 0.44 & 0.62 & 0.024 & (16, 19) & 0.78\\
& 18 & 118.3  & 104.6  & 5.097 & $-0.3444$ & 1.02 & 0.56 & 0.79 & 0.019 & (17, 18) & 1.19\\
& 19 & 112.14 & 25.47  & 5.002 & $-0.3449$ & 5.34 & 13.90 & 9.62 & 0.028 & (18, 21) & 1.31\\
& 20 & 124.65 & 19.87  & 5.187 & $-0.3408$ & 0.74 & 0.60 & 0.67 & 0.019 & (19, 20) & 0.90\\
& 21 & 69.31  & 17.05  & 5.274 & $-0.3408$ & 0.52 & 0.56 & 0.54 & 0.023 & (19, 22) & 0.82\\
& 22 & 85.69  & 36.12  & 5.127 & $-0.3433$ & 3.82 & 4.22 & 4.02 & 0.028 & (21, 23) & 0.53\\
& 23 & 139.81 & 52.81  & 5.138 & $-0.3431$ & 0.48 & 0.52 & 0.50 & 0.018 & (22, 25) & 0.78\\
& 24 & 108.62 & 76.32  & 5.005 & $-0.3459$ & 0.94 & 1.00 & 0.97 & 0.014 & (23, 24) & 0.42\\
& 25 & 146.49 & 105.09 & 4.921 & $-0.3473$ & 1.02 & 0.40 & 0.71 & 0.028 & (24, 25) & 100\\
\end{tabular}
\end{ruledtabular}
\end{table*}

\begin{table*}[tb]
\caption{\label{tab:properties_fig13_ehningen}%
Calibration data for the $n$-qubit QAOA demonstration on ibmq\_ehningen in Figs.~\ref{fig:ibmq_run_ar_all} and \ref{fig:ibmq_run_sp_all}.}
\resizebox{\textwidth}{!}{
\begin{tabular}{lllllllllllll}
\hline\hline
$n$& Method & Qubit & $T_1~(\rm{\mu s})$ & $T_2~(\rm{\mu s})$ & Freq. (GHz) & Anh. (GHz) & $P_{01}$ (\%) & $P_{10}$ (\%) & RO err. (\%) & Gate err. (\%) & CX gate & CX err. (\%)\\
\colrule
\multirow{12}{*}{3} & \multirow{3}{*}{AOQMAP} 
  & 1 & 185.32 & 219.79 & 5.182 & $-0.34$   & 0.82 & 1.00 & 0.91 & 0.023 & (1, 4) & 0.44\\
&  & 4 & 119.66 & 198.52 & 5.054 & $-0.3426$ & 0.90 & 0.52 & 0.71 & 0.017 & (4, 7) & 0.55\\
&  & 7 & 175.82 & 243.50 & 4.978 & $-0.344$  & 0.76 & 0.66 & 0.71 & 0.017 &       &     \\
\cline{2-13}
& \multirow{3}{*}{Tket} 
  & 1 & 185.32 & 219.79 & 5.182 & $-0.34$   & 0.82 & 1.00 & 0.91 & 0.023 & (1, 4) & 0.44\\
&   & 4 & 114.45 & 198.52 & 5.054 & $-0.3426$ & 0.90 & 0.52 & 0.71 & 0.017 & (4, 7) & 0.55\\
&   & 7 & 175.82 & 243.50 & 4.978 & $-0.344$  & 0.76 & 0.66 & 0.71 & 0.017 &       &     \\
\cline{2-13}
& \multirow{3}{*}{Qiskit} 
  & 1 & 185.32 & 219.79 & 5.182 & $-0.34$   & 0.82 & 1.00 & 0.91 & 0.023 & (1, 4) & 0.44\\
&   & 4 & 119.66 & 198.52 & 5.054 & $-0.3426$ & 0.90 & 0.52 & 0.71 & 0.017 & (4, 7) & 0.55\\
&   & 7 & 175.82 & 243.50 & 4.978 & $-0.344$  & 0.76 & 0.66 & 0.71 & 0.017 &       &     \\
\cline{2-13}
& \multirow{3}{*}{SWAPNK} 
  & 1 & 185.32 & 219.79 & 5.182 & $-0.34$   & 0.82 & 1.00 & 0.91 & 0.023 & (1, 4) & 0.44\\
&   & 4 & 114.45 & 198.52 & 5.054 & $-0.3426$ & 0.90 & 0.52 & 0.71 & 0.017 & (4, 7) & 0.55\\
&   & 7 & 175.82 & 243.50 & 4.978 & $-0.344$  & 0.76 & 0.66 & 0.71 & 0.017 &       &     \\
\colrule
\multirow{20}{*}{5} & \multirow{5}{*}{AOQMAP} 
 & 10 & 115.74 & 57.65  & 4.835 & $-0.3471$ & 0.74 & 0.46 & 0.60 & 0.017 & (10, 12) & 0.50 \\
&  & 11 & 127.56 & 74.74  & 5.119 & $-0.3405$ & 1.54 & 1.34 & 1.44 & 0.025 & (11, 14) & 0.84 \\
&  & 12 & 183.56 & 442.33 & 4.725 & $-0.3484$ & 1.00 & 0.90 & 0.95 & 0.022 & (12, 13) & 0.66 \\
&  & 13 & 147.90 & 271.77 & 4.926 & $-0.3440$ & 1.74 & 0.96 & 1.35 & 0.018 & (13, 14) & 0.59 \\
&  & 14 & 107.65 & 196.10 & 5.177 & $-0.3408$ & 1.08 & 0.44 & 0.76 & 0.031 &         &      \\
\cline{2-13}
& \multirow{5}{*}{Tket} 
 & 13 & 127.27 & 257.59 & 4.926 & $-0.3440$ & 1.34 & 0.78 & 1.06 & 0.023 & (13, 14) & 0.59 \\
&  & 14 & 219.32 & 242.17 & 5.177 & $-0.3408$ & 1.14 & 0.32 & 0.73 & 0.029 & (14, 16) & 0.59 \\
&  & 16 & 139.95 & 209.99 & 5.022 & $-0.3435$ & 0.60 & 0.44 & 0.52 & 0.016 & (16, 19) & 0.56 \\
&  & 19 & 137.67 & 81.26  & 4.784 & $-0.3485$ & 1.06 & 0.64 & 0.85 & 0.021 & (19, 20) & 0.47 \\
&  & 20 & 219.74 & 227.51 & 5.042 & $-0.3426$ & 1.76 & 2.62 & 2.19 & 0.031 &         &      \\
\cline{2-13}
& \multirow{5}{*}{Qiskit} 
 & 16 & 168.67 & 213.73 & 5.022 & $-0.3435$ & 0.70 & 0.42 & 0.56 & 0.016 & (16, 19) & 0.80 \\
&  & 19 & 130.23 & 79.19  & 4.784 & $-0.3485$ & 1.10 & 0.52 & 0.81 & 0.022 & (19, 20) & 0.54 \\
&  & 20 & 153.35 & 275.77 & 5.042 & $-0.3426$ & 1.36 & 2.26 & 1.81 & 0.036 & (19, 22) & 0.70 \\
&  & 22 & 187.89 & 32.12  & 4.725 & $-0.3464$ & 2.00 & 0.64 & 1.32 & 0.020 & (22, 25) & 0.51 \\
&  & 25 & 210.17 & 76.32  & 4.950 & $-0.3457$ & 0.90 & 0.80 & 0.85 & 0.016 &         &      \\
\cline{2-13}
& \multirow{5}{*}{SWAPNK}
 & 19 & 130.23 & 79.19  & 4.784 & $-0.3485$ & 1.10 & 0.52 & 0.81 & 0.022 & (19, 20) & 0.54 \\
&  & 20 & 153.35 & 275.77 & 5.042 & $-0.3426$ & 1.36 & 2.26 & 1.81 & 0.036 & (19, 22) & 0.70 \\
&  & 22 & 187.89 & 32.12  & 4.725 & $-0.3464$ & 2.00 & 0.64 & 1.32 & 0.020 & (22, 25) & 0.51 \\
&  & 24 & 197.66 & 280.74 & 5.074 & $-0.3416$ & 1.04 & 0.64 & 0.84 & 0.013 & (24, 25) & 0.76 \\
&  & 25 & 210.17 & 76.32  & 4.950 & $-0.3457$ & 0.90 & 0.80 & 0.85 & 0.016 &         &      \\
\colrule
\multirow{24}{*}{6} & \multirow{6}{*}{AOQMAP}
 & 1  & 112.64 & 141.73 & 5.182 & $-0.3400$ & 1.06 & 0.56 & 0.81 & 0.024 & (1, 4)   & 0.76 \\
&  & 4  & 106.71 & 151.78 & 5.054 & $-0.3426$ & 0.58 & 0.52 & 0.55 & 0.018 & (4, 7)   & 0.40 \\
&  & 7  & 175.57 & 105.71 & 4.978 & $-0.3440$ & 0.96 & 0.48 & 0.72 & 0.016 & (7, 10)  & 1.59 \\
&  & 10 & 88.47  & 62.04  & 4.835 & $-0.3471$ & 0.86 & 0.42 & 0.64 & 0.022 & (10, 12) & 0.60 \\
&  & 12 & 186.44 & 239.62 & 4.725 & $-0.3484$ & 0.88 & 0.98 & 0.93 & 0.021 & (12, 13) & 0.52 \\
&  & 13 & 157.18 & 224.98 & 4.926 & $-0.3440$ & 1.42 & 1.10 & 1.26 & 0.021 &         &      \\
\cline{2-13}
& \multirow{6}{*}{Tket}
 & 12 & 219.63 & 317.95 & 4.725 & $-0.3484$ & 1.04 & 0.88 & 0.96 & 0.020 & (12, 13) & 0.66 \\
&  & 13 & 127.27 & 257.59 & 4.926 & $-0.3440$ & 1.34 & 0.78 & 1.06 & 0.023 & (13, 14) & 0.59 \\
&  & 14 & 219.32 & 242.17 & 5.177 & $-0.3408$ & 1.14 & 0.32 & 0.73 & 0.029 & (14, 16) & 0.59 \\
&  & 16 & 139.95 & 209.99 & 5.022 & $-0.3435$ & 0.60 & 0.44 & 0.52 & 0.016 & (16, 19) & 0.56 \\
&  & 19 & 137.67 & 81.26  & 4.784 & $-0.3485$ & 1.06 & 0.64 & 0.85 & 0.021 & (19, 20) & 0.47 \\
&  & 20 & 219.74 & 227.51 & 5.042 & $-0.3426$ & 1.76 & 2.62 & 2.19 & 0.031 &         &      \\
\cline{2-13}
& \multirow{6}{*}{Qiskit}
 & 14 & 217.62 & 256.21 & 5.177 & $-0.3408$ & 1.00 & 0.44 & 0.72 & 0.030 & (14, 16) & 0.76 \\
&  & 16 & 168.67 & 213.73 & 5.022 & $-0.3435$ & 0.70 & 0.42 & 0.56 & 0.016 & (16, 19) & 0.80 \\
&  & 19 & 130.23 & 79.19  & 4.784 & $-0.3485$ & 1.10 & 0.52 & 0.81 & 0.022 & (19, 20) & 0.54 \\
&  & 20 & 153.35 & 275.77 & 5.042 & $-0.3426$ & 1.36 & 2.26 & 1.81 & 0.036 & (19, 22) & 0.70 \\
&  & 22 & 187.89 & 32.12  & 4.725 & $-0.3464$ & 2.00 & 0.64 & 1.32 & 0.020 & (22, 25) & 0.51 \\
&  & 25 & 210.17 & 76.32  & 4.950 & $-0.3457$ & 0.90 & 0.80 & 0.85 & 0.016 &         &      \\
\cline{2-13}
& \multirow{6}{*}{SWAPNK}
 & 19 & 130.23 & 79.19  & 4.784 & $-0.3485$ & 1.10 & 0.52 & 0.81 & 0.022 & (19, 20) & 0.54 \\
&  & 20 & 140.45 & 275.77 & 5.042 & $-0.3426$ & 1.36 & 2.26 & 1.81 & 0.036 & (19, 22) & 0.70 \\
&  & 22 & 306.37 & 32.12  & 4.725 & $-0.3464$ & 2.00 & 0.64 & 1.32 & 0.020 & (22, 25) & 0.51 \\
&  & 23 & 173.97 & 259.46 & 4.805 & $-0.3471$ & 0.92 & 0.76 & 0.84 & 0.024 & (23, 24) & 0.59 \\
&  & 24 & 139.76 & 280.74 & 5.074 & $-0.3416$ & 1.04 & 0.64 & 0.84 & 0.013 & (24, 25) & 0.76 \\
&  & 25 & 210.17 & 76.32  & 4.950 & $-0.3457$ & 0.90 & 0.80 & 0.85 & 0.016 &         &      \\
 \hline\hline
\end{tabular}}
\end{table*}

\begin{table*}[tb]
\caption{\label{tab:properties_fig13_nairobi}%
Calibration data for the 3-qubit QAOA demonstration on ibm\_nairobi in Figs.~\ref{fig:ibmq_run_ar_all} and \ref{fig:ibmq_run_sp_all}.}
\begin{ruledtabular}
\begin{tabular}{lllllllllll}
Qubit & $T_1~(\rm{\mu s})$ & $T_2~(\rm{\mu s})$ & Freq. (GHz) & Anh. (GHz) & $P_{01}$ (\%) & $P_{10}$ (\%) & RO err. (\%) & Gate err. (\%) & CX gate & CX err. (\%)\\
\colrule
0 & 118.2  & 30.97  & 5.26  & $-0.3398$ & 3.10 & 1.56 & 2.33 & 0.028 & (0, 1) & 1.03\\
1 & 115.27 & 108.96 & 5.17  & $-0.3406$ & 3.42 & 1.18 & 2.30 & 0.030 & (1, 2) & 0.80\\
2 & 109.8  & 131.81 & 5.274 & $-0.3389$ & 4.80 & 0.72 & 2.76 & 0.053 & (1, 3) & 0.60\\
3 & 100.25 & 66.01  & 5.027 & $-0.3425$ & 4.02 & 0.82 & 2.42 & 0.042 &  
 & \\
\end{tabular}
\end{ruledtabular}

\caption{\label{tab:properties_fig13_cusco}%
Calibration data for the $n$-qubit QAOA demonstration on ibm\_cusco in Figs.~\ref{fig:ibmq_run_ar_all} and \ref{fig:ibmq_run_sp_all}.}
\begin{ruledtabular}
\begin{tabular}{lllllllllllll}
$n$ & Qubit & $T_1~(\rm{\mu s})$ & $T_2~(\rm{\mu s})$ & Freq. (GHz) & Anh. (GHz) & $P_{01}$ (\%) & $P_{10}$ (\%) & RO err. (\%) & Gate err. (\%) & ECR gate & ECR err. (\%)\\
\colrule
\multirow{14}{*}{5} & 0  & 189.11 & 132.77 & 5.015  & $-0.3074$ & 0.94  & 1.14  & 1.04  & 0.016 & (0,1)   & 0.90\\
& 1  & 263.60 & 279.60 & 4.956  & $-0.3078$ & 1.74  & 1.90  & 1.82  & 0.087 & (2,1)   & 0.78\\
& 2  & 281.81 & 378.48 & 4.807  & $-0.3104$ & 0.58  & 0.16  & 0.37  & 0.013 & (3,2)   & 0.70\\
& 3  & 186.89 & 218.49 & 5.232  & $-0.3037$ & 5.56  & 6.38  & 5.97  & 0.019 & (4,3)   & 0.77\\
& 4  & 97.79  & 145.27 & 5.138  & $-0.3054$ & 0.72  & 1.26  & 0.99  & 0.013 & (4,15)  & 0.42\\
& 5  & 125.46 & 153.12 & 4.988  & $-0.3076$ & 1.40  & 2.44  & 1.92  & 0.020 & (5,4)   & 0.52\\
& 6  & 293.68 & 109.06 & 4.899  & $-0.3075$ & 2.06  & 1.54  & 1.80  & 0.017 & (5,6)   & 0.47\\
& 7  & 123.62 & 65.29  & 5.023  & $-0.3068$ & 4.78  & 6.52  & 5.65  & 0.069 & (7,6)   & 0.65\\
& 15 & 192.38 & 63.09  & 5.015  & $-0.3070$ & 1.50  & 1.38  & 1.44  & 0.015 & (44,45) & 0.85\\
& 44 & 205.90 & 252.15 & 5.387  & $-0.3018$ & 3.50  & 3.54  & 3.52  & 0.014 & (46,45) & 1.55\\
& 45 & 211.66 & 263.52 & 4.960  & $-0.3071$ & 9.32  & 3.20  & 6.26  & 0.047 & (54,45) & 0.57\\
& 46 & 94.46  & 111.22 & 5.220  & $-0.3042$ & 1.22  & 4.28  & 2.75  & 0.036 & (54,64) & 0.89\\
& 54 & 232.49 & 235.38 & 5.127  & $-0.3045$ & 1.64  & 0.88  & 1.26  & 0.015 & & \\
& 64 & 103.23 & 126.04 & 5.225  & $-0.3044$ & 2.44  & 2.18  & 2.31  & 0.067 &  &  \\
\hline
\multirow{31}{*}{10} & 0  & 166.52 & 153.00 & 5.015 & $-0.3074$ & 0.94 & 1.14 & 1.04 & 0.019 & (0, 1)  & 1.34\\
& 1  & 46.79  & 92.70  & 4.956 & $-0.3078$ & 1.74 & 1.90 & 1.82 & 0.029 & (2, 1)  & 1.34\\
& 2  & 206.46 & 303.31 & 4.807 & $-0.3104$ & 0.58 & 0.16 & 0.37 & 0.015 & (3, 2)  & 0.80\\
& 3  & 106.73 & 166.80 & 5.232 & $-0.3037$ & 5.56 & 6.38 & 5.97 & 0.017 & (4, 3)  & 0.82\\
& 4  & 97.79  & 170.31 & 5.138 & $-0.3054$ & 0.72 & 1.26 & 0.99 & 0.015 & (4, 15) & 0.44\\
& 5  & 127.30 & 151.30 & 4.988 & $-0.3076$ & 1.40 & 2.44 & 1.92 & 0.018 & (5, 4)  & 0.57\\
& 6  & 193.88 & 103.67 & 4.899 & $-0.3075$ & 2.06 & 1.54 & 1.80 & 0.026 & (5, 6)  & 0.51\\
& 7  & 94.32  & 86.65  & 5.023 & $-0.3068$ & 4.78 & 6.52 & 5.65 & 0.025 & (7, 6)  & 0.54\\
& 8  & 188.77 & 74.32  & 5.112 & $-0.3057$ & 3.98 & 3.20 & 3.59 & 0.050 & (8, 7)  & 0.80\\
& 9  & 109.93 & 18.87  & 5.153 & $-0.3057$ & 7.00 & 7.30 & 7.15 & 0.056 & (9, 8)  & 1.34\\
& 14 & 228.15 & 96.76  & 4.930 & $-0.3080$ & 1.68 & 0.98 & 1.33 & 0.018 & (14, 0) & 0.56\\
& 15 & 217.31 & 63.08  & 5.015 & $-0.3070$ & 1.50 & 1.38 & 1.44 & 0.018 & (16, 8) & 2.46\\
& 16 & 154.00 & 11.77  & 5.279 & $-0.3024$ & 2.04 & 2.16 & 2.10 & 0.110 & (20, 19)& 0.77\\
& 19 & 218.98 & 246.21 & 4.958 & $-0.3080$ & 1.22 & 0.46 & 0.84 & 0.017 & (21, 20)& 1.51\\
& 20 & 110.45 & 122.17 & 5.044 & $-0.3059$ & 1.36 & 0.96 & 1.16 & 0.059 & (21, 22)& 0.83\\
& 21 & 194.17 & 165.34 & 5.160 & $-0.3032$ & 7.82 & 6.90 & 7.36 & 0.051 & (22, 15)& 1.04\\
& 22 & 170.53 & 92.47  & 5.282 & $-0.3030$ & 1.20 & 1.04 & 1.12 & 0.025 & (23, 22)& 1.49\\
& 23 & 125.61 & 76.39  & 5.163 & $-0.3052$ & 11.00& 9.90 & 10.45& 0.029 & (23, 24)& 0.71\\
& 24 & 99.75  & 93.21  & 5.080 & $-0.3049$ & 4.50 & 4.24 & 4.37 & 0.119 & (25, 24)& 0.75\\
& 25 & 145.41 & 115.54 & 5.137 & $-0.3052$ & 1.26 & 1.28 & 1.27 & 0.026 & (60, 61)& 1.69\\
& 60 & 146.71 & 140.25 & 5.215 & $-0.3038$ & 8.42 & 7.32 & 7.87 & 0.030 & (62, 61)& 0.69\\
& 61 & 173.45 & 43.52  & 5.172 & $-0.2923$ & 9.72 & 10.76& 10.24& 0.031 & (62, 63)& 2.35\\
& 62 & 105.67 & 51.10  & 5.042 & $-0.3063$ & 2.52 & 2.02 & 2.27 & 0.034 & (64, 63)& 4.04\\
& 63 & 100.05 & 52.86  & 5.190 & $-0.3137$ & 4.94 & 5.46 & 5.20 & 0.144 & (72, 62)& 0.95\\
& 64 & 109.50 & 114.47 & 5.225 & $-0.3044$ & 2.44 & 2.18 & 2.31 & 0.068 & (72, 81)& 2.20\\
& 72 & 89.54  & 51.99  & 5.306 & $-0.3021$ & 7.22 & 41.82& 24.52& 0.041 & (80, 79)& 1.14\\
& 79 & 154.87 & 158.37 & 5.161 & $-0.3048$ & 5.72 & 3.44 & 4.58 & 0.158 & (80, 81)& 4.26\\
& 80 & 151.14 & 147.43 & 5.061 & $-0.3063$ & 1.40 & 0.76 & 1.08 & 0.026 & (81, 82)& 2.02\\
& 81 & 118.18 & 169.45 & 5.129 & $-0.3054$ & 12.67& 0.33 & 6.50 & 0.287 & (83, 82)& 1.29\\
& 82 & 68.75  & 95.77  & 5.204 & $-0.3046$ & 6.00 & 6.33 & 6.17 & 0.097 &        &    \\
& 83 & 116.90 & 193.49 & 5.098 & $-0.3061$ & 4.46 & 0.56 & 2.51 & 0.035 &        &    \\
\end{tabular}
\end{ruledtabular}
\end{table*}

\bibliography{refs}

\begin{thebibliography}{101}%
\makeatletter
\providecommand \@ifxundefined [1]{%
 \@ifx{#1\undefined}
}%
\providecommand \@ifnum [1]{%
 \ifnum #1\expandafter \@firstoftwo
 \else \expandafter \@secondoftwo
 \fi
}%
\providecommand \@ifx [1]{%
 \ifx #1\expandafter \@firstoftwo
 \else \expandafter \@secondoftwo
 \fi
}%
\providecommand \natexlab [1]{#1}%
\providecommand \enquote  [1]{``#1''}%
\providecommand \bibnamefont  [1]{#1}%
\providecommand \bibfnamefont [1]{#1}%
\providecommand \citenamefont [1]{#1}%
\providecommand \href@noop [0]{\@secondoftwo}%
\providecommand \href [0]{\begingroup \@sanitize@url \@href}%
\providecommand \@href[1]{\@@startlink{#1}\@@href}%
\providecommand \@@href[1]{\endgroup#1\@@endlink}%
\providecommand \@sanitize@url [0]{\catcode `\\12\catcode `\$12\catcode `\&12\catcode `\#12\catcode `\^12\catcode `\_12\catcode `\%12\relax}%
\providecommand \@@startlink[1]{}%
\providecommand \@@endlink[0]{}%
\providecommand \url  [0]{\begingroup\@sanitize@url \@url }%
\providecommand \@url [1]{\endgroup\@href {#1}{\urlprefix }}%
\providecommand \urlprefix  [0]{URL }%
\providecommand \Eprint [0]{\href }%
\providecommand \doibase [0]{https://doi.org/}%
\providecommand \selectlanguage [0]{\@gobble}%
\providecommand \bibinfo  [0]{\@secondoftwo}%
\providecommand \bibfield  [0]{\@secondoftwo}%
\providecommand \translation [1]{[#1]}%
\providecommand \BibitemOpen [0]{}%
\providecommand \bibitemStop [0]{}%
\providecommand \bibitemNoStop [0]{.\EOS\space}%
\providecommand \EOS [0]{\spacefactor3000\relax}%
\providecommand \BibitemShut  [1]{\csname bibitem#1\endcsname}%
\let\auto@bib@innerbib\@empty
\bibitem [{\citenamefont {Cerezo}\ \emph {et~al.}(2021)\citenamefont {Cerezo}, \citenamefont {Arrasmith}, \citenamefont {Babbush}, \citenamefont {Benjamin}, \citenamefont {Endo}, \citenamefont {Fujii}, \citenamefont {McClean}, \citenamefont {Mitarai}, \citenamefont {Yuan}, \citenamefont {Cincio} \emph {et~al.}}]{cerezo2021variational}%
  \BibitemOpen
  \bibfield  {author} {\bibinfo {author} {\bibfnamefont {M.}~\bibnamefont {Cerezo}}, \bibinfo {author} {\bibfnamefont {A.}~\bibnamefont {Arrasmith}}, \bibinfo {author} {\bibfnamefont {R.}~\bibnamefont {Babbush}}, \bibinfo {author} {\bibfnamefont {S.~C.}\ \bibnamefont {Benjamin}}, \bibinfo {author} {\bibfnamefont {S.}~\bibnamefont {Endo}}, \bibinfo {author} {\bibfnamefont {K.}~\bibnamefont {Fujii}}, \bibinfo {author} {\bibfnamefont {J.~R.}\ \bibnamefont {McClean}}, \bibinfo {author} {\bibfnamefont {K.}~\bibnamefont {Mitarai}}, \bibinfo {author} {\bibfnamefont {X.}~\bibnamefont {Yuan}}, \bibinfo {author} {\bibfnamefont {L.}~\bibnamefont {Cincio}}, \emph {et~al.},\ }\bibfield  {title} {\bibinfo {title} {Variational quantum algorithms},\ }\href@noop {} {\bibfield  {journal} {\bibinfo  {journal} {Nature Reviews Physics}\ }\textbf {\bibinfo {volume} {3}},\ \bibinfo {pages} {625} (\bibinfo {year} {2021})}\BibitemShut {NoStop}%
\bibitem [{\citenamefont {Bharti}\ \emph {et~al.}(2022)\citenamefont {Bharti}, \citenamefont {Cervera-Lierta}, \citenamefont {Kyaw}, \citenamefont {Haug}, \citenamefont {Alperin-Lea}, \citenamefont {Anand}, \citenamefont {Degroote}, \citenamefont {Heimonen}, \citenamefont {Kottmann}, \citenamefont {Menke} \emph {et~al.}}]{bharti2022noisy}%
  \BibitemOpen
  \bibfield  {author} {\bibinfo {author} {\bibfnamefont {K.}~\bibnamefont {Bharti}}, \bibinfo {author} {\bibfnamefont {A.}~\bibnamefont {Cervera-Lierta}}, \bibinfo {author} {\bibfnamefont {T.~H.}\ \bibnamefont {Kyaw}}, \bibinfo {author} {\bibfnamefont {T.}~\bibnamefont {Haug}}, \bibinfo {author} {\bibfnamefont {S.}~\bibnamefont {Alperin-Lea}}, \bibinfo {author} {\bibfnamefont {A.}~\bibnamefont {Anand}}, \bibinfo {author} {\bibfnamefont {M.}~\bibnamefont {Degroote}}, \bibinfo {author} {\bibfnamefont {H.}~\bibnamefont {Heimonen}}, \bibinfo {author} {\bibfnamefont {J.~S.}\ \bibnamefont {Kottmann}}, \bibinfo {author} {\bibfnamefont {T.}~\bibnamefont {Menke}}, \emph {et~al.},\ }\bibfield  {title} {\bibinfo {title} {Noisy intermediate-scale quantum algorithms},\ }\href {https://doi.org/10.1103/RevModPhys.94.015004} {\bibfield  {journal} {\bibinfo  {journal} {Reviews of Modern Physics}\ }\textbf {\bibinfo {volume} {94}},\ \bibinfo {pages} {015004} (\bibinfo {year} {2022})}\BibitemShut {NoStop}%
\bibitem [{\citenamefont {Farhi}\ \emph {et~al.}(2014)\citenamefont {Farhi}, \citenamefont {Goldstone},\ and\ \citenamefont {Gutmann}}]{farhi2014quantum}%
  \BibitemOpen
  \bibfield  {author} {\bibinfo {author} {\bibfnamefont {E.}~\bibnamefont {Farhi}}, \bibinfo {author} {\bibfnamefont {J.}~\bibnamefont {Goldstone}},\ and\ \bibinfo {author} {\bibfnamefont {S.}~\bibnamefont {Gutmann}},\ }\bibfield  {title} {\bibinfo {title} {A quantum approximate optimization algorithm},\ }\href@noop {} {\bibfield  {journal} {\bibinfo  {journal} {arXiv preprint arXiv:1411.4028}\ } (\bibinfo {year} {2014})}\BibitemShut {NoStop}%
\bibitem [{\citenamefont {Ma}\ \emph {et~al.}(2020)\citenamefont {Ma}, \citenamefont {Govoni},\ and\ \citenamefont {Galli}}]{ma2020quantum}%
  \BibitemOpen
  \bibfield  {author} {\bibinfo {author} {\bibfnamefont {H.}~\bibnamefont {Ma}}, \bibinfo {author} {\bibfnamefont {M.}~\bibnamefont {Govoni}},\ and\ \bibinfo {author} {\bibfnamefont {G.}~\bibnamefont {Galli}},\ }\bibfield  {title} {\bibinfo {title} {Quantum simulations of materials on near-term quantum computers},\ }\href@noop {} {\bibfield  {journal} {\bibinfo  {journal} {npj Computational Materials}\ }\textbf {\bibinfo {volume} {6}},\ \bibinfo {pages} {85} (\bibinfo {year} {2020})}\BibitemShut {NoStop}%
\bibitem [{\citenamefont {Hu}\ \emph {et~al.}(2022)\citenamefont {Hu}, \citenamefont {Meng}, \citenamefont {Wang}, \citenamefont {Luan}, \citenamefont {Fu}, \citenamefont {Zhang}, \citenamefont {Zhang},\ and\ \citenamefont {Yu}}]{hu2022greedy}%
  \BibitemOpen
  \bibfield  {author} {\bibinfo {author} {\bibfnamefont {Y.}~\bibnamefont {Hu}}, \bibinfo {author} {\bibfnamefont {F.}~\bibnamefont {Meng}}, \bibinfo {author} {\bibfnamefont {X.}~\bibnamefont {Wang}}, \bibinfo {author} {\bibfnamefont {T.}~\bibnamefont {Luan}}, \bibinfo {author} {\bibfnamefont {Y.}~\bibnamefont {Fu}}, \bibinfo {author} {\bibfnamefont {Z.}~\bibnamefont {Zhang}}, \bibinfo {author} {\bibfnamefont {X.}~\bibnamefont {Zhang}},\ and\ \bibinfo {author} {\bibfnamefont {X.}~\bibnamefont {Yu}},\ }\bibfield  {title} {\bibinfo {title} {Greedy algorithm based circuit optimization for near-term quantum simulation},\ }\href@noop {} {\bibfield  {journal} {\bibinfo  {journal} {Quantum Science and Technology}\ }\textbf {\bibinfo {volume} {7}},\ \bibinfo {pages} {045001} (\bibinfo {year} {2022})}\BibitemShut {NoStop}%
\bibitem [{\citenamefont {Li}\ \emph {et~al.}(2019)\citenamefont {Li}, \citenamefont {Ding},\ and\ \citenamefont {Xie}}]{li2019tackling}%
  \BibitemOpen
  \bibfield  {author} {\bibinfo {author} {\bibfnamefont {G.}~\bibnamefont {Li}}, \bibinfo {author} {\bibfnamefont {Y.}~\bibnamefont {Ding}},\ and\ \bibinfo {author} {\bibfnamefont {Y.}~\bibnamefont {Xie}},\ }\bibfield  {title} {\bibinfo {title} {Tackling the qubit mapping problem for nisq-era quantum devices},\ }in\ \href {https://doi.org/10.1145/3297858.3304023} {\emph {\bibinfo {booktitle} {Proceedings of the Twenty-Fourth International Conference on Architectural Support for Programming Languages and Operating Systems}}},\ \bibinfo {series and number} {ASPLOS '19}\ (\bibinfo  {publisher} {Association for Computing Machinery},\ \bibinfo {address} {New York, NY, USA},\ \bibinfo {year} {2019})\ p.\ \bibinfo {pages} {1001–1014}\BibitemShut {NoStop}%
\bibitem [{\citenamefont {Cowtan}\ \emph {et~al.}(2019)\citenamefont {Cowtan}, \citenamefont {Dilkes}, \citenamefont {Duncan}, \citenamefont {Krajenbrink}, \citenamefont {Simmons},\ and\ \citenamefont {Sivarajah}}]{cowtan2019qubit}%
  \BibitemOpen
  \bibfield  {author} {\bibinfo {author} {\bibfnamefont {A.}~\bibnamefont {Cowtan}}, \bibinfo {author} {\bibfnamefont {S.}~\bibnamefont {Dilkes}}, \bibinfo {author} {\bibfnamefont {R.}~\bibnamefont {Duncan}}, \bibinfo {author} {\bibfnamefont {A.}~\bibnamefont {Krajenbrink}}, \bibinfo {author} {\bibfnamefont {W.}~\bibnamefont {Simmons}},\ and\ \bibinfo {author} {\bibfnamefont {S.}~\bibnamefont {Sivarajah}},\ }\bibfield  {title} {\bibinfo {title} {{On the Qubit Routing Problem}},\ }in\ \href {https://doi.org/10.4230/LIPIcs.TQC.2019.5} {\emph {\bibinfo {booktitle} {14th Conference on the Theory of Quantum Computation, Communication and Cryptography (TQC 2019)}}},\ \bibinfo {series} {Leibniz International Proceedings in Informatics (LIPIcs)}, Vol.\ \bibinfo {volume} {135},\ \bibinfo {editor} {edited by\ \bibinfo {editor} {\bibfnamefont {W.}~\bibnamefont {van Dam}}\ and\ \bibinfo {editor} {\bibfnamefont {L.}~\bibnamefont {Mancinska}}}\ (\bibinfo  {publisher} {Schloss Dagstuhl--Leibniz-Zentrum fuer Informatik},\
  \bibinfo {address} {Dagstuhl, Germany},\ \bibinfo {year} {2019})\ pp.\ \bibinfo {pages} {5:1--5:32}\BibitemShut {NoStop}%
\bibitem [{\citenamefont {Zhu}\ \emph {et~al.}(2022)\citenamefont {Zhu}, \citenamefont {Feng},\ and\ \citenamefont {Guan}}]{zhu2022an}%
  \BibitemOpen
  \bibfield  {author} {\bibinfo {author} {\bibfnamefont {P.}~\bibnamefont {Zhu}}, \bibinfo {author} {\bibfnamefont {S.}~\bibnamefont {Feng}},\ and\ \bibinfo {author} {\bibfnamefont {Z.}~\bibnamefont {Guan}},\ }\bibfield  {title} {\bibinfo {title} {An iterated local search methodology for the qubit mapping problem},\ }\href {https://doi.org/10.1109/TCAD.2021.3112143} {\bibfield  {journal} {\bibinfo  {journal} {IEEE Transactions on Computer-Aided Design of Integrated Circuits and Systems}\ }\textbf {\bibinfo {volume} {41}},\ \bibinfo {pages} {2587} (\bibinfo {year} {2022})}\BibitemShut {NoStop}%
\bibitem [{\citenamefont {Niu}\ \emph {et~al.}(2020)\citenamefont {Niu}, \citenamefont {Suau}, \citenamefont {Staffelbach},\ and\ \citenamefont {Todri-Sanial}}]{niu2020hardware}%
  \BibitemOpen
  \bibfield  {author} {\bibinfo {author} {\bibfnamefont {S.}~\bibnamefont {Niu}}, \bibinfo {author} {\bibfnamefont {A.}~\bibnamefont {Suau}}, \bibinfo {author} {\bibfnamefont {G.}~\bibnamefont {Staffelbach}},\ and\ \bibinfo {author} {\bibfnamefont {A.}~\bibnamefont {Todri-Sanial}},\ }\bibfield  {title} {\bibinfo {title} {A hardware-aware heuristic for the qubit mapping problem in the {NISQ} era},\ }\href@noop {} {\bibfield  {journal} {\bibinfo  {journal} {IEEE Transactions on Quantum Engineering}\ }\textbf {\bibinfo {volume} {1}},\ \bibinfo {pages} {1} (\bibinfo {year} {2020})}\BibitemShut {NoStop}%
\bibitem [{\citenamefont {Zhang}\ \emph {et~al.}(2021)\citenamefont {Zhang}, \citenamefont {Hayes}, \citenamefont {Qiu}, \citenamefont {Jin}, \citenamefont {Chen},\ and\ \citenamefont {Zhang}}]{zhang2021time}%
  \BibitemOpen
  \bibfield  {author} {\bibinfo {author} {\bibfnamefont {C.}~\bibnamefont {Zhang}}, \bibinfo {author} {\bibfnamefont {A.~B.}\ \bibnamefont {Hayes}}, \bibinfo {author} {\bibfnamefont {L.}~\bibnamefont {Qiu}}, \bibinfo {author} {\bibfnamefont {Y.}~\bibnamefont {Jin}}, \bibinfo {author} {\bibfnamefont {Y.}~\bibnamefont {Chen}},\ and\ \bibinfo {author} {\bibfnamefont {E.~Z.}\ \bibnamefont {Zhang}},\ }\bibfield  {title} {\bibinfo {title} {Time-optimal qubit mapping},\ }in\ \href {https://doi.org/10.1145/3445814.3446706} {\emph {\bibinfo {booktitle} {Proceedings of the 26th ACM International Conference on Architectural Support for Programming Languages and Operating Systems}}},\ \bibinfo {series and number} {ASPLOS '21}\ (\bibinfo  {publisher} {Association for Computing Machinery},\ \bibinfo {address} {New York, NY, USA},\ \bibinfo {year} {2021})\ p.\ \bibinfo {pages} {360–374}\BibitemShut {NoStop}%
\bibitem [{\citenamefont {Siraichi}\ \emph {et~al.}(2018)\citenamefont {Siraichi}, \citenamefont {Santos}, \citenamefont {Collange},\ and\ \citenamefont {Pereira}}]{siraichi2018qubit}%
  \BibitemOpen
  \bibfield  {author} {\bibinfo {author} {\bibfnamefont {M.~Y.}\ \bibnamefont {Siraichi}}, \bibinfo {author} {\bibfnamefont {V.~F.~d.}\ \bibnamefont {Santos}}, \bibinfo {author} {\bibfnamefont {C.}~\bibnamefont {Collange}},\ and\ \bibinfo {author} {\bibfnamefont {F.~M.~Q.}\ \bibnamefont {Pereira}},\ }\bibfield  {title} {\bibinfo {title} {Qubit allocation},\ }in\ \href {https://doi.org/10.1145/3168822} {\emph {\bibinfo {booktitle} {Proceedings of the 2018 International Symposium on Code Generation and Optimization}}},\ \bibinfo {series and number} {CGO 2018}\ (\bibinfo  {publisher} {Association for Computing Machinery},\ \bibinfo {address} {New York, NY, USA},\ \bibinfo {year} {2018})\ p.\ \bibinfo {pages} {113–125}\BibitemShut {NoStop}%
\bibitem [{\citenamefont {Bhattacharjee}\ and\ \citenamefont {Chattopadhyay}(2017)}]{bhattacharjee2017depth}%
  \BibitemOpen
  \bibfield  {author} {\bibinfo {author} {\bibfnamefont {D.}~\bibnamefont {Bhattacharjee}}\ and\ \bibinfo {author} {\bibfnamefont {A.}~\bibnamefont {Chattopadhyay}},\ }\bibfield  {title} {\bibinfo {title} {Depth-optimal quantum circuit placement for arbitrary topologies},\ }\href@noop {} {\bibfield  {journal} {\bibinfo  {journal} {arXiv preprint arXiv:1703.08540}\ } (\bibinfo {year} {2017})}\BibitemShut {NoStop}%
\bibitem [{\citenamefont {Shafaei}\ \emph {et~al.}(2014)\citenamefont {Shafaei}, \citenamefont {Saeedi},\ and\ \citenamefont {Pedram}}]{ShafaeiSP14}%
  \BibitemOpen
  \bibfield  {author} {\bibinfo {author} {\bibfnamefont {A.}~\bibnamefont {Shafaei}}, \bibinfo {author} {\bibfnamefont {M.}~\bibnamefont {Saeedi}},\ and\ \bibinfo {author} {\bibfnamefont {M.}~\bibnamefont {Pedram}},\ }\bibfield  {title} {\bibinfo {title} {Qubit placement to minimize communication overhead in 2d quantum architectures},\ }in\ \href {https://doi.org/10.1109/ASPDAC.2014.6742940} {\emph {\bibinfo {booktitle} {2014 19th Asia and South Pacific Design Automation Conference (ASP-DAC)}}}\ (\bibinfo  {publisher} {{IEEE}},\ \bibinfo {address} {Piscataway, NJ, USA},\ \bibinfo {year} {2014})\ pp.\ \bibinfo {pages} {495--500}\BibitemShut {NoStop}%
\bibitem [{\citenamefont {Murali}\ \emph {et~al.}(2019)\citenamefont {Murali}, \citenamefont {Baker}, \citenamefont {Javadi-Abhari}, \citenamefont {Chong},\ and\ \citenamefont {Martonosi}}]{murali2019noise}%
  \BibitemOpen
  \bibfield  {author} {\bibinfo {author} {\bibfnamefont {P.}~\bibnamefont {Murali}}, \bibinfo {author} {\bibfnamefont {J.~M.}\ \bibnamefont {Baker}}, \bibinfo {author} {\bibfnamefont {A.}~\bibnamefont {Javadi-Abhari}}, \bibinfo {author} {\bibfnamefont {F.~T.}\ \bibnamefont {Chong}},\ and\ \bibinfo {author} {\bibfnamefont {M.}~\bibnamefont {Martonosi}},\ }\bibfield  {title} {\bibinfo {title} {Noise-adaptive compiler mappings for noisy intermediate-scale quantum computers},\ }in\ \href {https://doi.org/10.1145/3297858.3304075} {\emph {\bibinfo {booktitle} {Proceedings of the Twenty-Fourth International Conference on Architectural Support for Programming Languages and Operating Systems}}},\ \bibinfo {series and number} {ASPLOS '19}\ (\bibinfo  {publisher} {Association for Computing Machinery},\ \bibinfo {address} {New York, NY, USA},\ \bibinfo {year} {2019})\ p.\ \bibinfo {pages} {1015–1029}\BibitemShut {NoStop}%
\bibitem [{\citenamefont {Siraichi}\ \emph {et~al.}(2019)\citenamefont {Siraichi}, \citenamefont {Santos}, \citenamefont {Collange},\ and\ \citenamefont {Pereira}}]{siraichi2019qubit}%
  \BibitemOpen
  \bibfield  {author} {\bibinfo {author} {\bibfnamefont {M.~Y.}\ \bibnamefont {Siraichi}}, \bibinfo {author} {\bibfnamefont {V.~F.~d.}\ \bibnamefont {Santos}}, \bibinfo {author} {\bibfnamefont {C.}~\bibnamefont {Collange}},\ and\ \bibinfo {author} {\bibfnamefont {F.~M. Q.~a.}\ \bibnamefont {Pereira}},\ }\bibfield  {title} {\bibinfo {title} {Qubit allocation as a combination of subgraph isomorphism and token swapping},\ }\bibfield  {journal} {\bibinfo  {journal} {Proc. ACM Program. Lang.}\ }\textbf {\bibinfo {volume} {3}},\ \href {https://doi.org/10.1145/3360546} {10.1145/3360546} (\bibinfo {year} {2019})\BibitemShut {NoStop}%
\bibitem [{\citenamefont {Tan}\ and\ \citenamefont {Cong}(2020)}]{tan2020optimal}%
  \BibitemOpen
  \bibfield  {author} {\bibinfo {author} {\bibfnamefont {B.}~\bibnamefont {Tan}}\ and\ \bibinfo {author} {\bibfnamefont {J.}~\bibnamefont {Cong}},\ }\bibfield  {title} {\bibinfo {title} {Optimal layout synthesis for quantum computing},\ }in\ \href {https://doi.org/10.1145/3400302.3415620} {\emph {\bibinfo {booktitle} {Proceedings of the 39th International Conference on Computer-Aided Design}}},\ \bibinfo {series and number} {ICCAD '20}\ (\bibinfo  {publisher} {Association for Computing Machinery},\ \bibinfo {address} {New York, NY, USA},\ \bibinfo {year} {2020})\BibitemShut {NoStop}%
\bibitem [{\citenamefont {Zulehner}\ \emph {et~al.}(2018)\citenamefont {Zulehner}, \citenamefont {Paler},\ and\ \citenamefont {Wille}}]{zulehner2018efficient}%
  \BibitemOpen
  \bibfield  {author} {\bibinfo {author} {\bibfnamefont {A.}~\bibnamefont {Zulehner}}, \bibinfo {author} {\bibfnamefont {A.}~\bibnamefont {Paler}},\ and\ \bibinfo {author} {\bibfnamefont {R.}~\bibnamefont {Wille}},\ }\bibfield  {title} {\bibinfo {title} {An efficient methodology for mapping quantum circuits to the {IBM QX} architectures},\ }\href@noop {} {\bibfield  {journal} {\bibinfo  {journal} {IEEE Transactions on Computer-Aided Design of Integrated Circuits and Systems}\ }\textbf {\bibinfo {volume} {38}},\ \bibinfo {pages} {1226} (\bibinfo {year} {2018})}\BibitemShut {NoStop}%
\bibitem [{\citenamefont {Childs}\ \emph {et~al.}(2019{\natexlab{a}})\citenamefont {Childs}, \citenamefont {Schoute},\ and\ \citenamefont {Unsal}}]{childs19}%
  \BibitemOpen
  \bibfield  {author} {\bibinfo {author} {\bibfnamefont {A.~M.}\ \bibnamefont {Childs}}, \bibinfo {author} {\bibfnamefont {E.}~\bibnamefont {Schoute}},\ and\ \bibinfo {author} {\bibfnamefont {C.~M.}\ \bibnamefont {Unsal}},\ }\bibfield  {title} {\bibinfo {title} {{Circuit Transformations for Quantum Architectures}},\ }in\ \href {https://doi.org/10.4230/LIPIcs.TQC.2019.3} {\emph {\bibinfo {booktitle} {14th Conference on the Theory of Quantum Computation, Communication and Cryptography (TQC 2019)}}},\ \bibinfo {series} {Leibniz International Proceedings in Informatics (LIPIcs)}, Vol.\ \bibinfo {volume} {135},\ \bibinfo {editor} {edited by\ \bibinfo {editor} {\bibfnamefont {W.}~\bibnamefont {van Dam}}\ and\ \bibinfo {editor} {\bibfnamefont {L.}~\bibnamefont {Mancinska}}}\ (\bibinfo  {publisher} {Schloss Dagstuhl--Leibniz-Zentrum fuer Informatik},\ \bibinfo {address} {Dagstuhl, Germany},\ \bibinfo {year} {2019})\ pp.\ \bibinfo {pages} {3:1--3:24}\BibitemShut {NoStop}%
\bibitem [{\citenamefont {Li}\ \emph {et~al.}(2020)\citenamefont {Li}, \citenamefont {Meng}, \citenamefont {Zhang},\ and\ \citenamefont {Yu}}]{li2020qubits}%
  \BibitemOpen
  \bibfield  {author} {\bibinfo {author} {\bibfnamefont {Z.-T.}\ \bibnamefont {Li}}, \bibinfo {author} {\bibfnamefont {F.-X.}\ \bibnamefont {Meng}}, \bibinfo {author} {\bibfnamefont {Z.-C.}\ \bibnamefont {Zhang}},\ and\ \bibinfo {author} {\bibfnamefont {X.-T.}\ \bibnamefont {Yu}},\ }\bibfield  {title} {\bibinfo {title} {Qubits’ mapping and routing for {NISQ} on variability of quantum gates},\ }\href@noop {} {\bibfield  {journal} {\bibinfo  {journal} {Quantum Information Processing}\ }\textbf {\bibinfo {volume} {19}},\ \bibinfo {pages} {1} (\bibinfo {year} {2020})}\BibitemShut {NoStop}%
\bibitem [{\citenamefont {Harrigan}\ \emph {et~al.}(2021)\citenamefont {Harrigan}, \citenamefont {Sung}, \citenamefont {Neeley}, \citenamefont {Satzinger}, \citenamefont {Arute}, \citenamefont {Arya}, \citenamefont {Atalaya}, \citenamefont {Bardin}, \citenamefont {Barends}, \citenamefont {Boixo} \emph {et~al.}}]{harrigan2021quantum}%
  \BibitemOpen
  \bibfield  {author} {\bibinfo {author} {\bibfnamefont {M.~P.}\ \bibnamefont {Harrigan}}, \bibinfo {author} {\bibfnamefont {K.~J.}\ \bibnamefont {Sung}}, \bibinfo {author} {\bibfnamefont {M.}~\bibnamefont {Neeley}}, \bibinfo {author} {\bibfnamefont {K.~J.}\ \bibnamefont {Satzinger}}, \bibinfo {author} {\bibfnamefont {F.}~\bibnamefont {Arute}}, \bibinfo {author} {\bibfnamefont {K.}~\bibnamefont {Arya}}, \bibinfo {author} {\bibfnamefont {J.}~\bibnamefont {Atalaya}}, \bibinfo {author} {\bibfnamefont {J.~C.}\ \bibnamefont {Bardin}}, \bibinfo {author} {\bibfnamefont {R.}~\bibnamefont {Barends}}, \bibinfo {author} {\bibfnamefont {S.}~\bibnamefont {Boixo}}, \emph {et~al.},\ }\bibfield  {title} {\bibinfo {title} {Quantum approximate optimization of non-planar graph problems on a planar superconducting processor},\ }\href@noop {} {\bibfield  {journal} {\bibinfo  {journal} {Nature Physics}\ }\textbf {\bibinfo {volume} {17}},\ \bibinfo {pages} {332} (\bibinfo {year} {2021})}\BibitemShut {NoStop}%
\bibitem [{\citenamefont {Kivlichan}\ \emph {et~al.}(2018)\citenamefont {Kivlichan}, \citenamefont {McClean}, \citenamefont {Wiebe}, \citenamefont {Gidney}, \citenamefont {Aspuru-Guzik}, \citenamefont {Chan},\ and\ \citenamefont {Babbush}}]{kivlichan2018quantum}%
  \BibitemOpen
  \bibfield  {author} {\bibinfo {author} {\bibfnamefont {I.~D.}\ \bibnamefont {Kivlichan}}, \bibinfo {author} {\bibfnamefont {J.}~\bibnamefont {McClean}}, \bibinfo {author} {\bibfnamefont {N.}~\bibnamefont {Wiebe}}, \bibinfo {author} {\bibfnamefont {C.}~\bibnamefont {Gidney}}, \bibinfo {author} {\bibfnamefont {A.}~\bibnamefont {Aspuru-Guzik}}, \bibinfo {author} {\bibfnamefont {G.~K.-L.}\ \bibnamefont {Chan}},\ and\ \bibinfo {author} {\bibfnamefont {R.}~\bibnamefont {Babbush}},\ }\bibfield  {title} {\bibinfo {title} {Quantum simulation of electronic structure with linear depth and connectivity},\ }\href {https://doi.org/10.1103/PhysRevLett.120.110501} {\bibfield  {journal} {\bibinfo  {journal} {Physical review letters}\ }\textbf {\bibinfo {volume} {120}},\ \bibinfo {pages} {110501} (\bibinfo {year} {2018})}\BibitemShut {NoStop}%
\bibitem [{\citenamefont {O'Gorman}\ \emph {et~al.}(2019)\citenamefont {O'Gorman}, \citenamefont {Huggins}, \citenamefont {Rieffel},\ and\ \citenamefont {Whaley}}]{o2019generalized}%
  \BibitemOpen
  \bibfield  {author} {\bibinfo {author} {\bibfnamefont {B.}~\bibnamefont {O'Gorman}}, \bibinfo {author} {\bibfnamefont {W.~J.}\ \bibnamefont {Huggins}}, \bibinfo {author} {\bibfnamefont {E.~G.}\ \bibnamefont {Rieffel}},\ and\ \bibinfo {author} {\bibfnamefont {K.~B.}\ \bibnamefont {Whaley}},\ }\bibfield  {title} {\bibinfo {title} {Generalized swap networks for near-term quantum computing},\ }\href@noop {} {\bibfield  {journal} {\bibinfo  {journal} {arXiv preprint arXiv:1905.05118}\ } (\bibinfo {year} {2019})}\BibitemShut {NoStop}%
\bibitem [{\citenamefont {Weidenfeller}\ \emph {et~al.}(2022)\citenamefont {Weidenfeller}, \citenamefont {Valor}, \citenamefont {Gacon}, \citenamefont {Tornow}, \citenamefont {Bello}, \citenamefont {Woerner},\ and\ \citenamefont {Egger}}]{weidenfeller2022scaling}%
  \BibitemOpen
  \bibfield  {author} {\bibinfo {author} {\bibfnamefont {J.}~\bibnamefont {Weidenfeller}}, \bibinfo {author} {\bibfnamefont {L.~C.}\ \bibnamefont {Valor}}, \bibinfo {author} {\bibfnamefont {J.}~\bibnamefont {Gacon}}, \bibinfo {author} {\bibfnamefont {C.}~\bibnamefont {Tornow}}, \bibinfo {author} {\bibfnamefont {L.}~\bibnamefont {Bello}}, \bibinfo {author} {\bibfnamefont {S.}~\bibnamefont {Woerner}},\ and\ \bibinfo {author} {\bibfnamefont {D.~J.}\ \bibnamefont {Egger}},\ }\bibfield  {title} {\bibinfo {title} {Scaling of the quantum approximate optimization algorithm on superconducting qubit based hardware},\ }\href@noop {} {\bibfield  {journal} {\bibinfo  {journal} {Quantum}\ }\textbf {\bibinfo {volume} {6}},\ \bibinfo {pages} {870} (\bibinfo {year} {2022})}\BibitemShut {NoStop}%
\bibitem [{\citenamefont {Hashim}\ \emph {et~al.}(2022)\citenamefont {Hashim}, \citenamefont {Rines}, \citenamefont {Omole}, \citenamefont {Naik}, \citenamefont {Kreikebaum}, \citenamefont {Santiago}, \citenamefont {Chong}, \citenamefont {Siddiqi},\ and\ \citenamefont {Gokhale}}]{hashim22optimized}%
  \BibitemOpen
  \bibfield  {author} {\bibinfo {author} {\bibfnamefont {A.}~\bibnamefont {Hashim}}, \bibinfo {author} {\bibfnamefont {R.}~\bibnamefont {Rines}}, \bibinfo {author} {\bibfnamefont {V.}~\bibnamefont {Omole}}, \bibinfo {author} {\bibfnamefont {R.~K.}\ \bibnamefont {Naik}}, \bibinfo {author} {\bibfnamefont {J.~M.}\ \bibnamefont {Kreikebaum}}, \bibinfo {author} {\bibfnamefont {D.~I.}\ \bibnamefont {Santiago}}, \bibinfo {author} {\bibfnamefont {F.~T.}\ \bibnamefont {Chong}}, \bibinfo {author} {\bibfnamefont {I.}~\bibnamefont {Siddiqi}},\ and\ \bibinfo {author} {\bibfnamefont {P.}~\bibnamefont {Gokhale}},\ }\bibfield  {title} {\bibinfo {title} {Optimized swap networks with equivalent circuit averaging for qaoa},\ }\href {https://doi.org/10.1103/PhysRevResearch.4.033028} {\bibfield  {journal} {\bibinfo  {journal} {Phys. Rev. Res.}\ }\textbf {\bibinfo {volume} {4}},\ \bibinfo {pages} {033028} (\bibinfo {year} {2022})}\BibitemShut {NoStop}%
\bibitem [{\citenamefont {Javadi-Abhari}\ \emph {et~al.}(2024)\citenamefont {Javadi-Abhari}, \citenamefont {Treinish}, \citenamefont {Krsulich}, \citenamefont {Wood}, \citenamefont {Lishman}, \citenamefont {Gacon}, \citenamefont {Martiel}, \citenamefont {Nation}, \citenamefont {Bishop}, \citenamefont {Cross}, \citenamefont {Johnson},\ and\ \citenamefont {Gambetta}}]{qiskit}%
  \BibitemOpen
  \bibfield  {author} {\bibinfo {author} {\bibfnamefont {A.}~\bibnamefont {Javadi-Abhari}}, \bibinfo {author} {\bibfnamefont {M.}~\bibnamefont {Treinish}}, \bibinfo {author} {\bibfnamefont {K.}~\bibnamefont {Krsulich}}, \bibinfo {author} {\bibfnamefont {C.~J.}\ \bibnamefont {Wood}}, \bibinfo {author} {\bibfnamefont {J.}~\bibnamefont {Lishman}}, \bibinfo {author} {\bibfnamefont {J.}~\bibnamefont {Gacon}}, \bibinfo {author} {\bibfnamefont {S.}~\bibnamefont {Martiel}}, \bibinfo {author} {\bibfnamefont {P.~D.}\ \bibnamefont {Nation}}, \bibinfo {author} {\bibfnamefont {L.~S.}\ \bibnamefont {Bishop}}, \bibinfo {author} {\bibfnamefont {A.~W.}\ \bibnamefont {Cross}}, \bibinfo {author} {\bibfnamefont {B.~R.}\ \bibnamefont {Johnson}},\ and\ \bibinfo {author} {\bibfnamefont {J.~M.}\ \bibnamefont {Gambetta}},\ }\bibfield  {title} {\bibinfo {title} {Quantum computing with {Q}iskit},\ }\href@noop {} {\bibfield  {journal} {\bibinfo  {journal} {arXiv preprint arXiv:2405.08810}\ } (\bibinfo {year} {2024})}\BibitemShut
  {NoStop}%
\bibitem [{\citenamefont {Sivarajah}\ \emph {et~al.}(2020)\citenamefont {Sivarajah}, \citenamefont {Dilkes}, \citenamefont {Cowtan}, \citenamefont {Simmons}, \citenamefont {Edgington},\ and\ \citenamefont {Duncan}}]{sivarajah2020tket}%
  \BibitemOpen
  \bibfield  {author} {\bibinfo {author} {\bibfnamefont {S.}~\bibnamefont {Sivarajah}}, \bibinfo {author} {\bibfnamefont {S.}~\bibnamefont {Dilkes}}, \bibinfo {author} {\bibfnamefont {A.}~\bibnamefont {Cowtan}}, \bibinfo {author} {\bibfnamefont {W.}~\bibnamefont {Simmons}}, \bibinfo {author} {\bibfnamefont {A.}~\bibnamefont {Edgington}},\ and\ \bibinfo {author} {\bibfnamefont {R.}~\bibnamefont {Duncan}},\ }\bibfield  {title} {\bibinfo {title} {t|ket⟩: a retargetable compiler for nisq devices},\ }\href {https://doi.org/10.1088/2058-9565/ab8e92} {\bibfield  {journal} {\bibinfo  {journal} {Quantum Science and Technology}\ }\textbf {\bibinfo {volume} {6}},\ \bibinfo {pages} {014003} (\bibinfo {year} {2020})}\BibitemShut {NoStop}%
\bibitem [{\citenamefont {Rosenberg}\ \emph {et~al.}(2020)\citenamefont {Rosenberg}, \citenamefont {Weber}, \citenamefont {Conway}, \citenamefont {Yost}, \citenamefont {Mallek}, \citenamefont {Calusine}, \citenamefont {Das}, \citenamefont {Kim}, \citenamefont {Schwartz}, \citenamefont {Woods}, \citenamefont {Yoder},\ and\ \citenamefont {Oliver}}]{rosenberg2020solid}%
  \BibitemOpen
  \bibfield  {author} {\bibinfo {author} {\bibfnamefont {D.}~\bibnamefont {Rosenberg}}, \bibinfo {author} {\bibfnamefont {S.~J.}\ \bibnamefont {Weber}}, \bibinfo {author} {\bibfnamefont {D.}~\bibnamefont {Conway}}, \bibinfo {author} {\bibfnamefont {D.-R.~W.}\ \bibnamefont {Yost}}, \bibinfo {author} {\bibfnamefont {J.}~\bibnamefont {Mallek}}, \bibinfo {author} {\bibfnamefont {G.}~\bibnamefont {Calusine}}, \bibinfo {author} {\bibfnamefont {R.}~\bibnamefont {Das}}, \bibinfo {author} {\bibfnamefont {D.}~\bibnamefont {Kim}}, \bibinfo {author} {\bibfnamefont {M.~E.}\ \bibnamefont {Schwartz}}, \bibinfo {author} {\bibfnamefont {W.}~\bibnamefont {Woods}}, \bibinfo {author} {\bibfnamefont {J.~L.}\ \bibnamefont {Yoder}},\ and\ \bibinfo {author} {\bibfnamefont {W.~D.}\ \bibnamefont {Oliver}},\ }\bibfield  {title} {\bibinfo {title} {Solid-state qubits: 3d integration and packaging},\ }\href {https://doi.org/10.1109/MMM.2020.2993478} {\bibfield  {journal} {\bibinfo  {journal} {IEEE Microwave Magazine}\ }\textbf {\bibinfo
  {volume} {21}},\ \bibinfo {pages} {72} (\bibinfo {year} {2020})}\BibitemShut {NoStop}%
\bibitem [{\citenamefont {Niedzielski}\ \emph {et~al.}(2019)\citenamefont {Niedzielski}, \citenamefont {Kim}, \citenamefont {Schwartz}, \citenamefont {Rosenberg}, \citenamefont {Calusine}, \citenamefont {Das}, \citenamefont {Melville}, \citenamefont {Plant}, \citenamefont {Racz}, \citenamefont {Yoder}, \citenamefont {Ruth-Yost},\ and\ \citenamefont {Oliver}}]{niedzielski2019silicon}%
  \BibitemOpen
  \bibfield  {author} {\bibinfo {author} {\bibfnamefont {B.~M.}\ \bibnamefont {Niedzielski}}, \bibinfo {author} {\bibfnamefont {D.~K.}\ \bibnamefont {Kim}}, \bibinfo {author} {\bibfnamefont {M.~E.}\ \bibnamefont {Schwartz}}, \bibinfo {author} {\bibfnamefont {D.}~\bibnamefont {Rosenberg}}, \bibinfo {author} {\bibfnamefont {G.}~\bibnamefont {Calusine}}, \bibinfo {author} {\bibfnamefont {R.}~\bibnamefont {Das}}, \bibinfo {author} {\bibfnamefont {A.~J.}\ \bibnamefont {Melville}}, \bibinfo {author} {\bibfnamefont {J.}~\bibnamefont {Plant}}, \bibinfo {author} {\bibfnamefont {L.}~\bibnamefont {Racz}}, \bibinfo {author} {\bibfnamefont {J.~L.}\ \bibnamefont {Yoder}}, \bibinfo {author} {\bibfnamefont {D.}~\bibnamefont {Ruth-Yost}},\ and\ \bibinfo {author} {\bibfnamefont {W.~D.}\ \bibnamefont {Oliver}},\ }\bibfield  {title} {\bibinfo {title} {Silicon hard-stop spacers for 3d integration of superconducting qubits},\ }in\ \href {https://doi.org/10.1109/IEDM19573.2019.8993515} {\emph {\bibinfo {booktitle} {2019 IEEE
  International Electron Devices Meeting (IEDM)}}}\ (\bibinfo  {publisher} {IEEE},\ \bibinfo {address} {Piscataway, NJ, USA},\ \bibinfo {year} {2019})\ pp.\ \bibinfo {pages} {31.3.1--31.3.4}\BibitemShut {NoStop}%
\bibitem [{\citenamefont {Rahamim}\ \emph {et~al.}(2017)\citenamefont {Rahamim}, \citenamefont {Behrle}, \citenamefont {Peterer}, \citenamefont {Patterson}, \citenamefont {Spring}, \citenamefont {Tsunoda}, \citenamefont {Manenti}, \citenamefont {Tancredi},\ and\ \citenamefont {Leek}}]{rahamim2017double}%
  \BibitemOpen
  \bibfield  {author} {\bibinfo {author} {\bibfnamefont {J.}~\bibnamefont {Rahamim}}, \bibinfo {author} {\bibfnamefont {T.}~\bibnamefont {Behrle}}, \bibinfo {author} {\bibfnamefont {M.~J.}\ \bibnamefont {Peterer}}, \bibinfo {author} {\bibfnamefont {A.}~\bibnamefont {Patterson}}, \bibinfo {author} {\bibfnamefont {P.~A.}\ \bibnamefont {Spring}}, \bibinfo {author} {\bibfnamefont {T.}~\bibnamefont {Tsunoda}}, \bibinfo {author} {\bibfnamefont {R.}~\bibnamefont {Manenti}}, \bibinfo {author} {\bibfnamefont {G.}~\bibnamefont {Tancredi}},\ and\ \bibinfo {author} {\bibfnamefont {P.~J.}\ \bibnamefont {Leek}},\ }\bibfield  {title} {\bibinfo {title} {{Double-sided coaxial circuit QED with out-of-plane wiring}},\ }\href {https://doi.org/10.1063/1.4984299} {\bibfield  {journal} {\bibinfo  {journal} {Applied Physics Letters}\ }\textbf {\bibinfo {volume} {110}},\ \bibinfo {pages} {222602} (\bibinfo {year} {2017})},\ \Eprint
  {https://arxiv.org/abs/https://pubs.aip.org/aip/apl/article-pdf/doi/10.1063/1.4984299/14500580/222602\_1\_online.pdf} {https://pubs.aip.org/aip/apl/article-pdf/doi/10.1063/1.4984299/14500580/222602\_1\_online.pdf} \BibitemShut {NoStop}%
\bibitem [{\citenamefont {Wang}\ \emph {et~al.}(2021)\citenamefont {Wang}, \citenamefont {Luan}, \citenamefont {Qiao}, \citenamefont {Um}, \citenamefont {Zhang}, \citenamefont {Wang}, \citenamefont {Yuan}, \citenamefont {Gu}, \citenamefont {Zhang},\ and\ \citenamefont {Kim}}]{wang2021single}%
  \BibitemOpen
  \bibfield  {author} {\bibinfo {author} {\bibfnamefont {P.}~\bibnamefont {Wang}}, \bibinfo {author} {\bibfnamefont {C.-Y.}\ \bibnamefont {Luan}}, \bibinfo {author} {\bibfnamefont {M.}~\bibnamefont {Qiao}}, \bibinfo {author} {\bibfnamefont {M.}~\bibnamefont {Um}}, \bibinfo {author} {\bibfnamefont {J.}~\bibnamefont {Zhang}}, \bibinfo {author} {\bibfnamefont {Y.}~\bibnamefont {Wang}}, \bibinfo {author} {\bibfnamefont {X.}~\bibnamefont {Yuan}}, \bibinfo {author} {\bibfnamefont {M.}~\bibnamefont {Gu}}, \bibinfo {author} {\bibfnamefont {J.}~\bibnamefont {Zhang}},\ and\ \bibinfo {author} {\bibfnamefont {K.}~\bibnamefont {Kim}},\ }\bibfield  {title} {\bibinfo {title} {Single ion qubit with estimated coherence time exceeding one hour},\ }\href@noop {} {\bibfield  {journal} {\bibinfo  {journal} {Nature communications}\ }\textbf {\bibinfo {volume} {12}},\ \bibinfo {pages} {233} (\bibinfo {year} {2021})}\BibitemShut {NoStop}%
\bibitem [{\citenamefont {Brida}\ \emph {et~al.}(2012)\citenamefont {Brida}, \citenamefont {Degiovanni}, \citenamefont {Genovese}, \citenamefont {Piacentini}, \citenamefont {Traina}, \citenamefont {Della~Frera}, \citenamefont {Tosi}, \citenamefont {Bahgat~Shehata}, \citenamefont {Scarcella}, \citenamefont {Gulinatti}, \citenamefont {Ghioni}, \citenamefont {Polyakov}, \citenamefont {Migdall},\ and\ \citenamefont {Giudice}}]{brida2012an}%
  \BibitemOpen
  \bibfield  {author} {\bibinfo {author} {\bibfnamefont {G.}~\bibnamefont {Brida}}, \bibinfo {author} {\bibfnamefont {I.~P.}\ \bibnamefont {Degiovanni}}, \bibinfo {author} {\bibfnamefont {M.}~\bibnamefont {Genovese}}, \bibinfo {author} {\bibfnamefont {F.}~\bibnamefont {Piacentini}}, \bibinfo {author} {\bibfnamefont {P.}~\bibnamefont {Traina}}, \bibinfo {author} {\bibfnamefont {A.}~\bibnamefont {Della~Frera}}, \bibinfo {author} {\bibfnamefont {A.}~\bibnamefont {Tosi}}, \bibinfo {author} {\bibfnamefont {A.}~\bibnamefont {Bahgat~Shehata}}, \bibinfo {author} {\bibfnamefont {C.}~\bibnamefont {Scarcella}}, \bibinfo {author} {\bibfnamefont {A.}~\bibnamefont {Gulinatti}}, \bibinfo {author} {\bibfnamefont {M.}~\bibnamefont {Ghioni}}, \bibinfo {author} {\bibfnamefont {S.~V.}\ \bibnamefont {Polyakov}}, \bibinfo {author} {\bibfnamefont {A.}~\bibnamefont {Migdall}},\ and\ \bibinfo {author} {\bibfnamefont {A.}~\bibnamefont {Giudice}},\ }\bibfield  {title} {\bibinfo {title} {{An extremely low-noise heralded single-photon
  source: A breakthrough for quantum technologies}},\ }\href {https://doi.org/10.1063/1.4768288} {\bibfield  {journal} {\bibinfo  {journal} {Applied Physics Letters}\ }\textbf {\bibinfo {volume} {101}},\ \bibinfo {pages} {221112} (\bibinfo {year} {2012})},\ \Eprint {https://arxiv.org/abs/https://pubs.aip.org/aip/apl/article-pdf/doi/10.1063/1.4768288/14259225/221112\_1\_online.pdf} {https://pubs.aip.org/aip/apl/article-pdf/doi/10.1063/1.4768288/14259225/221112\_1\_online.pdf} \BibitemShut {NoStop}%
\bibitem [{\citenamefont {Preskill}(2018)}]{preskill2018quantum}%
  \BibitemOpen
  \bibfield  {author} {\bibinfo {author} {\bibfnamefont {J.}~\bibnamefont {Preskill}},\ }\bibfield  {title} {\bibinfo {title} {Quantum computing in the nisq era and beyond},\ }\href@noop {} {\bibfield  {journal} {\bibinfo  {journal} {Quantum}\ }\textbf {\bibinfo {volume} {2}},\ \bibinfo {pages} {79} (\bibinfo {year} {2018})}\BibitemShut {NoStop}%
\bibitem [{\citenamefont {Chandarana}\ \emph {et~al.}(2023)\citenamefont {Chandarana}, \citenamefont {Vieites}, \citenamefont {Hegade}, \citenamefont {Solano}, \citenamefont {Ban},\ and\ \citenamefont {Chen}}]{chandarana2023}%
  \BibitemOpen
  \bibfield  {author} {\bibinfo {author} {\bibfnamefont {P.}~\bibnamefont {Chandarana}}, \bibinfo {author} {\bibfnamefont {P.~S.}\ \bibnamefont {Vieites}}, \bibinfo {author} {\bibfnamefont {N.~N.}\ \bibnamefont {Hegade}}, \bibinfo {author} {\bibfnamefont {E.}~\bibnamefont {Solano}}, \bibinfo {author} {\bibfnamefont {Y.}~\bibnamefont {Ban}},\ and\ \bibinfo {author} {\bibfnamefont {X.}~\bibnamefont {Chen}},\ }\bibfield  {title} {\bibinfo {title} {Meta-learning digitized-counterdiabatic quantum optimization},\ }\href {https://doi.org/10.1088/2058-9565/ace54a} {\bibfield  {journal} {\bibinfo  {journal} {Quantum Science and Technology}\ }\textbf {\bibinfo {volume} {8}},\ \bibinfo {pages} {045007} (\bibinfo {year} {2023})}\BibitemShut {NoStop}%
\bibitem [{\citenamefont {Kandala}\ \emph {et~al.}(2017)\citenamefont {Kandala}, \citenamefont {Mezzacapo}, \citenamefont {Temme}, \citenamefont {Takita}, \citenamefont {Brink}, \citenamefont {Chow},\ and\ \citenamefont {Gambetta}}]{kandala2017hardware}%
  \BibitemOpen
  \bibfield  {author} {\bibinfo {author} {\bibfnamefont {A.}~\bibnamefont {Kandala}}, \bibinfo {author} {\bibfnamefont {A.}~\bibnamefont {Mezzacapo}}, \bibinfo {author} {\bibfnamefont {K.}~\bibnamefont {Temme}}, \bibinfo {author} {\bibfnamefont {M.}~\bibnamefont {Takita}}, \bibinfo {author} {\bibfnamefont {M.}~\bibnamefont {Brink}}, \bibinfo {author} {\bibfnamefont {J.~M.}\ \bibnamefont {Chow}},\ and\ \bibinfo {author} {\bibfnamefont {J.~M.}\ \bibnamefont {Gambetta}},\ }\bibfield  {title} {\bibinfo {title} {Hardware-efficient variational quantum eigensolver for small molecules and quantum magnets},\ }\href@noop {} {\bibfield  {journal} {\bibinfo  {journal} {Nature}\ }\textbf {\bibinfo {volume} {549}},\ \bibinfo {pages} {242} (\bibinfo {year} {2017})}\BibitemShut {NoStop}%
\bibitem [{\citenamefont {Ji}\ \emph {et~al.}(2022)\citenamefont {Ji}, \citenamefont {Brandhofer},\ and\ \citenamefont {Polian}}]{ji2022calibration}%
  \BibitemOpen
  \bibfield  {author} {\bibinfo {author} {\bibfnamefont {Y.}~\bibnamefont {Ji}}, \bibinfo {author} {\bibfnamefont {S.}~\bibnamefont {Brandhofer}},\ and\ \bibinfo {author} {\bibfnamefont {I.}~\bibnamefont {Polian}},\ }\bibfield  {title} {\bibinfo {title} {Calibration-aware transpilation for variational quantum optimization},\ }in\ \href {https://doi.org/10.1109/QCE53715.2022.00040} {\emph {\bibinfo {booktitle} {2022 IEEE International Conference on Quantum Computing and Engineering (QCE)}}}\ (\bibinfo  {publisher} {IEEE Computer Society},\ \bibinfo {address} {Los Alamitos, CA, USA},\ \bibinfo {year} {2022})\ pp.\ \bibinfo {pages} {204--214}\BibitemShut {NoStop}%
\bibitem [{\citenamefont {Gokhale}\ \emph {et~al.}(2020)\citenamefont {Gokhale}, \citenamefont {Javadi-Abhari}, \citenamefont {Earnest}, \citenamefont {Shi},\ and\ \citenamefont {Chong}}]{gokhale2020optimized}%
  \BibitemOpen
  \bibfield  {author} {\bibinfo {author} {\bibfnamefont {P.}~\bibnamefont {Gokhale}}, \bibinfo {author} {\bibfnamefont {A.}~\bibnamefont {Javadi-Abhari}}, \bibinfo {author} {\bibfnamefont {N.}~\bibnamefont {Earnest}}, \bibinfo {author} {\bibfnamefont {Y.}~\bibnamefont {Shi}},\ and\ \bibinfo {author} {\bibfnamefont {F.~T.}\ \bibnamefont {Chong}},\ }\bibfield  {title} {\bibinfo {title} {Optimized quantum compilation for near-term algorithms with openpulse},\ }in\ \href {https://doi.org/10.1109/MICRO50266.2020.00027} {\emph {\bibinfo {booktitle} {2020 53rd Annual IEEE/ACM International Symposium on Microarchitecture (MICRO)}}}\ (\bibinfo  {publisher} {IEEE Computer Society},\ \bibinfo {address} {Los Alamitos, CA, USA},\ \bibinfo {year} {2020})\ pp.\ \bibinfo {pages} {186--200}\BibitemShut {NoStop}%
\bibitem [{\citenamefont {Smith}\ \emph {et~al.}(2022)\citenamefont {Smith}, \citenamefont {Ravi}, \citenamefont {Alexander}, \citenamefont {Bronn}, \citenamefont {Carvalho}, \citenamefont {Cervera-Lierta}, \citenamefont {Chong}, \citenamefont {Chow}, \citenamefont {Cubeddu}, \citenamefont {Hashim}, \citenamefont {Jiang}, \citenamefont {Lanes}, \citenamefont {Otten}, \citenamefont {Schuster}, \citenamefont {Gokhale}, \citenamefont {Earnest},\ and\ \citenamefont {Galda}}]{smith2022summary}%
  \BibitemOpen
  \bibfield  {author} {\bibinfo {author} {\bibfnamefont {K.~N.}\ \bibnamefont {Smith}}, \bibinfo {author} {\bibfnamefont {G.~S.}\ \bibnamefont {Ravi}}, \bibinfo {author} {\bibfnamefont {T.}~\bibnamefont {Alexander}}, \bibinfo {author} {\bibfnamefont {N.~T.}\ \bibnamefont {Bronn}}, \bibinfo {author} {\bibfnamefont {A.}~\bibnamefont {Carvalho}}, \bibinfo {author} {\bibfnamefont {A.}~\bibnamefont {Cervera-Lierta}}, \bibinfo {author} {\bibfnamefont {F.~T.}\ \bibnamefont {Chong}}, \bibinfo {author} {\bibfnamefont {J.~M.}\ \bibnamefont {Chow}}, \bibinfo {author} {\bibfnamefont {M.}~\bibnamefont {Cubeddu}}, \bibinfo {author} {\bibfnamefont {A.}~\bibnamefont {Hashim}}, \bibinfo {author} {\bibfnamefont {L.}~\bibnamefont {Jiang}}, \bibinfo {author} {\bibfnamefont {O.}~\bibnamefont {Lanes}}, \bibinfo {author} {\bibfnamefont {M.~J.}\ \bibnamefont {Otten}}, \bibinfo {author} {\bibfnamefont {D.~I.}\ \bibnamefont {Schuster}}, \bibinfo {author} {\bibfnamefont {P.}~\bibnamefont {Gokhale}}, \bibinfo {author} {\bibfnamefont
  {N.}~\bibnamefont {Earnest}},\ and\ \bibinfo {author} {\bibfnamefont {A.}~\bibnamefont {Galda}},\ }\href {https://doi.org/10.48550/arXiv.2202.13600} {\bibinfo {title} {Summary: Chicago quantum exchange (cqe) pulse-level quantum control workshop}} (\bibinfo {year} {2022}),\ \Eprint {https://arxiv.org/abs/2202.13600} {arXiv:2202.13600} \BibitemShut {NoStop}%
\bibitem [{\citenamefont {Gokhale}\ \emph {et~al.}(2019)\citenamefont {Gokhale}, \citenamefont {Ding}, \citenamefont {Propson}, \citenamefont {Winkler}, \citenamefont {Leung}, \citenamefont {Shi}, \citenamefont {Schuster}, \citenamefont {Hoffmann},\ and\ \citenamefont {Chong}}]{gokhale2019partial}%
  \BibitemOpen
  \bibfield  {author} {\bibinfo {author} {\bibfnamefont {P.}~\bibnamefont {Gokhale}}, \bibinfo {author} {\bibfnamefont {Y.}~\bibnamefont {Ding}}, \bibinfo {author} {\bibfnamefont {T.}~\bibnamefont {Propson}}, \bibinfo {author} {\bibfnamefont {C.}~\bibnamefont {Winkler}}, \bibinfo {author} {\bibfnamefont {N.}~\bibnamefont {Leung}}, \bibinfo {author} {\bibfnamefont {Y.}~\bibnamefont {Shi}}, \bibinfo {author} {\bibfnamefont {D.~I.}\ \bibnamefont {Schuster}}, \bibinfo {author} {\bibfnamefont {H.}~\bibnamefont {Hoffmann}},\ and\ \bibinfo {author} {\bibfnamefont {F.~T.}\ \bibnamefont {Chong}},\ }\bibfield  {title} {\bibinfo {title} {Partial compilation of variational algorithms for noisy intermediate-scale quantum machines},\ }in\ \href {https://doi.org/10.1145/3352460.3358313} {\emph {\bibinfo {booktitle} {Proceedings of the 52nd Annual IEEE/ACM International Symposium on Microarchitecture}}},\ \bibinfo {series and number} {MICRO '52}\ (\bibinfo  {publisher} {Association for Computing Machinery},\ \bibinfo
  {address} {New York, NY, USA},\ \bibinfo {year} {2019})\ p.\ \bibinfo {pages} {266–278}\BibitemShut {NoStop}%
\bibitem [{\citenamefont {Shi}\ \emph {et~al.}(2019)\citenamefont {Shi}, \citenamefont {Leung}, \citenamefont {Gokhale}, \citenamefont {Rossi}, \citenamefont {Schuster}, \citenamefont {Hoffmann},\ and\ \citenamefont {Chong}}]{shi2019optimized}%
  \BibitemOpen
  \bibfield  {author} {\bibinfo {author} {\bibfnamefont {Y.}~\bibnamefont {Shi}}, \bibinfo {author} {\bibfnamefont {N.}~\bibnamefont {Leung}}, \bibinfo {author} {\bibfnamefont {P.}~\bibnamefont {Gokhale}}, \bibinfo {author} {\bibfnamefont {Z.}~\bibnamefont {Rossi}}, \bibinfo {author} {\bibfnamefont {D.~I.}\ \bibnamefont {Schuster}}, \bibinfo {author} {\bibfnamefont {H.}~\bibnamefont {Hoffmann}},\ and\ \bibinfo {author} {\bibfnamefont {F.~T.}\ \bibnamefont {Chong}},\ }\bibfield  {title} {\bibinfo {title} {Optimized compilation of aggregated instructions for realistic quantum computers},\ }in\ \href {https://doi.org/10.1145/3297858.3304018} {\emph {\bibinfo {booktitle} {Proceedings of the Twenty-Fourth International Conference on Architectural Support for Programming Languages and Operating Systems}}},\ \bibinfo {series and number} {ASPLOS '19}\ (\bibinfo  {publisher} {Association for Computing Machinery},\ \bibinfo {address} {New York, NY, USA},\ \bibinfo {year} {2019})\ p.\ \bibinfo {pages}
  {1031–1044}\BibitemShut {NoStop}%
\bibitem [{\citenamefont {Gokhale}\ \emph {et~al.}(2024)\citenamefont {Gokhale}, \citenamefont {Tomesh}, \citenamefont {Suchara},\ and\ \citenamefont {Chong}}]{gokhale2021faster}%
  \BibitemOpen
  \bibfield  {author} {\bibinfo {author} {\bibfnamefont {P.}~\bibnamefont {Gokhale}}, \bibinfo {author} {\bibfnamefont {T.}~\bibnamefont {Tomesh}}, \bibinfo {author} {\bibfnamefont {M.}~\bibnamefont {Suchara}},\ and\ \bibinfo {author} {\bibfnamefont {F.}~\bibnamefont {Chong}},\ }\bibfield  {title} {\bibinfo {title} {Faster and more reliable quantum swaps via native gates},\ }in\ \href {https://doi.org/10.1145/3656019.3689818} {\emph {\bibinfo {booktitle} {Proceedings of the 2024 International Conference on Parallel Architectures and Compilation Techniques}}},\ \bibinfo {series and number} {PACT '24}\ (\bibinfo  {publisher} {Association for Computing Machinery},\ \bibinfo {address} {New York, NY, USA},\ \bibinfo {year} {2024})\ p.\ \bibinfo {pages} {351–362}\BibitemShut {NoStop}%
\bibitem [{\citenamefont {Nation}\ and\ \citenamefont {Treinish}(2023)}]{nation2023suppressing}%
  \BibitemOpen
  \bibfield  {author} {\bibinfo {author} {\bibfnamefont {P.~D.}\ \bibnamefont {Nation}}\ and\ \bibinfo {author} {\bibfnamefont {M.}~\bibnamefont {Treinish}},\ }\bibfield  {title} {\bibinfo {title} {Suppressing quantum circuit errors due to system variability},\ }\href {https://doi.org/10.1103/PRXQuantum.4.010327} {\bibfield  {journal} {\bibinfo  {journal} {PRX Quantum}\ }\textbf {\bibinfo {volume} {4}},\ \bibinfo {pages} {010327} (\bibinfo {year} {2023})}\BibitemShut {NoStop}%
\bibitem [{\citenamefont {Ji}\ \emph {et~al.}(2024)\citenamefont {Ji}, \citenamefont {Koenig},\ and\ \citenamefont {Polian}}]{ji2023improving}%
  \BibitemOpen
  \bibfield  {author} {\bibinfo {author} {\bibfnamefont {Y.}~\bibnamefont {Ji}}, \bibinfo {author} {\bibfnamefont {K.~F.}\ \bibnamefont {Koenig}},\ and\ \bibinfo {author} {\bibfnamefont {I.}~\bibnamefont {Polian}},\ }\bibfield  {title} {\bibinfo {title} {Improving the performance of digitized counterdiabatic quantum optimization via algorithm-oriented qubit mapping},\ }\href {https://doi.org/10.1103/PhysRevA.110.032421} {\bibfield  {journal} {\bibinfo  {journal} {Phys. Rev. A}\ }\textbf {\bibinfo {volume} {110}},\ \bibinfo {pages} {032421} (\bibinfo {year} {2024})}\BibitemShut {NoStop}%
\bibitem [{\citenamefont {Brandhofer}\ \emph {et~al.}(2023)\citenamefont {Brandhofer}, \citenamefont {Braun}, \citenamefont {Dehn}, \citenamefont {Hellstern}, \citenamefont {H{\"u}ls}, \citenamefont {Ji}, \citenamefont {Polian}, \citenamefont {Bhatia},\ and\ \citenamefont {Wellens}}]{brandhofer2023benchmarking}%
  \BibitemOpen
  \bibfield  {author} {\bibinfo {author} {\bibfnamefont {S.}~\bibnamefont {Brandhofer}}, \bibinfo {author} {\bibfnamefont {D.}~\bibnamefont {Braun}}, \bibinfo {author} {\bibfnamefont {V.}~\bibnamefont {Dehn}}, \bibinfo {author} {\bibfnamefont {G.}~\bibnamefont {Hellstern}}, \bibinfo {author} {\bibfnamefont {M.}~\bibnamefont {H{\"u}ls}}, \bibinfo {author} {\bibfnamefont {Y.}~\bibnamefont {Ji}}, \bibinfo {author} {\bibfnamefont {I.}~\bibnamefont {Polian}}, \bibinfo {author} {\bibfnamefont {A.~S.}\ \bibnamefont {Bhatia}},\ and\ \bibinfo {author} {\bibfnamefont {T.}~\bibnamefont {Wellens}},\ }\bibfield  {title} {\bibinfo {title} {Benchmarking the performance of portfolio optimization with qaoa},\ }\href {https://doi.org/10.1007/s11128-022-03766-5} {\bibfield  {journal} {\bibinfo  {journal} {Quantum Information Processing}\ }\textbf {\bibinfo {volume} {22}},\ \bibinfo {pages} {1} (\bibinfo {year} {2023})}\BibitemShut {NoStop}%
\bibitem [{\citenamefont {Earnest}\ \emph {et~al.}(2021)\citenamefont {Earnest}, \citenamefont {Tornow},\ and\ \citenamefont {Egger}}]{earnest2021pulse}%
  \BibitemOpen
  \bibfield  {author} {\bibinfo {author} {\bibfnamefont {N.}~\bibnamefont {Earnest}}, \bibinfo {author} {\bibfnamefont {C.}~\bibnamefont {Tornow}},\ and\ \bibinfo {author} {\bibfnamefont {D.~J.}\ \bibnamefont {Egger}},\ }\bibfield  {title} {\bibinfo {title} {Pulse-efficient circuit transpilation for quantum applications on cross-resonance-based hardware},\ }\href {https://doi.org/10.1103/PhysRevResearch.3.043088} {\bibfield  {journal} {\bibinfo  {journal} {Phys. Rev. Res.}\ }\textbf {\bibinfo {volume} {3}},\ \bibinfo {pages} {043088} (\bibinfo {year} {2021})}\BibitemShut {NoStop}%
\bibitem [{\citenamefont {Ji}\ \emph {et~al.}(2023)\citenamefont {Ji}, \citenamefont {Koenig},\ and\ \citenamefont {Polian}}]{ji2023optimizing}%
  \BibitemOpen
  \bibfield  {author} {\bibinfo {author} {\bibfnamefont {Y.}~\bibnamefont {Ji}}, \bibinfo {author} {\bibfnamefont {K.~F.}\ \bibnamefont {Koenig}},\ and\ \bibinfo {author} {\bibfnamefont {I.}~\bibnamefont {Polian}},\ }\bibfield  {title} {\bibinfo {title} {Optimizing quantum algorithms on bipotent architectures},\ }\href {https://doi.org/10.1103/PhysRevA.108.022610} {\bibfield  {journal} {\bibinfo  {journal} {Phys. Rev. A}\ }\textbf {\bibinfo {volume} {108}},\ \bibinfo {pages} {022610} (\bibinfo {year} {2023})}\BibitemShut {NoStop}%
\bibitem [{\citenamefont {Childs}\ \emph {et~al.}(2019{\natexlab{b}})\citenamefont {Childs}, \citenamefont {Ostrander},\ and\ \citenamefont {Su}}]{childs2019faster}%
  \BibitemOpen
  \bibfield  {author} {\bibinfo {author} {\bibfnamefont {A.~M.}\ \bibnamefont {Childs}}, \bibinfo {author} {\bibfnamefont {A.}~\bibnamefont {Ostrander}},\ and\ \bibinfo {author} {\bibfnamefont {Y.}~\bibnamefont {Su}},\ }\bibfield  {title} {\bibinfo {title} {Faster quantum simulation by randomization},\ }\href@noop {} {\bibfield  {journal} {\bibinfo  {journal} {Quantum}\ }\textbf {\bibinfo {volume} {3}},\ \bibinfo {pages} {182} (\bibinfo {year} {2019}{\natexlab{b}})}\BibitemShut {NoStop}%
\bibitem [{\citenamefont {Faehrmann}\ \emph {et~al.}(2022)\citenamefont {Faehrmann}, \citenamefont {Steudtner}, \citenamefont {Kueng}, \citenamefont {Kieferova},\ and\ \citenamefont {Eisert}}]{faehrmann2022randomizing}%
  \BibitemOpen
  \bibfield  {author} {\bibinfo {author} {\bibfnamefont {P.~K.}\ \bibnamefont {Faehrmann}}, \bibinfo {author} {\bibfnamefont {M.}~\bibnamefont {Steudtner}}, \bibinfo {author} {\bibfnamefont {R.}~\bibnamefont {Kueng}}, \bibinfo {author} {\bibfnamefont {M.}~\bibnamefont {Kieferova}},\ and\ \bibinfo {author} {\bibfnamefont {J.}~\bibnamefont {Eisert}},\ }\bibfield  {title} {\bibinfo {title} {Randomizing multi-product formulas for hamiltonian simulation},\ }\href@noop {} {\bibfield  {journal} {\bibinfo  {journal} {Quantum}\ }\textbf {\bibinfo {volume} {6}},\ \bibinfo {pages} {806} (\bibinfo {year} {2022})}\BibitemShut {NoStop}%
\bibitem [{\citenamefont {Hagge}(2020)}]{hagge2020optimal}%
  \BibitemOpen
  \bibfield  {author} {\bibinfo {author} {\bibfnamefont {T.}~\bibnamefont {Hagge}},\ }\bibfield  {title} {\bibinfo {title} {Optimal fermionic swap networks for hubbard models},\ }\href@noop {} {\bibfield  {journal} {\bibinfo  {journal} {arXiv preprint arXiv:2001.08324}\ } (\bibinfo {year} {2020})}\BibitemShut {NoStop}%
\bibitem [{\citenamefont {Peham}\ \emph {et~al.}(2023)\citenamefont {Peham}, \citenamefont {Burgholzer},\ and\ \citenamefont {Wille}}]{peham2023equivalence}%
  \BibitemOpen
  \bibfield  {author} {\bibinfo {author} {\bibfnamefont {T.}~\bibnamefont {Peham}}, \bibinfo {author} {\bibfnamefont {L.}~\bibnamefont {Burgholzer}},\ and\ \bibinfo {author} {\bibfnamefont {R.}~\bibnamefont {Wille}},\ }\bibfield  {title} {\bibinfo {title} {Equivalence checking of parameterized quantum circuits: Verifying the compilation of variational quantum algorithms},\ }in\ \href {https://doi.org/10.1145/3566097.3567932} {\emph {\bibinfo {booktitle} {Proceedings of the 28th Asia and South Pacific Design Automation Conference}}},\ \bibinfo {series and number} {ASPDAC '23}\ (\bibinfo  {publisher} {Association for Computing Machinery},\ \bibinfo {address} {New York, NY, USA},\ \bibinfo {year} {2023})\ p.\ \bibinfo {pages} {702–708}\BibitemShut {NoStop}%
\bibitem [{\citenamefont {Mohar}\ and\ \citenamefont {Poljak}(1990)}]{mohar1990eigenvalues}%
  \BibitemOpen
  \bibfield  {author} {\bibinfo {author} {\bibfnamefont {B.}~\bibnamefont {Mohar}}\ and\ \bibinfo {author} {\bibfnamefont {S.}~\bibnamefont {Poljak}},\ }\bibfield  {title} {\bibinfo {title} {Eigenvalues and the max-cut problem},\ }\href@noop {} {\bibfield  {journal} {\bibinfo  {journal} {Czechoslovak Mathematical Journal}\ }\textbf {\bibinfo {volume} {40}},\ \bibinfo {pages} {343} (\bibinfo {year} {1990})}\BibitemShut {NoStop}%
\bibitem [{\citenamefont {Peruzzo}\ \emph {et~al.}(2014)\citenamefont {Peruzzo}, \citenamefont {McClean}, \citenamefont {Shadbolt}, \citenamefont {Yung}, \citenamefont {Zhou}, \citenamefont {Love}, \citenamefont {Aspuru-Guzik},\ and\ \citenamefont {O’brien}}]{peruzzo2014variational}%
  \BibitemOpen
  \bibfield  {author} {\bibinfo {author} {\bibfnamefont {A.}~\bibnamefont {Peruzzo}}, \bibinfo {author} {\bibfnamefont {J.}~\bibnamefont {McClean}}, \bibinfo {author} {\bibfnamefont {P.}~\bibnamefont {Shadbolt}}, \bibinfo {author} {\bibfnamefont {M.-H.}\ \bibnamefont {Yung}}, \bibinfo {author} {\bibfnamefont {X.-Q.}\ \bibnamefont {Zhou}}, \bibinfo {author} {\bibfnamefont {P.~J.}\ \bibnamefont {Love}}, \bibinfo {author} {\bibfnamefont {A.}~\bibnamefont {Aspuru-Guzik}},\ and\ \bibinfo {author} {\bibfnamefont {J.~L.}\ \bibnamefont {O’brien}},\ }\bibfield  {title} {\bibinfo {title} {A variational eigenvalue solver on a photonic quantum processor},\ }\href@noop {} {\bibfield  {journal} {\bibinfo  {journal} {Nature communications}\ }\textbf {\bibinfo {volume} {5}},\ \bibinfo {pages} {1} (\bibinfo {year} {2014})}\BibitemShut {NoStop}%
\bibitem [{\citenamefont {Tilly}\ \emph {et~al.}(2022)\citenamefont {Tilly}, \citenamefont {Chen}, \citenamefont {Cao}, \citenamefont {Picozzi}, \citenamefont {Setia}, \citenamefont {Li}, \citenamefont {Grant}, \citenamefont {Wossnig}, \citenamefont {Rungger}, \citenamefont {Booth} \emph {et~al.}}]{tilly2022variational}%
  \BibitemOpen
  \bibfield  {author} {\bibinfo {author} {\bibfnamefont {J.}~\bibnamefont {Tilly}}, \bibinfo {author} {\bibfnamefont {H.}~\bibnamefont {Chen}}, \bibinfo {author} {\bibfnamefont {S.}~\bibnamefont {Cao}}, \bibinfo {author} {\bibfnamefont {D.}~\bibnamefont {Picozzi}}, \bibinfo {author} {\bibfnamefont {K.}~\bibnamefont {Setia}}, \bibinfo {author} {\bibfnamefont {Y.}~\bibnamefont {Li}}, \bibinfo {author} {\bibfnamefont {E.}~\bibnamefont {Grant}}, \bibinfo {author} {\bibfnamefont {L.}~\bibnamefont {Wossnig}}, \bibinfo {author} {\bibfnamefont {I.}~\bibnamefont {Rungger}}, \bibinfo {author} {\bibfnamefont {G.~H.}\ \bibnamefont {Booth}}, \emph {et~al.},\ }\bibfield  {title} {\bibinfo {title} {The variational quantum eigensolver: a review of methods and best practices},\ }\href {https://doi.org/https://doi.org/10.1016/j.physrep.2022.08.003} {\bibfield  {journal} {\bibinfo  {journal} {Physics Reports}\ }\textbf {\bibinfo {volume} {986}},\ \bibinfo {pages} {1} (\bibinfo {year} {2022})}\BibitemShut {NoStop}%
\bibitem [{\citenamefont {Fedorov}\ \emph {et~al.}(2022)\citenamefont {Fedorov}, \citenamefont {Peng}, \citenamefont {Govind},\ and\ \citenamefont {Alexeev}}]{fedorov2022vqe}%
  \BibitemOpen
  \bibfield  {author} {\bibinfo {author} {\bibfnamefont {D.~A.}\ \bibnamefont {Fedorov}}, \bibinfo {author} {\bibfnamefont {B.}~\bibnamefont {Peng}}, \bibinfo {author} {\bibfnamefont {N.}~\bibnamefont {Govind}},\ and\ \bibinfo {author} {\bibfnamefont {Y.}~\bibnamefont {Alexeev}},\ }\bibfield  {title} {\bibinfo {title} {Vqe method: A short survey and recent developments},\ }\href@noop {} {\bibfield  {journal} {\bibinfo  {journal} {Materials Theory}\ }\textbf {\bibinfo {volume} {6}},\ \bibinfo {pages} {1} (\bibinfo {year} {2022})}\BibitemShut {NoStop}%
\bibitem [{\citenamefont {Cao}\ \emph {et~al.}(2019)\citenamefont {Cao}, \citenamefont {Romero}, \citenamefont {Olson}, \citenamefont {Degroote}, \citenamefont {Johnson}, \citenamefont {Kieferov{\'a}}, \citenamefont {Kivlichan}, \citenamefont {Menke}, \citenamefont {Peropadre}, \citenamefont {Sawaya} \emph {et~al.}}]{cao2019quantum}%
  \BibitemOpen
  \bibfield  {author} {\bibinfo {author} {\bibfnamefont {Y.}~\bibnamefont {Cao}}, \bibinfo {author} {\bibfnamefont {J.}~\bibnamefont {Romero}}, \bibinfo {author} {\bibfnamefont {J.~P.}\ \bibnamefont {Olson}}, \bibinfo {author} {\bibfnamefont {M.}~\bibnamefont {Degroote}}, \bibinfo {author} {\bibfnamefont {P.~D.}\ \bibnamefont {Johnson}}, \bibinfo {author} {\bibfnamefont {M.}~\bibnamefont {Kieferov{\'a}}}, \bibinfo {author} {\bibfnamefont {I.~D.}\ \bibnamefont {Kivlichan}}, \bibinfo {author} {\bibfnamefont {T.}~\bibnamefont {Menke}}, \bibinfo {author} {\bibfnamefont {B.}~\bibnamefont {Peropadre}}, \bibinfo {author} {\bibfnamefont {N.~P.}\ \bibnamefont {Sawaya}}, \emph {et~al.},\ }\bibfield  {title} {\bibinfo {title} {Quantum chemistry in the age of quantum computing},\ }\href@noop {} {\bibfield  {journal} {\bibinfo  {journal} {Chemical reviews}\ }\textbf {\bibinfo {volume} {119}},\ \bibinfo {pages} {10856} (\bibinfo {year} {2019})}\BibitemShut {NoStop}%
\bibitem [{\citenamefont {Bauer}\ \emph {et~al.}(2020)\citenamefont {Bauer}, \citenamefont {Bravyi}, \citenamefont {Motta},\ and\ \citenamefont {Chan}}]{bauer2020quantum}%
  \BibitemOpen
  \bibfield  {author} {\bibinfo {author} {\bibfnamefont {B.}~\bibnamefont {Bauer}}, \bibinfo {author} {\bibfnamefont {S.}~\bibnamefont {Bravyi}}, \bibinfo {author} {\bibfnamefont {M.}~\bibnamefont {Motta}},\ and\ \bibinfo {author} {\bibfnamefont {G.~K.-L.}\ \bibnamefont {Chan}},\ }\bibfield  {title} {\bibinfo {title} {Quantum algorithms for quantum chemistry and quantum materials science},\ }\href@noop {} {\bibfield  {journal} {\bibinfo  {journal} {Chemical Reviews}\ }\textbf {\bibinfo {volume} {120}},\ \bibinfo {pages} {12685} (\bibinfo {year} {2020})}\BibitemShut {NoStop}%
\bibitem [{\citenamefont {Amaro}\ \emph {et~al.}(2022)\citenamefont {Amaro}, \citenamefont {Modica}, \citenamefont {Rosenkranz}, \citenamefont {Fiorentini}, \citenamefont {Benedetti},\ and\ \citenamefont {Lubasch}}]{amaro2022filtering}%
  \BibitemOpen
  \bibfield  {author} {\bibinfo {author} {\bibfnamefont {D.}~\bibnamefont {Amaro}}, \bibinfo {author} {\bibfnamefont {C.}~\bibnamefont {Modica}}, \bibinfo {author} {\bibfnamefont {M.}~\bibnamefont {Rosenkranz}}, \bibinfo {author} {\bibfnamefont {M.}~\bibnamefont {Fiorentini}}, \bibinfo {author} {\bibfnamefont {M.}~\bibnamefont {Benedetti}},\ and\ \bibinfo {author} {\bibfnamefont {M.}~\bibnamefont {Lubasch}},\ }\bibfield  {title} {\bibinfo {title} {Filtering variational quantum algorithms for combinatorial optimization},\ }\href {https://doi.org/10.1088/2058-9565/ac3e54} {\bibfield  {journal} {\bibinfo  {journal} {Quantum Science and Technology}\ }\textbf {\bibinfo {volume} {7}},\ \bibinfo {pages} {015021} (\bibinfo {year} {2022})}\BibitemShut {NoStop}%
\bibitem [{\citenamefont {Romero}\ \emph {et~al.}(2018)\citenamefont {Romero}, \citenamefont {Babbush}, \citenamefont {McClean}, \citenamefont {Hempel}, \citenamefont {Love},\ and\ \citenamefont {Aspuru-Guzik}}]{romero2018strategies}%
  \BibitemOpen
  \bibfield  {author} {\bibinfo {author} {\bibfnamefont {J.}~\bibnamefont {Romero}}, \bibinfo {author} {\bibfnamefont {R.}~\bibnamefont {Babbush}}, \bibinfo {author} {\bibfnamefont {J.~R.}\ \bibnamefont {McClean}}, \bibinfo {author} {\bibfnamefont {C.}~\bibnamefont {Hempel}}, \bibinfo {author} {\bibfnamefont {P.~J.}\ \bibnamefont {Love}},\ and\ \bibinfo {author} {\bibfnamefont {A.}~\bibnamefont {Aspuru-Guzik}},\ }\bibfield  {title} {\bibinfo {title} {Strategies for quantum computing molecular energies using the unitary coupled cluster ansatz},\ }\href {https://doi.org/10.1088/2058-9565/aad3e4} {\bibfield  {journal} {\bibinfo  {journal} {Quantum Science and Technology}\ }\textbf {\bibinfo {volume} {4}},\ \bibinfo {pages} {014008} (\bibinfo {year} {2018})}\BibitemShut {NoStop}%
\bibitem [{\citenamefont {Wecker}\ \emph {et~al.}(2015)\citenamefont {Wecker}, \citenamefont {Hastings},\ and\ \citenamefont {Troyer}}]{wecker2015progress}%
  \BibitemOpen
  \bibfield  {author} {\bibinfo {author} {\bibfnamefont {D.}~\bibnamefont {Wecker}}, \bibinfo {author} {\bibfnamefont {M.~B.}\ \bibnamefont {Hastings}},\ and\ \bibinfo {author} {\bibfnamefont {M.}~\bibnamefont {Troyer}},\ }\bibfield  {title} {\bibinfo {title} {Progress towards practical quantum variational algorithms},\ }\href@noop {} {\bibfield  {journal} {\bibinfo  {journal} {Physical Review A}\ }\textbf {\bibinfo {volume} {92}},\ \bibinfo {pages} {042303} (\bibinfo {year} {2015})}\BibitemShut {NoStop}%
\bibitem [{\citenamefont {Nannicini}(2019)}]{nannicini2019performance}%
  \BibitemOpen
  \bibfield  {author} {\bibinfo {author} {\bibfnamefont {G.}~\bibnamefont {Nannicini}},\ }\bibfield  {title} {\bibinfo {title} {Performance of hybrid quantum-classical variational heuristics for combinatorial optimization},\ }\href {https://doi.org/10.1103/PhysRevE.99.013304} {\bibfield  {journal} {\bibinfo  {journal} {Phys. Rev. E}\ }\textbf {\bibinfo {volume} {99}},\ \bibinfo {pages} {013304} (\bibinfo {year} {2019})}\BibitemShut {NoStop}%
\bibitem [{\citenamefont {Ji}\ and\ \citenamefont {Polian}(2024)}]{ji2024synergistic}%
  \BibitemOpen
  \bibfield  {author} {\bibinfo {author} {\bibfnamefont {Y.}~\bibnamefont {Ji}}\ and\ \bibinfo {author} {\bibfnamefont {I.}~\bibnamefont {Polian}},\ }\bibfield  {title} {\bibinfo {title} {Synergistic dynamical decoupling and circuit design for enhanced algorithm performance on near-term quantum devices},\ }\bibfield  {journal} {\bibinfo  {journal} {Entropy}\ }\textbf {\bibinfo {volume} {26}},\ \href {https://doi.org/10.3390/e26070586} {10.3390/e26070586} (\bibinfo {year} {2024})\BibitemShut {NoStop}%
\bibitem [{\citenamefont {Arute}\ \emph {et~al.}(2019)\citenamefont {Arute}, \citenamefont {Arya}, \citenamefont {Babbush}, \citenamefont {Bacon}, \citenamefont {Bardin}, \citenamefont {Barends}, \citenamefont {Biswas}, \citenamefont {Boixo}, \citenamefont {Brandao}, \citenamefont {Buell} \emph {et~al.}}]{arute2019quantum}%
  \BibitemOpen
  \bibfield  {author} {\bibinfo {author} {\bibfnamefont {F.}~\bibnamefont {Arute}}, \bibinfo {author} {\bibfnamefont {K.}~\bibnamefont {Arya}}, \bibinfo {author} {\bibfnamefont {R.}~\bibnamefont {Babbush}}, \bibinfo {author} {\bibfnamefont {D.}~\bibnamefont {Bacon}}, \bibinfo {author} {\bibfnamefont {J.~C.}\ \bibnamefont {Bardin}}, \bibinfo {author} {\bibfnamefont {R.}~\bibnamefont {Barends}}, \bibinfo {author} {\bibfnamefont {R.}~\bibnamefont {Biswas}}, \bibinfo {author} {\bibfnamefont {S.}~\bibnamefont {Boixo}}, \bibinfo {author} {\bibfnamefont {F.~G.}\ \bibnamefont {Brandao}}, \bibinfo {author} {\bibfnamefont {D.~A.}\ \bibnamefont {Buell}}, \emph {et~al.},\ }\bibfield  {title} {\bibinfo {title} {Quantum supremacy using a programmable superconducting processor},\ }\href@noop {} {\bibfield  {journal} {\bibinfo  {journal} {Nature}\ }\textbf {\bibinfo {volume} {574}},\ \bibinfo {pages} {505} (\bibinfo {year} {2019})}\BibitemShut {NoStop}%
\bibitem [{\citenamefont {Otterbach}\ \emph {et~al.}(2017)\citenamefont {Otterbach}, \citenamefont {Manenti}, \citenamefont {Alidoust}, \citenamefont {Bestwick}, \citenamefont {Block}, \citenamefont {Bloom}, \citenamefont {Caldwell}, \citenamefont {Didier}, \citenamefont {Fried}, \citenamefont {Hong} \emph {et~al.}}]{otterbach2017unsupervised}%
  \BibitemOpen
  \bibfield  {author} {\bibinfo {author} {\bibfnamefont {J.~S.}\ \bibnamefont {Otterbach}}, \bibinfo {author} {\bibfnamefont {R.}~\bibnamefont {Manenti}}, \bibinfo {author} {\bibfnamefont {N.}~\bibnamefont {Alidoust}}, \bibinfo {author} {\bibfnamefont {A.}~\bibnamefont {Bestwick}}, \bibinfo {author} {\bibfnamefont {M.}~\bibnamefont {Block}}, \bibinfo {author} {\bibfnamefont {B.}~\bibnamefont {Bloom}}, \bibinfo {author} {\bibfnamefont {S.}~\bibnamefont {Caldwell}}, \bibinfo {author} {\bibfnamefont {N.}~\bibnamefont {Didier}}, \bibinfo {author} {\bibfnamefont {E.~S.}\ \bibnamefont {Fried}}, \bibinfo {author} {\bibfnamefont {S.}~\bibnamefont {Hong}}, \emph {et~al.},\ }\bibfield  {title} {\bibinfo {title} {Unsupervised machine learning on a hybrid quantum computer},\ }\href@noop {} {\bibfield  {journal} {\bibinfo  {journal} {arXiv preprint arXiv:1712.05771}\ } (\bibinfo {year} {2017})}\BibitemShut {NoStop}%
\bibitem [{\citenamefont {Monroe}\ and\ \citenamefont {Kim}(2013)}]{monroe2013scaling}%
  \BibitemOpen
  \bibfield  {author} {\bibinfo {author} {\bibfnamefont {C.}~\bibnamefont {Monroe}}\ and\ \bibinfo {author} {\bibfnamefont {J.}~\bibnamefont {Kim}},\ }\bibfield  {title} {\bibinfo {title} {Scaling the ion trap quantum processor},\ }\href {https://doi.org/10.1126/science.1231298} {\bibfield  {journal} {\bibinfo  {journal} {Science}\ }\textbf {\bibinfo {volume} {339}},\ \bibinfo {pages} {1164} (\bibinfo {year} {2013})}\BibitemShut {NoStop}%
\bibitem [{\citenamefont {Pogorelov}\ \emph {et~al.}(2021)\citenamefont {Pogorelov}, \citenamefont {Feldker}, \citenamefont {Marciniak}, \citenamefont {Postler}, \citenamefont {Jacob}, \citenamefont {Krieglsteiner}, \citenamefont {Podlesnic}, \citenamefont {Meth}, \citenamefont {Negnevitsky}, \citenamefont {Stadler}, \citenamefont {H\"ofer}, \citenamefont {W\"achter}, \citenamefont {Lakhmanskiy}, \citenamefont {Blatt}, \citenamefont {Schindler},\ and\ \citenamefont {Monz}}]{pogorelov2021compact}%
  \BibitemOpen
  \bibfield  {author} {\bibinfo {author} {\bibfnamefont {I.}~\bibnamefont {Pogorelov}}, \bibinfo {author} {\bibfnamefont {T.}~\bibnamefont {Feldker}}, \bibinfo {author} {\bibfnamefont {C.~D.}\ \bibnamefont {Marciniak}}, \bibinfo {author} {\bibfnamefont {L.}~\bibnamefont {Postler}}, \bibinfo {author} {\bibfnamefont {G.}~\bibnamefont {Jacob}}, \bibinfo {author} {\bibfnamefont {O.}~\bibnamefont {Krieglsteiner}}, \bibinfo {author} {\bibfnamefont {V.}~\bibnamefont {Podlesnic}}, \bibinfo {author} {\bibfnamefont {M.}~\bibnamefont {Meth}}, \bibinfo {author} {\bibfnamefont {V.}~\bibnamefont {Negnevitsky}}, \bibinfo {author} {\bibfnamefont {M.}~\bibnamefont {Stadler}}, \bibinfo {author} {\bibfnamefont {B.}~\bibnamefont {H\"ofer}}, \bibinfo {author} {\bibfnamefont {C.}~\bibnamefont {W\"achter}}, \bibinfo {author} {\bibfnamefont {K.}~\bibnamefont {Lakhmanskiy}}, \bibinfo {author} {\bibfnamefont {R.}~\bibnamefont {Blatt}}, \bibinfo {author} {\bibfnamefont {P.}~\bibnamefont {Schindler}},\ and\ \bibinfo {author}
  {\bibfnamefont {T.}~\bibnamefont {Monz}},\ }\bibfield  {title} {\bibinfo {title} {Compact ion-trap quantum computing demonstrator},\ }\href {https://doi.org/10.1103/PRXQuantum.2.020343} {\bibfield  {journal} {\bibinfo  {journal} {PRX Quantum}\ }\textbf {\bibinfo {volume} {2}},\ \bibinfo {pages} {020343} (\bibinfo {year} {2021})}\BibitemShut {NoStop}%
\bibitem [{\citenamefont {Ramette}\ \emph {et~al.}(2022)\citenamefont {Ramette}, \citenamefont {Sinclair}, \citenamefont {Vendeiro}, \citenamefont {Rudelis}, \citenamefont {Cetina},\ and\ \citenamefont {Vuleti\ifmmode~\acute{c}\else \'{c}\fi{}}}]{ramette2022any}%
  \BibitemOpen
  \bibfield  {author} {\bibinfo {author} {\bibfnamefont {J.}~\bibnamefont {Ramette}}, \bibinfo {author} {\bibfnamefont {J.}~\bibnamefont {Sinclair}}, \bibinfo {author} {\bibfnamefont {Z.}~\bibnamefont {Vendeiro}}, \bibinfo {author} {\bibfnamefont {A.}~\bibnamefont {Rudelis}}, \bibinfo {author} {\bibfnamefont {M.}~\bibnamefont {Cetina}},\ and\ \bibinfo {author} {\bibfnamefont {V.}~\bibnamefont {Vuleti\ifmmode~\acute{c}\else \'{c}\fi{}}},\ }\bibfield  {title} {\bibinfo {title} {Any-to-any connected cavity-mediated architecture for quantum computing with trapped ions or rydberg arrays},\ }\href {https://doi.org/10.1103/PRXQuantum.3.010344} {\bibfield  {journal} {\bibinfo  {journal} {PRX Quantum}\ }\textbf {\bibinfo {volume} {3}},\ \bibinfo {pages} {010344} (\bibinfo {year} {2022})}\BibitemShut {NoStop}%
\bibitem [{\citenamefont {Cetina}\ \emph {et~al.}(2022)\citenamefont {Cetina}, \citenamefont {Egan}, \citenamefont {Noel}, \citenamefont {Goldman}, \citenamefont {Biswas}, \citenamefont {Risinger}, \citenamefont {Zhu},\ and\ \citenamefont {Monroe}}]{cetina2022control}%
  \BibitemOpen
  \bibfield  {author} {\bibinfo {author} {\bibfnamefont {M.}~\bibnamefont {Cetina}}, \bibinfo {author} {\bibfnamefont {L.}~\bibnamefont {Egan}}, \bibinfo {author} {\bibfnamefont {C.}~\bibnamefont {Noel}}, \bibinfo {author} {\bibfnamefont {M.}~\bibnamefont {Goldman}}, \bibinfo {author} {\bibfnamefont {D.}~\bibnamefont {Biswas}}, \bibinfo {author} {\bibfnamefont {A.}~\bibnamefont {Risinger}}, \bibinfo {author} {\bibfnamefont {D.}~\bibnamefont {Zhu}},\ and\ \bibinfo {author} {\bibfnamefont {C.}~\bibnamefont {Monroe}},\ }\bibfield  {title} {\bibinfo {title} {Control of transverse motion for quantum gates on individually addressed atomic qubits},\ }\href {https://doi.org/10.1103/PRXQuantum.3.010334} {\bibfield  {journal} {\bibinfo  {journal} {PRX Quantum}\ }\textbf {\bibinfo {volume} {3}},\ \bibinfo {pages} {010334} (\bibinfo {year} {2022})}\BibitemShut {NoStop}%
\bibitem [{\citenamefont {Politi}\ \emph {et~al.}(2008)\citenamefont {Politi}, \citenamefont {Cryan}, \citenamefont {Rarity}, \citenamefont {Yu},\ and\ \citenamefont {O'Brien}}]{politi2008silica}%
  \BibitemOpen
  \bibfield  {author} {\bibinfo {author} {\bibfnamefont {A.}~\bibnamefont {Politi}}, \bibinfo {author} {\bibfnamefont {M.~J.}\ \bibnamefont {Cryan}}, \bibinfo {author} {\bibfnamefont {J.~G.}\ \bibnamefont {Rarity}}, \bibinfo {author} {\bibfnamefont {S.}~\bibnamefont {Yu}},\ and\ \bibinfo {author} {\bibfnamefont {J.~L.}\ \bibnamefont {O'Brien}},\ }\bibfield  {title} {\bibinfo {title} {Silica-on-silicon waveguide quantum circuits},\ }\href {https://doi.org/10.1126/science.1155441} {\bibfield  {journal} {\bibinfo  {journal} {Science}\ }\textbf {\bibinfo {volume} {320}},\ \bibinfo {pages} {646} (\bibinfo {year} {2008})},\ \Eprint {https://arxiv.org/abs/https://www.science.org/doi/pdf/10.1126/science.1155441} {https://www.science.org/doi/pdf/10.1126/science.1155441} \BibitemShut {NoStop}%
\bibitem [{\citenamefont {O'brien}\ \emph {et~al.}(2009)\citenamefont {O'brien}, \citenamefont {Furusawa},\ and\ \citenamefont {Vu{\v{c}}kovi{\'c}}}]{o2009photonic}%
  \BibitemOpen
  \bibfield  {author} {\bibinfo {author} {\bibfnamefont {J.~L.}\ \bibnamefont {O'brien}}, \bibinfo {author} {\bibfnamefont {A.}~\bibnamefont {Furusawa}},\ and\ \bibinfo {author} {\bibfnamefont {J.}~\bibnamefont {Vu{\v{c}}kovi{\'c}}},\ }\bibfield  {title} {\bibinfo {title} {Photonic quantum technologies},\ }\href@noop {} {\bibfield  {journal} {\bibinfo  {journal} {Nature Photonics}\ }\textbf {\bibinfo {volume} {3}},\ \bibinfo {pages} {687} (\bibinfo {year} {2009})}\BibitemShut {NoStop}%
\bibitem [{\citenamefont {Wang}\ \emph {et~al.}(2020)\citenamefont {Wang}, \citenamefont {Sciarrino}, \citenamefont {Laing},\ and\ \citenamefont {Thompson}}]{wang2020integrated}%
  \BibitemOpen
  \bibfield  {author} {\bibinfo {author} {\bibfnamefont {J.}~\bibnamefont {Wang}}, \bibinfo {author} {\bibfnamefont {F.}~\bibnamefont {Sciarrino}}, \bibinfo {author} {\bibfnamefont {A.}~\bibnamefont {Laing}},\ and\ \bibinfo {author} {\bibfnamefont {M.~G.}\ \bibnamefont {Thompson}},\ }\bibfield  {title} {\bibinfo {title} {Integrated photonic quantum technologies},\ }\href@noop {} {\bibfield  {journal} {\bibinfo  {journal} {Nature Photonics}\ }\textbf {\bibinfo {volume} {14}},\ \bibinfo {pages} {273} (\bibinfo {year} {2020})}\BibitemShut {NoStop}%
\bibitem [{\citenamefont {Luo}\ \emph {et~al.}(2023)\citenamefont {Luo}, \citenamefont {Cao}, \citenamefont {Shi}, \citenamefont {Wan}, \citenamefont {Zhang}, \citenamefont {Li}, \citenamefont {Chen}, \citenamefont {Li}, \citenamefont {Li}, \citenamefont {Wang}, \citenamefont {Sun}, \citenamefont {Karim}, \citenamefont {Cai}, \citenamefont {Kwek},\ and\ \citenamefont {Liu}}]{luo2023recent}%
  \BibitemOpen
  \bibfield  {author} {\bibinfo {author} {\bibfnamefont {W.}~\bibnamefont {Luo}}, \bibinfo {author} {\bibfnamefont {L.}~\bibnamefont {Cao}}, \bibinfo {author} {\bibfnamefont {Y.}~\bibnamefont {Shi}}, \bibinfo {author} {\bibfnamefont {L.}~\bibnamefont {Wan}}, \bibinfo {author} {\bibfnamefont {H.}~\bibnamefont {Zhang}}, \bibinfo {author} {\bibfnamefont {S.}~\bibnamefont {Li}}, \bibinfo {author} {\bibfnamefont {G.}~\bibnamefont {Chen}}, \bibinfo {author} {\bibfnamefont {Y.}~\bibnamefont {Li}}, \bibinfo {author} {\bibfnamefont {S.}~\bibnamefont {Li}}, \bibinfo {author} {\bibfnamefont {Y.}~\bibnamefont {Wang}}, \bibinfo {author} {\bibfnamefont {S.}~\bibnamefont {Sun}}, \bibinfo {author} {\bibfnamefont {M.~F.}\ \bibnamefont {Karim}}, \bibinfo {author} {\bibfnamefont {H.}~\bibnamefont {Cai}}, \bibinfo {author} {\bibfnamefont {L.~C.}\ \bibnamefont {Kwek}},\ and\ \bibinfo {author} {\bibfnamefont {A.~Q.}\ \bibnamefont {Liu}},\ }\bibfield  {title} {\bibinfo {title} {Recent progress in quantum photonic chips for quantum
  communication and internet},\ }\bibfield  {journal} {\bibinfo  {journal} {Light, Science \& Applications}\ }\textbf {\bibinfo {volume} {12}},\ \href {https://doi.org/10.1038/s41377-023-01173-8} {10.1038/s41377-023-01173-8} (\bibinfo {year} {2023})\BibitemShut {NoStop}%
\bibitem [{\citenamefont {Knill}\ \emph {et~al.}(2001)\citenamefont {Knill}, \citenamefont {Laflamme},\ and\ \citenamefont {Milburn}}]{knill2001scheme}%
  \BibitemOpen
  \bibfield  {author} {\bibinfo {author} {\bibfnamefont {E.}~\bibnamefont {Knill}}, \bibinfo {author} {\bibfnamefont {R.}~\bibnamefont {Laflamme}},\ and\ \bibinfo {author} {\bibfnamefont {G.~J.}\ \bibnamefont {Milburn}},\ }\bibfield  {title} {\bibinfo {title} {A scheme for efficient quantum computation with linear optics},\ }\href@noop {} {\bibfield  {journal} {\bibinfo  {journal} {nature}\ }\textbf {\bibinfo {volume} {409}},\ \bibinfo {pages} {46} (\bibinfo {year} {2001})}\BibitemShut {NoStop}%
\bibitem [{\citenamefont {Raussendorf}\ and\ \citenamefont {Briegel}(2001)}]{raussendorf2001one}%
  \BibitemOpen
  \bibfield  {author} {\bibinfo {author} {\bibfnamefont {R.}~\bibnamefont {Raussendorf}}\ and\ \bibinfo {author} {\bibfnamefont {H.~J.}\ \bibnamefont {Briegel}},\ }\bibfield  {title} {\bibinfo {title} {A one-way quantum computer},\ }\href {https://doi.org/10.1103/PhysRevLett.86.5188} {\bibfield  {journal} {\bibinfo  {journal} {Phys. Rev. Lett.}\ }\textbf {\bibinfo {volume} {86}},\ \bibinfo {pages} {5188} (\bibinfo {year} {2001})}\BibitemShut {NoStop}%
\bibitem [{\citenamefont {Walther}\ \emph {et~al.}(2005)\citenamefont {Walther}, \citenamefont {Resch}, \citenamefont {Rudolph}, \citenamefont {Schenck}, \citenamefont {Weinfurter}, \citenamefont {Vedral}, \citenamefont {Aspelmeyer},\ and\ \citenamefont {Zeilinger}}]{walther2005experimental}%
  \BibitemOpen
  \bibfield  {author} {\bibinfo {author} {\bibfnamefont {P.}~\bibnamefont {Walther}}, \bibinfo {author} {\bibfnamefont {K.~J.}\ \bibnamefont {Resch}}, \bibinfo {author} {\bibfnamefont {T.}~\bibnamefont {Rudolph}}, \bibinfo {author} {\bibfnamefont {E.}~\bibnamefont {Schenck}}, \bibinfo {author} {\bibfnamefont {H.}~\bibnamefont {Weinfurter}}, \bibinfo {author} {\bibfnamefont {V.}~\bibnamefont {Vedral}}, \bibinfo {author} {\bibfnamefont {M.}~\bibnamefont {Aspelmeyer}},\ and\ \bibinfo {author} {\bibfnamefont {A.}~\bibnamefont {Zeilinger}},\ }\bibfield  {title} {\bibinfo {title} {Experimental one-way quantum computing},\ }\href@noop {} {\bibfield  {journal} {\bibinfo  {journal} {Nature}\ }\textbf {\bibinfo {volume} {434}},\ \bibinfo {pages} {169} (\bibinfo {year} {2005})}\BibitemShut {NoStop}%
\bibitem [{\citenamefont {Menicucci}\ \emph {et~al.}(2008)\citenamefont {Menicucci}, \citenamefont {Flammia},\ and\ \citenamefont {Pfister}}]{menicucci2008one}%
  \BibitemOpen
  \bibfield  {author} {\bibinfo {author} {\bibfnamefont {N.~C.}\ \bibnamefont {Menicucci}}, \bibinfo {author} {\bibfnamefont {S.~T.}\ \bibnamefont {Flammia}},\ and\ \bibinfo {author} {\bibfnamefont {O.}~\bibnamefont {Pfister}},\ }\bibfield  {title} {\bibinfo {title} {One-way quantum computing in the optical frequency comb},\ }\href {https://doi.org/10.1103/PhysRevLett.101.130501} {\bibfield  {journal} {\bibinfo  {journal} {Phys. Rev. Lett.}\ }\textbf {\bibinfo {volume} {101}},\ \bibinfo {pages} {130501} (\bibinfo {year} {2008})}\BibitemShut {NoStop}%
\bibitem [{\citenamefont {Zilk}\ \emph {et~al.}(2022)\citenamefont {Zilk}, \citenamefont {Staudacher}, \citenamefont {Guggemos}, \citenamefont {Fürlinger}, \citenamefont {Kranzlmüller},\ and\ \citenamefont {Walther}}]{zilk2022compiler}%
  \BibitemOpen
  \bibfield  {author} {\bibinfo {author} {\bibfnamefont {F.}~\bibnamefont {Zilk}}, \bibinfo {author} {\bibfnamefont {K.}~\bibnamefont {Staudacher}}, \bibinfo {author} {\bibfnamefont {T.}~\bibnamefont {Guggemos}}, \bibinfo {author} {\bibfnamefont {K.}~\bibnamefont {Fürlinger}}, \bibinfo {author} {\bibfnamefont {D.}~\bibnamefont {Kranzlmüller}},\ and\ \bibinfo {author} {\bibfnamefont {P.}~\bibnamefont {Walther}},\ }\bibfield  {title} {\bibinfo {title} {A compiler for universal photonic quantum computers},\ }in\ \href {https://doi.org/10.1109/QCS56647.2022.00012} {\emph {\bibinfo {booktitle} {2022 IEEE/ACM Third International Workshop on Quantum Computing Software (QCS)}}}\ (\bibinfo  {publisher} {IEEE},\ \bibinfo {address} {Los Alamitos, CA, USA},\ \bibinfo {year} {2022})\ pp.\ \bibinfo {pages} {57--67}\BibitemShut {NoStop}%
\bibitem [{\citenamefont {Zhang}\ \emph {et~al.}(2023)\citenamefont {Zhang}, \citenamefont {Wu}, \citenamefont {Wang}, \citenamefont {Li}, \citenamefont {Shapourian}, \citenamefont {Shabani},\ and\ \citenamefont {Ding}}]{zhang2023oneq}%
  \BibitemOpen
  \bibfield  {author} {\bibinfo {author} {\bibfnamefont {H.}~\bibnamefont {Zhang}}, \bibinfo {author} {\bibfnamefont {A.}~\bibnamefont {Wu}}, \bibinfo {author} {\bibfnamefont {Y.}~\bibnamefont {Wang}}, \bibinfo {author} {\bibfnamefont {G.}~\bibnamefont {Li}}, \bibinfo {author} {\bibfnamefont {H.}~\bibnamefont {Shapourian}}, \bibinfo {author} {\bibfnamefont {A.}~\bibnamefont {Shabani}},\ and\ \bibinfo {author} {\bibfnamefont {Y.}~\bibnamefont {Ding}},\ }\bibfield  {title} {\bibinfo {title} {Oneq: A compilation framework for photonic one-way quantum computation},\ }in\ \href {https://doi.org/10.1145/3579371.3589047} {\emph {\bibinfo {booktitle} {Proceedings of the 50th Annual International Symposium on Computer Architecture}}},\ \bibinfo {series and number} {ISCA '23}\ (\bibinfo  {publisher} {Association for Computing Machinery},\ \bibinfo {address} {New York, NY, USA},\ \bibinfo {year} {2023})\BibitemShut {NoStop}%
\bibitem [{\citenamefont {Zhang}\ \emph {et~al.}(2024)\citenamefont {Zhang}, \citenamefont {Ruan}, \citenamefont {Shapourian}, \citenamefont {Kompella},\ and\ \citenamefont {Ding}}]{zhang2024oneperc}%
  \BibitemOpen
  \bibfield  {author} {\bibinfo {author} {\bibfnamefont {H.}~\bibnamefont {Zhang}}, \bibinfo {author} {\bibfnamefont {J.}~\bibnamefont {Ruan}}, \bibinfo {author} {\bibfnamefont {H.}~\bibnamefont {Shapourian}}, \bibinfo {author} {\bibfnamefont {R.~R.}\ \bibnamefont {Kompella}},\ and\ \bibinfo {author} {\bibfnamefont {Y.}~\bibnamefont {Ding}},\ }\bibfield  {title} {\bibinfo {title} {Oneperc: A randomness-aware compiler for photonic quantum computing},\ }in\ \href {https://doi.org/10.1145/3620666.3651372} {\emph {\bibinfo {booktitle} {Proceedings of the 29th ACM International Conference on Architectural Support for Programming Languages and Operating Systems, Volume 3}}},\ \bibinfo {series and number} {ASPLOS '24}\ (\bibinfo  {publisher} {Association for Computing Machinery},\ \bibinfo {address} {New York, NY, USA},\ \bibinfo {year} {2024})\ p.\ \bibinfo {pages} {738–754}\BibitemShut {NoStop}%
\bibitem [{\citenamefont {Agresti}\ \emph {et~al.}(2024)\citenamefont {Agresti}, \citenamefont {Paul}, \citenamefont {Schiansky}, \citenamefont {Steiner}, \citenamefont {Yin}, \citenamefont {Pentangelo}, \citenamefont {Piacentini}, \citenamefont {Crespi}, \citenamefont {Ban}, \citenamefont {Ceccarelli} \emph {et~al.}}]{agresti2024demonstration}%
  \BibitemOpen
  \bibfield  {author} {\bibinfo {author} {\bibfnamefont {I.}~\bibnamefont {Agresti}}, \bibinfo {author} {\bibfnamefont {K.}~\bibnamefont {Paul}}, \bibinfo {author} {\bibfnamefont {P.}~\bibnamefont {Schiansky}}, \bibinfo {author} {\bibfnamefont {S.}~\bibnamefont {Steiner}}, \bibinfo {author} {\bibfnamefont {Z.}~\bibnamefont {Yin}}, \bibinfo {author} {\bibfnamefont {C.}~\bibnamefont {Pentangelo}}, \bibinfo {author} {\bibfnamefont {S.}~\bibnamefont {Piacentini}}, \bibinfo {author} {\bibfnamefont {A.}~\bibnamefont {Crespi}}, \bibinfo {author} {\bibfnamefont {Y.}~\bibnamefont {Ban}}, \bibinfo {author} {\bibfnamefont {F.}~\bibnamefont {Ceccarelli}}, \emph {et~al.},\ }\bibfield  {title} {\bibinfo {title} {Demonstration of hardware efficient photonic variational quantum algorithm},\ }\href@noop {} {\bibfield  {journal} {\bibinfo  {journal} {arXiv preprint arXiv:2408.10339}\ } (\bibinfo {year} {2024})}\BibitemShut {NoStop}%
\bibitem [{\citenamefont {Mitarai}\ \emph {et~al.}(2018)\citenamefont {Mitarai}, \citenamefont {Negoro}, \citenamefont {Kitagawa},\ and\ \citenamefont {Fujii}}]{mitarai2018quantum}%
  \BibitemOpen
  \bibfield  {author} {\bibinfo {author} {\bibfnamefont {K.}~\bibnamefont {Mitarai}}, \bibinfo {author} {\bibfnamefont {M.}~\bibnamefont {Negoro}}, \bibinfo {author} {\bibfnamefont {M.}~\bibnamefont {Kitagawa}},\ and\ \bibinfo {author} {\bibfnamefont {K.}~\bibnamefont {Fujii}},\ }\bibfield  {title} {\bibinfo {title} {Quantum circuit learning},\ }\href {https://doi.org/10.1103/PhysRevA.98.032309} {\bibfield  {journal} {\bibinfo  {journal} {Phys. Rev. A}\ }\textbf {\bibinfo {volume} {98}},\ \bibinfo {pages} {032309} (\bibinfo {year} {2018})}\BibitemShut {NoStop}%
\bibitem [{\citenamefont {Benedetti}\ \emph {et~al.}(2019)\citenamefont {Benedetti}, \citenamefont {Lloyd}, \citenamefont {Sack},\ and\ \citenamefont {Fiorentini}}]{benedetti2019parameterized}%
  \BibitemOpen
  \bibfield  {author} {\bibinfo {author} {\bibfnamefont {M.}~\bibnamefont {Benedetti}}, \bibinfo {author} {\bibfnamefont {E.}~\bibnamefont {Lloyd}}, \bibinfo {author} {\bibfnamefont {S.}~\bibnamefont {Sack}},\ and\ \bibinfo {author} {\bibfnamefont {M.}~\bibnamefont {Fiorentini}},\ }\bibfield  {title} {\bibinfo {title} {Parameterized quantum circuits as machine learning models},\ }\href {https://doi.org/10.1088/2058-9565/ab4eb5} {\bibfield  {journal} {\bibinfo  {journal} {Quantum Science and Technology}\ }\textbf {\bibinfo {volume} {4}},\ \bibinfo {pages} {043001} (\bibinfo {year} {2019})}\BibitemShut {NoStop}%
\bibitem [{\citenamefont {Schuld}\ \emph {et~al.}(2020)\citenamefont {Schuld}, \citenamefont {Bocharov}, \citenamefont {Svore},\ and\ \citenamefont {Wiebe}}]{schuld2020circuit}%
  \BibitemOpen
  \bibfield  {author} {\bibinfo {author} {\bibfnamefont {M.}~\bibnamefont {Schuld}}, \bibinfo {author} {\bibfnamefont {A.}~\bibnamefont {Bocharov}}, \bibinfo {author} {\bibfnamefont {K.~M.}\ \bibnamefont {Svore}},\ and\ \bibinfo {author} {\bibfnamefont {N.}~\bibnamefont {Wiebe}},\ }\bibfield  {title} {\bibinfo {title} {Circuit-centric quantum classifiers},\ }\href {https://doi.org/10.1103/PhysRevA.101.032308} {\bibfield  {journal} {\bibinfo  {journal} {Phys. Rev. A}\ }\textbf {\bibinfo {volume} {101}},\ \bibinfo {pages} {032308} (\bibinfo {year} {2020})}\BibitemShut {NoStop}%
\bibitem [{\citenamefont {Meyer}\ \emph {et~al.}(2023)\citenamefont {Meyer}, \citenamefont {Mularski}, \citenamefont {Gil-Fuster}, \citenamefont {Mele}, \citenamefont {Arzani}, \citenamefont {Wilms},\ and\ \citenamefont {Eisert}}]{meyer2023exploiting}%
  \BibitemOpen
  \bibfield  {author} {\bibinfo {author} {\bibfnamefont {J.~J.}\ \bibnamefont {Meyer}}, \bibinfo {author} {\bibfnamefont {M.}~\bibnamefont {Mularski}}, \bibinfo {author} {\bibfnamefont {E.}~\bibnamefont {Gil-Fuster}}, \bibinfo {author} {\bibfnamefont {A.~A.}\ \bibnamefont {Mele}}, \bibinfo {author} {\bibfnamefont {F.}~\bibnamefont {Arzani}}, \bibinfo {author} {\bibfnamefont {A.}~\bibnamefont {Wilms}},\ and\ \bibinfo {author} {\bibfnamefont {J.}~\bibnamefont {Eisert}},\ }\bibfield  {title} {\bibinfo {title} {Exploiting symmetry in variational quantum machine learning},\ }\href {https://doi.org/10.1103/PRXQuantum.4.010328} {\bibfield  {journal} {\bibinfo  {journal} {PRX Quantum}\ }\textbf {\bibinfo {volume} {4}},\ \bibinfo {pages} {010328} (\bibinfo {year} {2023})}\BibitemShut {NoStop}%
\bibitem [{\citenamefont {Grover}(1996)}]{grover1996fast}%
  \BibitemOpen
  \bibfield  {author} {\bibinfo {author} {\bibfnamefont {L.~K.}\ \bibnamefont {Grover}},\ }\bibfield  {title} {\bibinfo {title} {A fast quantum mechanical algorithm for database search},\ }in\ \href {https://doi.org/10.1145/237814.237866} {\emph {\bibinfo {booktitle} {Proceedings of the Twenty-Eighth Annual ACM Symposium on Theory of Computing}}},\ \bibinfo {series and number} {STOC '96}\ (\bibinfo  {publisher} {Association for Computing Machinery},\ \bibinfo {address} {New York, NY, USA},\ \bibinfo {year} {1996})\ p.\ \bibinfo {pages} {212–219}\BibitemShut {NoStop}%
\bibitem [{\citenamefont {Powell}(1994)}]{powell1994direct}%
  \BibitemOpen
  \bibfield  {author} {\bibinfo {author} {\bibfnamefont {M.~J.~D.}\ \bibnamefont {Powell}},\ }\bibinfo {title} {A direct search optimization method that models the objective and constraint functions by linear interpolation},\ in\ \href {https://doi.org/10.1007/978-94-015-8330-5_4} {\emph {\bibinfo {booktitle} {Advances in Optimization and Numerical Analysis}}},\ \bibinfo {editor} {edited by\ \bibinfo {editor} {\bibfnamefont {S.}~\bibnamefont {Gomez}}\ and\ \bibinfo {editor} {\bibfnamefont {J.-P.}\ \bibnamefont {Hennart}}}\ (\bibinfo  {publisher} {Springer Netherlands},\ \bibinfo {address} {Dordrecht},\ \bibinfo {year} {1994})\ pp.\ \bibinfo {pages} {51--67}\BibitemShut {NoStop}%
\bibitem [{\citenamefont {Georgopoulos}\ \emph {et~al.}(2021)\citenamefont {Georgopoulos}, \citenamefont {Emary},\ and\ \citenamefont {Zuliani}}]{georgopoulos2021modeling}%
  \BibitemOpen
  \bibfield  {author} {\bibinfo {author} {\bibfnamefont {K.}~\bibnamefont {Georgopoulos}}, \bibinfo {author} {\bibfnamefont {C.}~\bibnamefont {Emary}},\ and\ \bibinfo {author} {\bibfnamefont {P.}~\bibnamefont {Zuliani}},\ }\bibfield  {title} {\bibinfo {title} {Modeling and simulating the noisy behavior of near-term quantum computers},\ }\href@noop {} {\bibfield  {journal} {\bibinfo  {journal} {Physical Review A}\ }\textbf {\bibinfo {volume} {104}},\ \bibinfo {pages} {062432} (\bibinfo {year} {2021})}\BibitemShut {NoStop}%
\bibitem [{\citenamefont {Zeng}\ \emph {et~al.}(2021)\citenamefont {Zeng}, \citenamefont {Wu}, \citenamefont {Cao}, \citenamefont {Zhang}, \citenamefont {Hou}, \citenamefont {Xu},\ and\ \citenamefont {Zeng}}]{zeng2021simulating}%
  \BibitemOpen
  \bibfield  {author} {\bibinfo {author} {\bibfnamefont {J.}~\bibnamefont {Zeng}}, \bibinfo {author} {\bibfnamefont {Z.}~\bibnamefont {Wu}}, \bibinfo {author} {\bibfnamefont {C.}~\bibnamefont {Cao}}, \bibinfo {author} {\bibfnamefont {C.}~\bibnamefont {Zhang}}, \bibinfo {author} {\bibfnamefont {S.-Y.}\ \bibnamefont {Hou}}, \bibinfo {author} {\bibfnamefont {P.}~\bibnamefont {Xu}},\ and\ \bibinfo {author} {\bibfnamefont {B.}~\bibnamefont {Zeng}},\ }\bibfield  {title} {\bibinfo {title} {Simulating noisy variational quantum eigensolver with local noise models},\ }\href@noop {} {\bibfield  {journal} {\bibinfo  {journal} {Quantum Engineering}\ }\textbf {\bibinfo {volume} {3}},\ \bibinfo {pages} {e77} (\bibinfo {year} {2021})}\BibitemShut {NoStop}%
\bibitem [{\citenamefont {Dahlhauser}\ and\ \citenamefont {Humble}(2021)}]{dahlhauser2021modeling}%
  \BibitemOpen
  \bibfield  {author} {\bibinfo {author} {\bibfnamefont {M.~L.}\ \bibnamefont {Dahlhauser}}\ and\ \bibinfo {author} {\bibfnamefont {T.~S.}\ \bibnamefont {Humble}},\ }\bibfield  {title} {\bibinfo {title} {Modeling noisy quantum circuits using experimental characterization},\ }\href {https://doi.org/10.1103/PhysRevA.103.042603} {\bibfield  {journal} {\bibinfo  {journal} {Phys. Rev. A}\ }\textbf {\bibinfo {volume} {103}},\ \bibinfo {pages} {042603} (\bibinfo {year} {2021})}\BibitemShut {NoStop}%
\bibitem [{\citenamefont {Dahlhauser}\ and\ \citenamefont {Humble}(2024)}]{dahlhauser2024benchmarking}%
  \BibitemOpen
  \bibfield  {author} {\bibinfo {author} {\bibfnamefont {M.~L.}\ \bibnamefont {Dahlhauser}}\ and\ \bibinfo {author} {\bibfnamefont {T.~S.}\ \bibnamefont {Humble}},\ }\bibfield  {title} {\bibinfo {title} {Benchmarking characterization methods for noisy quantum circuits},\ }\href {https://doi.org/10.1103/PhysRevA.109.042620} {\bibfield  {journal} {\bibinfo  {journal} {Phys. Rev. A}\ }\textbf {\bibinfo {volume} {109}},\ \bibinfo {pages} {042620} (\bibinfo {year} {2024})}\BibitemShut {NoStop}%
\bibitem [{aoq()}]{aoqmap2024}%
  \BibitemOpen
  \href@noop {} {}\bibinfo {howpublished} {\url{https://github.com/QuantumYanjunJi/AOQMAP}}\BibitemShut {NoStop}%
\bibitem [{\citenamefont {Hellinger}(1909)}]{hellinger1909neue}%
  \BibitemOpen
  \bibfield  {author} {\bibinfo {author} {\bibfnamefont {E.}~\bibnamefont {Hellinger}},\ }\bibfield  {title} {\bibinfo {title} {Neue begr{\"u}ndung der theorie quadratischer formen von unendlichvielen ver{\"a}nderlichen.},\ }\href@noop {} {\bibfield  {journal} {\bibinfo  {journal} {Journal f{\"u}r die reine und angewandte Mathematik}\ }\textbf {\bibinfo {volume} {1909}},\ \bibinfo {pages} {210} (\bibinfo {year} {1909})}\BibitemShut {NoStop}%
\bibitem [{\citenamefont {Leibfried}\ \emph {et~al.}(1996)\citenamefont {Leibfried}, \citenamefont {Meekhof}, \citenamefont {King}, \citenamefont {Monroe}, \citenamefont {Itano},\ and\ \citenamefont {Wineland}}]{leibfried1996experimental}%
  \BibitemOpen
  \bibfield  {author} {\bibinfo {author} {\bibfnamefont {D.}~\bibnamefont {Leibfried}}, \bibinfo {author} {\bibfnamefont {D.~M.}\ \bibnamefont {Meekhof}}, \bibinfo {author} {\bibfnamefont {B.~E.}\ \bibnamefont {King}}, \bibinfo {author} {\bibfnamefont {C.}~\bibnamefont {Monroe}}, \bibinfo {author} {\bibfnamefont {W.~M.}\ \bibnamefont {Itano}},\ and\ \bibinfo {author} {\bibfnamefont {D.~J.}\ \bibnamefont {Wineland}},\ }\bibfield  {title} {\bibinfo {title} {Experimental determination of the motional quantum state of a trapped atom},\ }\href {https://doi.org/10.1103/PhysRevLett.77.4281} {\bibfield  {journal} {\bibinfo  {journal} {Phys. Rev. Lett.}\ }\textbf {\bibinfo {volume} {77}},\ \bibinfo {pages} {4281} (\bibinfo {year} {1996})}\BibitemShut {NoStop}%
\bibitem [{\citenamefont {Cramer}\ \emph {et~al.}(2010)\citenamefont {Cramer}, \citenamefont {Plenio}, \citenamefont {Flammia}, \citenamefont {Somma}, \citenamefont {Gross}, \citenamefont {Bartlett}, \citenamefont {Landon-Cardinal}, \citenamefont {Poulin},\ and\ \citenamefont {Liu}}]{cramer2010efficient}%
  \BibitemOpen
  \bibfield  {author} {\bibinfo {author} {\bibfnamefont {M.}~\bibnamefont {Cramer}}, \bibinfo {author} {\bibfnamefont {M.~B.}\ \bibnamefont {Plenio}}, \bibinfo {author} {\bibfnamefont {S.~T.}\ \bibnamefont {Flammia}}, \bibinfo {author} {\bibfnamefont {R.}~\bibnamefont {Somma}}, \bibinfo {author} {\bibfnamefont {D.}~\bibnamefont {Gross}}, \bibinfo {author} {\bibfnamefont {S.~D.}\ \bibnamefont {Bartlett}}, \bibinfo {author} {\bibfnamefont {O.}~\bibnamefont {Landon-Cardinal}}, \bibinfo {author} {\bibfnamefont {D.}~\bibnamefont {Poulin}},\ and\ \bibinfo {author} {\bibfnamefont {Y.-K.}\ \bibnamefont {Liu}},\ }\bibfield  {title} {\bibinfo {title} {Efficient quantum state tomography},\ }\href@noop {} {\bibfield  {journal} {\bibinfo  {journal} {Nature communications}\ }\textbf {\bibinfo {volume} {1}},\ \bibinfo {pages} {149} (\bibinfo {year} {2010})}\BibitemShut {NoStop}%
\bibitem [{\citenamefont {Gross}\ \emph {et~al.}(2010)\citenamefont {Gross}, \citenamefont {Liu}, \citenamefont {Flammia}, \citenamefont {Becker},\ and\ \citenamefont {Eisert}}]{gross2010quantum}%
  \BibitemOpen
  \bibfield  {author} {\bibinfo {author} {\bibfnamefont {D.}~\bibnamefont {Gross}}, \bibinfo {author} {\bibfnamefont {Y.-K.}\ \bibnamefont {Liu}}, \bibinfo {author} {\bibfnamefont {S.~T.}\ \bibnamefont {Flammia}}, \bibinfo {author} {\bibfnamefont {S.}~\bibnamefont {Becker}},\ and\ \bibinfo {author} {\bibfnamefont {J.}~\bibnamefont {Eisert}},\ }\bibfield  {title} {\bibinfo {title} {Quantum state tomography via compressed sensing},\ }\href {https://doi.org/10.1103/PhysRevLett.105.150401} {\bibfield  {journal} {\bibinfo  {journal} {Phys. Rev. Lett.}\ }\textbf {\bibinfo {volume} {105}},\ \bibinfo {pages} {150401} (\bibinfo {year} {2010})}\BibitemShut {NoStop}%
\bibitem [{\citenamefont {Christandl}\ and\ \citenamefont {Renner}(2012)}]{christandl2012reliable}%
  \BibitemOpen
  \bibfield  {author} {\bibinfo {author} {\bibfnamefont {M.}~\bibnamefont {Christandl}}\ and\ \bibinfo {author} {\bibfnamefont {R.}~\bibnamefont {Renner}},\ }\bibfield  {title} {\bibinfo {title} {Reliable quantum state tomography},\ }\href {https://doi.org/10.1103/PhysRevLett.109.120403} {\bibfield  {journal} {\bibinfo  {journal} {Phys. Rev. Lett.}\ }\textbf {\bibinfo {volume} {109}},\ \bibinfo {pages} {120403} (\bibinfo {year} {2012})}\BibitemShut {NoStop}%
\bibitem [{\citenamefont {Poyatos}\ \emph {et~al.}(1997)\citenamefont {Poyatos}, \citenamefont {Cirac},\ and\ \citenamefont {Zoller}}]{poyatos1997complete}%
  \BibitemOpen
  \bibfield  {author} {\bibinfo {author} {\bibfnamefont {J.~F.}\ \bibnamefont {Poyatos}}, \bibinfo {author} {\bibfnamefont {J.~I.}\ \bibnamefont {Cirac}},\ and\ \bibinfo {author} {\bibfnamefont {P.}~\bibnamefont {Zoller}},\ }\bibfield  {title} {\bibinfo {title} {Complete characterization of a quantum process: The two-bit quantum gate},\ }\href {https://doi.org/10.1103/PhysRevLett.78.390} {\bibfield  {journal} {\bibinfo  {journal} {Phys. Rev. Lett.}\ }\textbf {\bibinfo {volume} {78}},\ \bibinfo {pages} {390} (\bibinfo {year} {1997})}\BibitemShut {NoStop}%
\bibitem [{\citenamefont {Mohseni}\ \emph {et~al.}(2008)\citenamefont {Mohseni}, \citenamefont {Rezakhani},\ and\ \citenamefont {Lidar}}]{mohseni2008quantum}%
  \BibitemOpen
  \bibfield  {author} {\bibinfo {author} {\bibfnamefont {M.}~\bibnamefont {Mohseni}}, \bibinfo {author} {\bibfnamefont {A.~T.}\ \bibnamefont {Rezakhani}},\ and\ \bibinfo {author} {\bibfnamefont {D.~A.}\ \bibnamefont {Lidar}},\ }\bibfield  {title} {\bibinfo {title} {Quantum-process tomography: Resource analysis of different strategies},\ }\href {https://doi.org/10.1103/PhysRevA.77.032322} {\bibfield  {journal} {\bibinfo  {journal} {Phys. Rev. A}\ }\textbf {\bibinfo {volume} {77}},\ \bibinfo {pages} {032322} (\bibinfo {year} {2008})}\BibitemShut {NoStop}%
\bibitem [{\citenamefont {Magesan}\ \emph {et~al.}(2011)\citenamefont {Magesan}, \citenamefont {Gambetta},\ and\ \citenamefont {Emerson}}]{magesan2011scalable}%
  \BibitemOpen
  \bibfield  {author} {\bibinfo {author} {\bibfnamefont {E.}~\bibnamefont {Magesan}}, \bibinfo {author} {\bibfnamefont {J.~M.}\ \bibnamefont {Gambetta}},\ and\ \bibinfo {author} {\bibfnamefont {J.}~\bibnamefont {Emerson}},\ }\bibfield  {title} {\bibinfo {title} {Scalable and robust randomized benchmarking of quantum processes},\ }\href {https://doi.org/10.1103/PhysRevLett.106.180504} {\bibfield  {journal} {\bibinfo  {journal} {Phys. Rev. Lett.}\ }\textbf {\bibinfo {volume} {106}},\ \bibinfo {pages} {180504} (\bibinfo {year} {2011})}\BibitemShut {NoStop}%
\bibitem [{\citenamefont {Knill}\ \emph {et~al.}(2008)\citenamefont {Knill}, \citenamefont {Leibfried}, \citenamefont {Reichle}, \citenamefont {Britton}, \citenamefont {Blakestad}, \citenamefont {Jost}, \citenamefont {Langer}, \citenamefont {Ozeri}, \citenamefont {Seidelin},\ and\ \citenamefont {Wineland}}]{knill2008randomized}%
  \BibitemOpen
  \bibfield  {author} {\bibinfo {author} {\bibfnamefont {E.}~\bibnamefont {Knill}}, \bibinfo {author} {\bibfnamefont {D.}~\bibnamefont {Leibfried}}, \bibinfo {author} {\bibfnamefont {R.}~\bibnamefont {Reichle}}, \bibinfo {author} {\bibfnamefont {J.}~\bibnamefont {Britton}}, \bibinfo {author} {\bibfnamefont {R.~B.}\ \bibnamefont {Blakestad}}, \bibinfo {author} {\bibfnamefont {J.~D.}\ \bibnamefont {Jost}}, \bibinfo {author} {\bibfnamefont {C.}~\bibnamefont {Langer}}, \bibinfo {author} {\bibfnamefont {R.}~\bibnamefont {Ozeri}}, \bibinfo {author} {\bibfnamefont {S.}~\bibnamefont {Seidelin}},\ and\ \bibinfo {author} {\bibfnamefont {D.~J.}\ \bibnamefont {Wineland}},\ }\bibfield  {title} {\bibinfo {title} {Randomized benchmarking of quantum gates},\ }\href {https://doi.org/10.1103/PhysRevA.77.012307} {\bibfield  {journal} {\bibinfo  {journal} {Phys. Rev. A}\ }\textbf {\bibinfo {volume} {77}},\ \bibinfo {pages} {012307} (\bibinfo {year} {2008})}\BibitemShut {NoStop}%
\bibitem [{\citenamefont {Helsen}\ \emph {et~al.}(2022)\citenamefont {Helsen}, \citenamefont {Roth}, \citenamefont {Onorati}, \citenamefont {Werner},\ and\ \citenamefont {Eisert}}]{helsen2022general}%
  \BibitemOpen
  \bibfield  {author} {\bibinfo {author} {\bibfnamefont {J.}~\bibnamefont {Helsen}}, \bibinfo {author} {\bibfnamefont {I.}~\bibnamefont {Roth}}, \bibinfo {author} {\bibfnamefont {E.}~\bibnamefont {Onorati}}, \bibinfo {author} {\bibfnamefont {A.}~\bibnamefont {Werner}},\ and\ \bibinfo {author} {\bibfnamefont {J.}~\bibnamefont {Eisert}},\ }\bibfield  {title} {\bibinfo {title} {General framework for randomized benchmarking},\ }\href {https://doi.org/10.1103/PRXQuantum.3.020357} {\bibfield  {journal} {\bibinfo  {journal} {PRX Quantum}\ }\textbf {\bibinfo {volume} {3}},\ \bibinfo {pages} {020357} (\bibinfo {year} {2022})}\BibitemShut {NoStop}%
\bibitem [{\citenamefont {Maciejewski}\ \emph {et~al.}(2020)\citenamefont {Maciejewski}, \citenamefont {Zimbor{\'{a}}s},\ and\ \citenamefont {Oszmaniec}}]{maciejewski2020mitigation}%
  \BibitemOpen
  \bibfield  {author} {\bibinfo {author} {\bibfnamefont {F.~B.}\ \bibnamefont {Maciejewski}}, \bibinfo {author} {\bibfnamefont {Z.}~\bibnamefont {Zimbor{\'{a}}s}},\ and\ \bibinfo {author} {\bibfnamefont {M.}~\bibnamefont {Oszmaniec}},\ }\bibfield  {title} {\bibinfo {title} {Mitigation of readout noise in near-term quantum devices by classical post-processing based on detector tomography},\ }\href {https://doi.org/10.22331/q-2020-04-24-257} {\bibfield  {journal} {\bibinfo  {journal} {{Quantum}}\ }\textbf {\bibinfo {volume} {4}},\ \bibinfo {pages} {257} (\bibinfo {year} {2020})}\BibitemShut {NoStop}%
\bibitem [{\citenamefont {Takeuchi}\ \emph {et~al.}(2022)\citenamefont {Takeuchi}, \citenamefont {Takahashi}, \citenamefont {Morimae},\ and\ \citenamefont {Tani}}]{takeuchi2022divide}%
  \BibitemOpen
  \bibfield  {author} {\bibinfo {author} {\bibfnamefont {Y.}~\bibnamefont {Takeuchi}}, \bibinfo {author} {\bibfnamefont {Y.}~\bibnamefont {Takahashi}}, \bibinfo {author} {\bibfnamefont {T.}~\bibnamefont {Morimae}},\ and\ \bibinfo {author} {\bibfnamefont {S.}~\bibnamefont {Tani}},\ }\bibfield  {title} {\bibinfo {title} {Divide-and-conquer verification method for noisy intermediate-scale quantum computation},\ }\href {https://doi.org/10.22331/q-2022-07-07-758} {\bibfield  {journal} {\bibinfo  {journal} {{Quantum}}\ }\textbf {\bibinfo {volume} {6}},\ \bibinfo {pages} {758} (\bibinfo {year} {2022})}\BibitemShut {NoStop}%
\end{thebibliography}%

\end{document}